\documentclass[11pt]{article}
\pdfoutput=1
\usepackage{amsmath,amssymb,mathtools}
\usepackage{graphicx}
\usepackage{units}
\usepackage{url}
\usepackage{booktabs}
\usepackage{subfigure}
\usepackage{hyperref}

\usepackage[sc,small]{caption}
\usepackage{color}

\DeclareFontFamily{U}{mathx}{\hyphenchar\font45}
\DeclareFontShape{U}{mathx}{m}{n}{<-> mathx10}{}
\DeclareSymbolFont{mathx}{U}{mathx}{m}{n}
\DeclareMathAccent{\wb}{0}{mathx}{"73}

\newcommand{\mathscr}{\mathcal}

\newcommand{\be}{\begin{eqnarray}}
\newcommand{\ee}{\end{eqnarray}}
\newcommand{\bea}{\begin{eqnarray}}
\newcommand{\eea}{\end{eqnarray}}

\newcommand{\J}{\mathbf{J}}
\newcommand{\tht}{\vartheta}

\newcommand{\A}{{\mathcal A}}
\newcommand{\K}{{\mathcal K}}
\newcommand{\M}{{\mathcal M}}
\newcommand{\N}{{\mathcal N}}

\newcommand{\Z}{{\mathcal Z}}
\newcommand{\I}{{\mathcal I}}

\newcommand{\thbw}[2]{\vartheta[\genfrac{}{}{0pt}{}{#1}{#2} ]}
\newcommand{\thbwp}[2]{\vartheta'[\genfrac{}{}{0pt}{}{#1}{#2} ]}

\newcommand{\thba}[2]{\vartheta[\!\!\begin{array}{c}{\phantom{}\vspace{-0.5mm}\scriptstyle#1}%
                        \\[-1.8mm]{\scriptstyle #2}\end{array}\!\!]}

\newcommand{\ba}[2]{[\!\!\begin{array}{c}{\scriptstyle#1}%
                        \\[-1.6mm]{\scriptstyle #2}\end{array}\!\!]}                         
\newcommand{\beqn}{\begin{eqnarray}}
\newcommand{\eeqn}{\end{eqnarray}}

\newcommand{\tq}{{\tilde q}}

\newcommand{\ca}{{\cal A}}
\newcommand{\ck}{{\cal K}}

\newcommand{\cn}{{\cal N}}
\newcommand{\cm}{{\cal M}}

\newcommand{\ct}{{\cal T}}

\newcommand{\T}{\mathcal{T}}
\newcommand{\ImT}{\tau}
\newcommand{\ImS}{\sigma}

\renewcommand{\t}[2]{\vartheta [{\tiny \begin{matrix} #1 \\ #2 \end{matrix} }]}

\newcommand{\p}{\partial}

\newcommand{\dd}{\text{d}}

\renewcommand{\ss}[2]{\Big[\genfrac{}{}{0pt}{1}{#1}{#2}\Big]}

\newcommand{\Ts}{\mathcal{T}}
\newcommand{\nn}{ \nonumber }
\newcommand{\hf}{ { \frac{1}{2} } }
\newcommand{\thf}{ { \textstyle \frac{1}{2} } }

\newcommand{\vev}[1]{\langle #1 \rangle}

\newcommand{\gammamatrix}{\boldsymbol{\gamma}}

\renewcommand{\sf}[1]{{\mathsf{#1}}}

\renewcommand{\refeq}[1]{\text{eq.~(\ref{#1})}}
\newcommand{\refEq}[1]{\text{Eq.~(\ref{#1})}}
\newcommand{\refeqs}[2]{\text{eqs.~(\ref{#1})--(\ref{#2})}}
\newcommand{\refsec}[1]{\S\ref{#1}}
\newcommand{\refsecs}[2]{\S\ref{#1}-\ref{#2}}
\newcommand{\refapp}[1]{\S\ref{#1}}
\newcommand{\reftab}[1]{table \ref{#1}}
\newcommand{\reffig}[1]{fig.~\ref{#1}}

\newcommand{\ts}[1]{{\textstyle #1}}

\newcommand{\Ref}[1]{(\ref{#1})}
\newcommand{\non}{\nonumber \\}
\newcommand{\zba}[2]{[\!\!\begin{array}{c}{\scriptstyle#1}%
                        \\[-1.6mm]{\scriptstyle #2}\end{array}\!\!]}

\newcommand{\fourfermion}{{\substack{\text{Four}~~~~\, \\ \text{fermions}}}}
\newcommand{\eightfermion}{{\substack{\text{Eight} ~~~\,\, \\ \text{fermions}}}}

\DeclareMathOperator{\Si}{\mathrm{Si}}

\DeclareMathOperator{\Cl}{\mathrm{Cl}}

\DeclareMathOperator{\tr}{\mathrm{tr}}

\let\Re\relax
\DeclareMathOperator{\Re}{\mathrm{Re}}
\let\Im\relax
\DeclareMathOperator{\Im}{\mathrm{Im}}
\DeclareMathOperator{\Res}{\mathrm{Res}}

\newcommand{\wt}[1]{\widetilde{#1}}
\newcommand{\bs}[1]{\boldsymbol{#1}}
\newcommand{\vt}{\vartheta}

\begin{document}

\hfill LMU-ASC 43/14\\
%\vspace{1cm}

\begin{center}
{\bf\LARGE
Towards the One-loop K\"ahler Metric\\ of 
Calabi-Yau Orientifolds \\  }

\vspace{1.5cm}

{\large
{\bf Marcus Berg$^{+}$, }
{\bf Michael Haack$^{\dag}$, }
{\bf Jin U Kang$^{\star}$}, 
{\bf Stefan Sj\"ors$^{\heartsuit}$}
\vspace{1cm}

{\it
$^{+}$
Department of Physics, Karlstad University\\
 651 88 Karlstad, Sweden \\[5mm]
%$^{\circ}$
% Oskar Klein Center for Cosmoparticle Physics \\   
%Department of Physics, 
%Stockholm University, Albanova\\ SE-106 91 Stockholm, Sweden\\ [5mm]
$^{\dag}$ 
 Arnold Sommerfeld Center for Theoretical Physics \\ 
Ludwig-Maximilians-Universit\"at M\"unchen \\ 
Theresienstrasse 37, 80333 M\"unchen, Germany\\ [5mm] 
$^{\star}$ 
Department of Physics, Kim Il Sung University\\ RyongNam Dong, TaeSong District, Pyongyang, DPR.\ Korea \\
and\\
Department of Physics, Nanjing University\\  22 Hankou Road, Nanjing 210093, PR.\ China \\and\\
Abdus Salam International Centre for Theoretical Physics\\  Strada Costiera 11, Trieste 34014, Italy \\[5mm]
$^{\heartsuit}$
Institut f\"ur Theoretische Physik, Universit\"at Heidelberg, \\
Philosophenweg 19, D-69120 Heidelberg \\[5mm]
}
}

\end{center}
\vspace{2mm}

%\abstract
\begin{center}
{\bf Abstract}\\
\end{center}
We evaluate 
string one-loop contributions
to the K\"ahler metric of closed string moduli
 in toroidal minimally supersymmetric (Calabi-Yau) orientifolds
with D-branes.
We focus on the poorly understood ${\mathcal N}=1$
sectors that  receive contributions from all massive string states.
\clearpage
 \tableofcontents

\section{Introduction}
Orientifolds and orbifolds with D-branes 
(see e.g.\ \cite{Blumenhagen:2006ci,Angelantonj:2002ct} for reviews)
are valuable theoretical laboratories for understanding string models of phenomenological relevance. They are the simplest nontrivial string models, in the sense that the worldsheet theories are free, but the symmetry is reduced compared to the simplest possible models. At the phenomenological level, too much symmetry can prevent a given string model from accommodating existing and future data, but  too little symmetry 
makes direct computations exceedingly difficult.
For example, if we orientifold Type IIB string theory to produce Type I string theory in 10 dimensions, then compactify on an orbifold that is a limit of 
the six-dimensional space $K3 \times T^2$, we are left with extended
${\mathcal N =2}$ supersymmetry in four dimensions. 
%This kind of model has been used for many purposes in string theory, 
%especially for furthering the understanding of duality relations between apparently different string models. 
In ${\mathcal N=2}$ supersymmetric models
%, or for models with ${\mathcal N=2}$ subsectors, 
it is typically  relatively straightforward to compute 
string effective actions from worldsheet conformal field theory including the first quantum corrections, in the approximation
of one string loop. This is because
only zero modes contribute to the string effective actions, as heavy string states 
do not preserve enough supersymmetry,
which effectively means that all dependence on worldsheet moduli 
cancels out of the integrands of string S-matrix elements, an enormous technical simplification.
However, these models are  not terribly attractive for   phenomenology. 

Instead one usually wants to compactify string theory on an orientifold that is a limit of a true Calabi-Yau manifold, to obtain minimal ${\mathcal N =1}$ supersymmetry in four dimensions. 
In these more generic models, the amount of supersymmetry is not enough to make the 
worldsheet conformal field theory calculations trivial. 
There remain integrals over both moduli of the worldsheet metric itself, and over
the positions of vertex operators. (For example, for a cylinder amplitude,
the worldsheet moduli are the length of the cylinder and all vertex positions. At tree level, one would  fix some positions by conformal symmetry, but at one-loop this 
is not as helpful and vertex operators are often left unfixed.) 
In simple terms, there is no reason for ${\mathcal N}=1$ contributions to simplify like ${\mathcal N}=2$ sectors do. 

On the good side, there is an extensive literature where somewhat related calculations
have been performed;
minimally supersymmetric string effective actions for the {\it heterotic} string
from the 1980s, and some later developments. It would not be useful or feasible to cite them all here,
but we wish to highlight papers by Lerche, Schellekens, Nilsson and Warner
from 1987  \cite{Lerche:1987qk} and by Stieberger and Taylor from 2002
\cite{Stieberger:2002wk}.  In these papers, very impressive calculations
of torus integrals are performed, where the integrands are highly nontrivial functions of both positions and worldsheet metric moduli. For orientifolds with D-branes, such calculations would need to be generalized to other worldsheet surfaces at Euler characteristic zero, with worldsheet boundaries and/or crosscaps (annulus, M\"obius strip, and Klein bottle, that we collectively denote by $\sigma$). At first sight this generalization seems impossible, since important details of torus integrals were used that simply do not apply in orientifolds with D-branes. 

In this paper, we will emphasize the existence of a natural mapping
between gravity-like amplitudes (closed strings) and gauge-theory-like amplitudes (open
and unoriented strings) at
one loop. The three worldsheet surfaces $\sigma$ with boundaries
and crosscaps are themselves defined as involutions
of a covering torus. This means that one can use the ``doubling trick'',
i.e.\ extend worldsheet fields to unphysical regions of the worldsheet surface
(``unphysical'' in the sense of the method of images). Including
these unphysical regions reconstitutes the covering torus. It is then in principle possible
to use methods similar
to those of closed strings, but it is not completely trivial.
For example, the alert
reader will have noted that this  could at most be possible for open and/or unoriented string amplitudes
where certain factors (such as Chan-Paton traces) have been stripped off. 

We will be interested in one-loop corrections to the K\"ahler metric
of closed string moduli in ${\mathcal N}=1$ supersymmetry. 
In  \cite{Berg:2011ij} we studied 
the one-loop correction to the K\"ahler metric
of {\it open string} scalars for D-branes intersecting at angles, and found that it vanishes in one special case.
This is not guaranteed by supersymmetric effective field theory,
and more generally, one does not expect it to vanish.
In this paper we take a step back from \cite{Berg:2011ij} in the sense that we do not allow nontrivial intersection angles between the D-branes, or D-brane worldvolume fluxes. This means that the moduli dependence will be simple and the essential output of our calculation can be reduced to a single computable finite constant for
each given model. This may be relevant for the phenomenology of these models, for example in determining the K\"ahler metrics for the moduli fields. 
We emphasized ``finite'' constant, since in analogous open-string calculations (e.g.\ \cite{Bain:2000fb}), there 
can be infrared-divergent field theory contributions that produce anomalous dimensions for the effective fields below the string scale, which
introduces possibly interesting renormalization group evolution. In the current
setting with closed string moduli, there
are no such infrared-divergent contributions. Provided the consistency conditions
from tadpole cancellation are fulfilled, ultraviolet divergences in
the string amplitude of interest also cancel between 
the various diagrams, so there are no divergences at all. 

Other work that is relevant to this general
kind of calculation include \cite{Benakli:2008ub,Conlon:2010xb,Anastasopoulos:2011kr}. We keep the general outline as generic as possible throughout this paper, but for concreteness we occasionally focus on a particular 
K\"ahler modulus for a particular orientifold. As example we use
 the K\"ahler modulus of the third two-torus in the $\mathbb{Z}_6'$ orientifold, whose orbifold generator is defined by  the twist vector 
$
\vec v = ( \frac16, -\frac12, \frac13)
$. This minimally supersymmetric four-dimensional orientifold
 contains D9-branes as well as D5-branes wrapped around the third torus. Without fluxes or angles, T-duality in all six internal directions exchanges the D-brane configuration 
 for one with D3-branes and D7-branes, respectively.

%%%%%%%%%%%%%%%%%%%%%%%%%%%%%%%%%%%%%%%%%%%%%%%%%%

\section{String effective action}
\label{sec:sea}

The moduli space of closed and open string moduli
is of great interest, both in its own right and in applications.  
Typically in Type IIB orientifolds, the moduli 
are K\"ahler moduli $T$, complex structure moduli $U$,
the complexified dilaton $S$, and D-brane moduli $\varphi$.
After moduli stabilization such as in flux compactifications, most of these fields will be heavy 
and should better be called ``would-be moduli''. Here, we
will not incorporate any direct effects of moduli stabilization.
In fact, experience
shows that  minimization of the
effective moduli potential  should  be attempted after 
the string effective action is understood at least at the one-loop level. Perhaps surprisingly,
certain questions may even require understanding of the string effective
action at higher-loop level, as we argue below.  

The string effective Lagrangian at low energy $\alpha ' E^2 \ll 1$, in the Einstein frame, will contain  the moduli metric of the K\"ahler moduli:
\be \label{Gtt}
{\mathcal L}_{\rm eff} \quad \supset \quad G_{T\bar{T}} \, \partial_{\mu} T \partial^{\mu} \bar{T} \; .
\ee
In the following, we are going to focus on the metric for $T_3$, the K\"ahler modulus of the 3rd two-torus of the orientifold under consideration. If $T$ does not carry a subscript, we
always implicitly mean this modulus.
Before we begin, a few comments about \refeq{Gtt}. We
employ ``moduli space power counting''\footnote{
see e.g.\  \cite{Polchinski:1998rr}, appendix B.}.
That is, we assign dimension zero  to all
classical field expectation values,
as opposed to the canonical dimensions dictated by the quantum fields (fluctuations). This means we can keep all powers of e.g.\ $T$ in $G_{T\bar{T}}$ without any tension with the low-energy truncation of the $\alpha'$ expansion, as long as we view the $T$ in $G_{T\bar{T}}$ as being the classical expectation value.
With this understanding, one may be misled to believe 
that we can always set $G_{T\bar{T}}\; \rightarrow \; \delta_{T\bar{T}}$ 
by canonically normalizing $T$. The metric can indeed  be transformed to the flat metric
at a point, but not when we probe the neighborhood of a point. For example, the Kaplunovsky-Louis soft terms in gravity mediated supersymmetry breaking \cite{Kaplunovsky:1993rd}
involve curvature tensors in field space. In general, 
we would like to know the moduli space metric $G_{T\bar{T}}$, including quantum corrections.

In minimal supersymmetry, the
moduli space metric $G_{T\bar{T}}$  is K\"ahler, but it is not protected from renormalization.
%and we will compute a one-loop contribution to this term. 
We remark that if we instead had considered the more symmetric models with extended ${\mathcal N}=2$ supersymmetry (or more),
the K\"ahler metric would have been given in terms of a prepotential, which is holomorphic
and obeys nonrenormalization theorems analogous to those of the gauge kinetic function
in minimal supersymmetry. 

With enough K\"ahler metric components computed, one can hope to 
``integrate'' the K\"ahler metric to obtain the K\"ahler potential
from which all K\"ahler metric components are derived, as in
\cite{Berg:2005ja}. We will not solve this problem in the present paper,
but rather focus on a single metric component. 

To compute a one-loop correction to $G_{T\bar{T}}$, we need
to calculate a two-point function of closed string vertex operators representing the volume of the 3rd torus measured in the string frame. This  is the objective of this paper. 
%This will be the leading contribution to the imaginary part of the K\"ahler modulus $T$. 
However, our result is not quite sufficient to construct the one-loop correction to the moduli metric of $T$ in the Einstein frame. We would also need to take into account the perturbative correction to the Einstein-Hilbert term. To see that this will have an influence on the moduli metric in the four-dimensional Einstein frame, let us
consider the analogous $\cn = 2$
supersymmetric model discussed in \cite{Antoniadis:1997eg}. (To be precise, those authors  only considered K\"ahler moduli that are transverse to the overall volume, which would not
include the K\"ahler modulus $T_3$ that we are considering in this paper. The following
discussion is only meant to illustrate possible  effects of corrections to the Einstein-Hilbert term.)
To one-loop order, the corrections to the Einstein-Hilbert term and the moduli metric are found to be
\be \label{s4}
S_4 &=& \frac{1}{2 \kappa_4^2} \int d^4 x \sqrt{-g} \left[ e^{-2 \Phi_4} + \frac{\chi}{(2 \pi)^3} \left( 2 \zeta(3) \frac{e^{-2 \Phi_4}}{{\cal V}} + \frac{2 \pi^2}{3}  \right) \right] R \nn \\
&& + \frac{1}{\kappa_4^2} \int d^4 x \sqrt{-g} \left[ e^{-2 \Phi_4} -  \frac{\chi}{(2 \pi)^3} \left( 2 \zeta(3) \frac{e^{-2 \Phi_4}}{{\cal V}} +  \frac{2 \pi^2}{3} \right) \right] G^{(0)}_{q \bar q}\, \partial_\mu q \partial^\mu \bar q + \ldots\ ,
\ee
where ${\cal V}$ is the volume of the Calabi-Yau measured in string frame and, following \cite{Antoniadis:1997eg}, we used the notation $q$ for the K\"ahler moduli, that sit in hypermultiplets in type IIB theory compactified on a Calabi-Yau manifold. All the corrections are proportional to the Euler number $\chi$ of the Calabi-Yau. The corrections proportional to $\zeta(3)$ arise from $\alpha'$ corrections at sphere level, whereas the other correction terms arise at 1-loop (i.e.\ from the torus diagram). Going to the four dimensional Einstein frame produces
\be \label{S4E}
S_4^{(E)} = \frac{1}{\kappa_4^2} \int d^4 x \sqrt{-g} \left\{ \frac{1}{2}R + \left[ 1 - \frac{2 \chi}{(2 \pi)^3} \left( 2 \frac{\zeta(3)}{{\cal V}} + e^{2 \Phi_4}  \frac{2 \pi^2}{3} \right) \right] G^{(0)}_{q \bar q}\, \partial_\mu q \partial^\mu \bar q \right\} + \ldots 
\ee
to leading order. Here ``leading order'' means that only the leading corrections in the small parameters $1/{\cal V}$ and $e^{2 \Phi_4}$ are kept. Thus, when writing down eq.\
 \eqref{S4E}, we consider $1/{\cal V}$ and $e^{2 \Phi_4}$ to
be of similar size. For example, terms proportional to $e^{2 \Phi_4}/{\cal V}$, which also arise from the Weyl rescaling, count as doubly suppressed
and are neglected, even though counting just the order in the dilaton,
they would have been of the same order as the second term in the round brackets of \eqref{S4E}. 
%Of course, in concrete applications one might be interested in considering a hierarchy between the two small parameters, like in the large volume scenario \cite{Balasubramanian:2005zx}. In that case it might be justified to neglect the term suppressed by $1/{\cal V}$ in the round brackets of \eqref{S4E}. 

Now we see why to determine the one-loop correction to the metric in the Einstein frame, it is not enough to calculate the one-loop correction in the string frame to one-loop order (given in the second line of \eqref{s4}). The correction to the Einstein-Hilbert term, multiplied by the zeroth order contribution to the moduli metric, gives an additional contribution to the correction to the moduli metric in Einstein frame to one loop order (again, neglecting doubly suppressed terms). 

%So much for the 
%situation with  four-dimensional
%$\cn = 2$ supersymmetry \cite{Antoniadis:1997eg}.
%With minimal $\cn = 1$ supersymmetry, 
We also need to deal with another complication: the  definitions of the K\"ahler moduli suffer modifications at one loop order,  in general. This was shown in \cite{Antoniadis:1996vw} for type I orientifolds, but it also holds in type II compactifications if one considers the mixing of the overall volume modulus with the dilaton, cf.\ \cite{Antoniadis:2003sw}, an effect which was circumvented by \cite{Antoniadis:1997eg} by only considering K\"ahler moduli orthogonal to the overall volume. For the following discussion we assume that the imaginary part of the $T$-modulus (which contains the actual volume of the 3rd torus and, thus, is the field that we are actually calculating a 2-point function of) takes the form 
\be \label{deltaImT}
\ImT = \ImT^{(0)}  + \delta \tau
\ee
with a one loop correction $\delta \tau$ that depends on the moduli fields. In general there could also be corrections from disk level, cf.\ \cite{Antoniadis:1996vw, Blumenhagen:2007ip, Camara:2009uv,Grimm:2013bha}, but in the absence of fluxes and neglecting open string scalars for the moment, we do not expect such corrections to arise.\footnote{The corrections found in \cite{Grimm:2013bha} do not obviously seem to require any non-vanishing background values for the open string fields and also do not depend on any fluxes. However,
here we assume that they do not arise in our toroidal orientifold example, since otherwise they should have been observed in the disk level calculations of the closed string K\"ahler metric in \cite{Lust:2004cx} or in the one loop calculation of the gauge couplings in \cite{Antoniadis:1999ge}. It would be interesting to check more directly if the corrections of \cite{Grimm:2013bha} really evaluate to zero for toroidal orientifolds. We thank T.\ Grimm for discussions on this point. 

For related work on higher order redefinition of the moduli fields in heterotic M-theory, see e.g.\ \cite{Lukas:1997fg,Anderson:2009nt}.} As we will show now, the precise knowledge of $\delta \tau$ is also necessary in order to read off the complete one-loop correction to the moduli metric of $T$. 

For $\cn=1$ supersymmetry, the effective action analogous to \eqref{s4} would be 
\be \label{kinetic_term}
S_4 =  \frac{1}{\kappa_4^2} \int d^4 x \sqrt{-g} \left[ \left( e^{-2 \Phi_4} + \delta E \right) {1 \over 2} R  + \left( \widetilde G^{(0)} + \widetilde G^{(1)} \right) \partial_\mu \ImT^{(0)} \partial^\mu \ImT^{(0)} \right]+ \ldots\ ,
\ee
where $\delta E$ contains both the sphere-level $\alpha'$ correction as well as the one-loop 
$g_{\rm s}$ correction, possibly including contributions from all one-loop amplitudes (torus, annulus, M\"obius and Klein bottle, see \reffig{diag} below), cf.\ \cite{Antoniadis:1996vw, Kohlprath:2003pu, Epple:2004ra}. On the other hand, $\widetilde G^{(0)}$ stands for the tree-level (sphere) contributions to the metric in the string frame (including $\alpha'$ corrections), while $\widetilde G^{(1)}$ contains all the one-loop contributions. There are no contributions from disk level here, cf.\ \cite{Lust:2004cx}. It is $\widetilde G^{(1)}$ that we are going to calculate in this paper, or rather the contribution to $\widetilde G^{(1)}$ arising from $\N=1$ sectors. Note that $\widetilde G^{(0)}$ and $\widetilde G^{(1)}$ are not quite the tree level and 1-loop contributions to the $T \bar T$-component of the K\"ahler metric in \eqref{Gtt}, but they are related as discussed below. Also note that both of them only depend on the imaginary parts of the K\"ahler moduli ${\rm Im}\, T_i =  \ImT_i$ given that the real parts of $T_i$ enjoy shift symmetries in string perturbation theory. They can of course also depend on the other moduli, like the complex structure and the dilaton. Moreover, the four-dimensional dilaton $e^{-2 \Phi_4}$ has to be understood as a function of the complex dilaton (or rather of the imaginary part of its tree level form $\ImS^{(0)}$) and the K\"ahler moduli (or rather the imaginary parts of their tree-level form $\ImT_i^{(0)}$), i.e.
\be \label{Phi}
e^{-2 \Phi_4} \equiv e^{-2 \Phi_{10}} t_1 t_2 t_3 = \sqrt{\ImS^{(0)} \ImT_1^{(0)} \ImT_2^{(0)} \ImT_3^{(0)}} \ ,
\ee
where $e^{-2 \Phi_{10}}$ is the ten-dimensional dilaton and
\be \label{Tt}
\ImS^{(0)}  =  e^{-\Phi_{10}} t_1 t_2 t_3\quad , \quad 
\ImT_i^{(0)} = e^{-\Phi_{10}} t_i
\ee
and the $t_i$ are 2-cycle volumes measured in the string frame metric, i.e.\ $t_i \sim \sqrt{G_i}$ with $G_i$ the determinant of the metric of the $i$-th torus.\footnote{Note that we did not include any mixed terms in the kinetic terms in \eqref{kinetic_term}, even though they could in principle contribute to the diagonal part of the kinetic terms of $\ImT_3$. That is, they could contribute to the coefficient of $\partial_\mu \ImT_3 \partial^\mu \ImT_3$, once the one-loop correction to the definition of the K\"ahler moduli, eq.\ \eqref{deltaImT}, is taken into account. However, due to the fact that to zeroth order (in $\alpha'$ and $g_s$) the moduli metric of the K\"ahler moduli is diagonal, this effect would necessarily be doubly suppressed (in ${\cal V}^{-1}$ and/or $g_s$). This justifies restricting to the diagonal part in \eqref{kinetic_term}.} 

In the ${\cal N}=2$ case considered in \cite{Antoniadis:1996vw}, $\delta \tau$ of \eqref{deltaImT} was not independent of $\delta E$, cf.\ the formula between their (4.12) and (4.13). If something analogous happens in our case, one would expect
\be \label{Tredefine}
\ImT_i^{(0)} \rightarrow \ImT_i^{(0)} + \frac{t_i}{2 \ImS^{(0)}} \delta E = \ImT_i^{(0)} + \left. \frac12 \left({\frac{\ImT_i^{(0)}}{\ImT_j^{(0)} \ImT_k^{(0)}}}\right)^{1/2}\!\!\! \frac{1}{\sqrt{\ImS^{(0)}}}\, \delta E \right|_{i \neq j \neq k}\ .
\ee 
Of course, this expectation would have to be confirmed via an explicit calculation. One might attempt to check the redefinition \eqref{Tredefine} for $\ImT_3$ by calculating the quantum corrected gauge coupling of the 5-brane gauge group, as this is given by $\ImT_3^{(0)}$ at lowest order in perturbation theory, i.e.\ at disk level. This should be completed by quantum corrections to make a term linear in the corrected $\ImT_3$. However, note that the correction in \eqref{Tredefine} is of order $e^{2 \Phi_{10}}$ relative to the leading disk level term. Thus, it would only arise from a genus-$3/2$ contribution to the gauge coupling. It would be interesting to calculate this higher-loop contribution along the lines of \cite{Antoniadis:2004qn}.

Now, introducing 
\be
e^{-2 \widetilde \Phi_4} \equiv e^{-2 \Phi_4} + \delta E
\ee 
and performing the Weyl rescaling to go to the four dimensional Einstein frame via 
\be
g_{\mu \nu} \rightarrow e^{2 \widetilde \Phi_4} g_{\mu \nu}
\ee
we end up with 
\be
S_4^{(E)} &=&  \frac{1}{\kappa_4^2} \int d^4 x \sqrt{-g} \left[ \frac{1}{2}R + \left(e^{2 \Phi_4} \widetilde G^{(0)} - \delta E e^{4 \Phi_4} \widetilde G^{(0)}+ e^{2 \Phi_4} \widetilde G^{(1)} \right) \partial_\mu \ImT^{(0)} \partial^\mu \ImT^{(0)} \right. \nn \\
&& \left. - 6 \partial_\mu \widetilde \Phi_4 \partial^\mu\widetilde \Phi_4 \right] + \ldots\ ,
\ee
where the dots contain again doubly suppressed terms. Expanding the second row to one loop order
\be
\partial_\mu \widetilde \Phi_4 \partial^\mu\widetilde \Phi_4 = \partial_\mu \Phi_4 \partial^\mu \Phi_4 (1- 2 \delta E e^{2 \Phi_4}) - \partial_\mu \Phi_4 \partial^\mu \delta E e^{2 \Phi_4} + \ldots
\ee
we obtain the kinetic term
\be
&& \left[ G^{(0)} - \delta E e^{4 \Phi_4} \widetilde G^{(0)}+ e^{2 \Phi_4} \widetilde G^{(1)}   \right. \\
&& \left. - 6 \left(\frac{\partial \Phi_4}{\partial \ImT^{(0)}}\right)^2 (- 2 \delta E e^{2 \Phi_4}) + 6 \frac{\partial \Phi_4}{\partial \ImT^{(0)}} \frac{\partial \delta E}{\partial \ImT^{(0)}} e^{2 \Phi_4} \right] \partial_\mu \ImT^{(0)} \partial^\mu \ImT^{(0)}+ \ldots \nn \\
&\equiv&  \left[ G^{(0)} + \overline G^{(1)} \right] \partial_\mu \ImT^{(0)} \partial^\mu \ImT^{(0)} + \ldots\ .
\ee
Here $G^{(0)}$ is  the tree level metric of the K\"ahler modulus $T_3$, i.e.
\be \label{Gzero}
G^{(0)} = G^{(0)}_{T \bar T}(T^{(0)}) = \frac{1}{4 (\ImT^{(0)})^2} \left( 1 + \chi \, \zeta(3) \frac{(\sigma^{(0)})^{3/2}}{{\cal V}}  + \ldots \right)\ ,
\ee
which follows from \cite{Becker:2002nn}. Note that this is not the same as $\widetilde G^{(0)}$. For instance, the tree level $\partial_\mu \Phi_4 \partial^\mu \Phi_4$ term is absorbed here in order to get from $\widetilde G^{(0)}$ to $G^{(0)}$.

Finally we obtain to one-loop order (making the dependence on $(\ImT^{(0)})$ explicit for clarity)
\be \label{total_kinetic}
&&\hspace{-2cm} \left[ G^{(0)}(\ImT^{(0)}) + \overline G^{(1)}(\ImT^{(0)}) \right] \partial_\mu (\ImT - \delta \tau) \partial^\mu (\ImT - \delta \tau) \nn \\
&=& \left[ G^{(0)}(\ImT) + \overline G^{(1)}(\ImT) + \frac{1}{2 \ImT^3} \delta \tau \right] \partial_\mu \ImT \partial^\mu \ImT - 2 G^{(0)}(\ImT) \partial_\mu \ImT \partial^\mu \delta \tau  + \ldots \nn \\
&=& \left[ G^{(0)}(\ImT) + \overline G^{(1)}(\ImT) + \frac{1}{2 \ImT^3} \delta \tau - 2 G^{(0)}(\ImT) \frac{\partial  \delta \tau}{\partial \ImT}  \right] \partial_\mu \ImT \partial^\mu \ImT \nn  + \ldots \\
& \equiv & \left[ G^{(0)}(\ImT) + G^{(1)}(\ImT) \right] \partial_\mu \ImT \partial^\mu \ImT + \ldots\ ,
\ee
where we neglected doubly suppressed terms and the dots in the third row also include potential off-diagonal kinetic terms mixing different $\tau_i$-moduli, resulting from the fact that $\delta \tau$ probably depends on all the $\tau_i$, cf.\ \eqref{Tredefine}. In the present discussion we focus on the diagonal kinetic term for $\tau = \tau_3$. Of course, in the term proportional to $G^{(0)}(\ImT) {\partial  \delta \tau}/{\partial \ImT}$ one could neglect the $\alpha'$-correction to $G^{(0)}$ given that it would be doubly suppressed. From \eqref{total_kinetic} we can read off the one-loop correction to the $T \bar T$ component of the moduli metric 
\be
G^{(1)}_{T \bar T}(T) &=& G^{(1)}(\ImT) = \overline G^{(1)}(\ImT)+ \frac{1}{2 \ImT^3} \delta \tau - \frac{1}{2 \ImT^2} \frac{\partial  \delta \tau}{\partial \ImT}  + \ldots \\
&=& e^{2 \Phi_4} \widetilde G^{(1)}(\ImT) + 12 \left(\frac{\partial \Phi_4}{\partial \ImT^{(0)}}\right)^2  \delta E e^{2 \Phi_4} + 6 \frac{\partial \Phi_4}{\partial \ImT^{(0)}} \frac{\partial \delta E}{\partial \ImT^{(0)}} e^{2 \Phi_4} \nn \\
&& - \delta E e^{4 \Phi_4} \widetilde G^{(0)}(\ImT) + \frac{1}{2 \ImT^3} \delta \tau - \frac{1}{2 \ImT^2} \frac{\partial  \delta \tau}{\partial \ImT} + \ldots  \label{finalmetric} \\[2mm]
& \equiv & e^{2 \Phi_4} \widetilde G^{(1)}(\ImT) + \Delta \ .
\ee
We see that to find the complete one-loop correction to the metric of 
the K\"ahler modulus $T$, we need  
not only the direct one-loop correction $\widetilde G^{(1)}_{T \bar T}(T)$ (whose determination is the focus of this paper) but also the additional correction $\Delta$, which requires knowledge of $\delta E$ and $\delta \tau$ to one-loop order. If the expectation \eqref{Tredefine} 
for $\delta \tau$ indeed bears out, it would remain to calculate
 the $\cn=1$ one-loop correction to the Einstein-Hilbert term $\delta E$. We leave this for future research.

%%%%%%%%%%%%%%%%%%%%%%%%%%%%%%%%%%%%%%%%%%%%%%%%%%

\section{Two-point function of closed strings}
\label{sec:two-point}
\begin{figure}
\begin{center}
\includegraphics[width=0.9\textwidth]{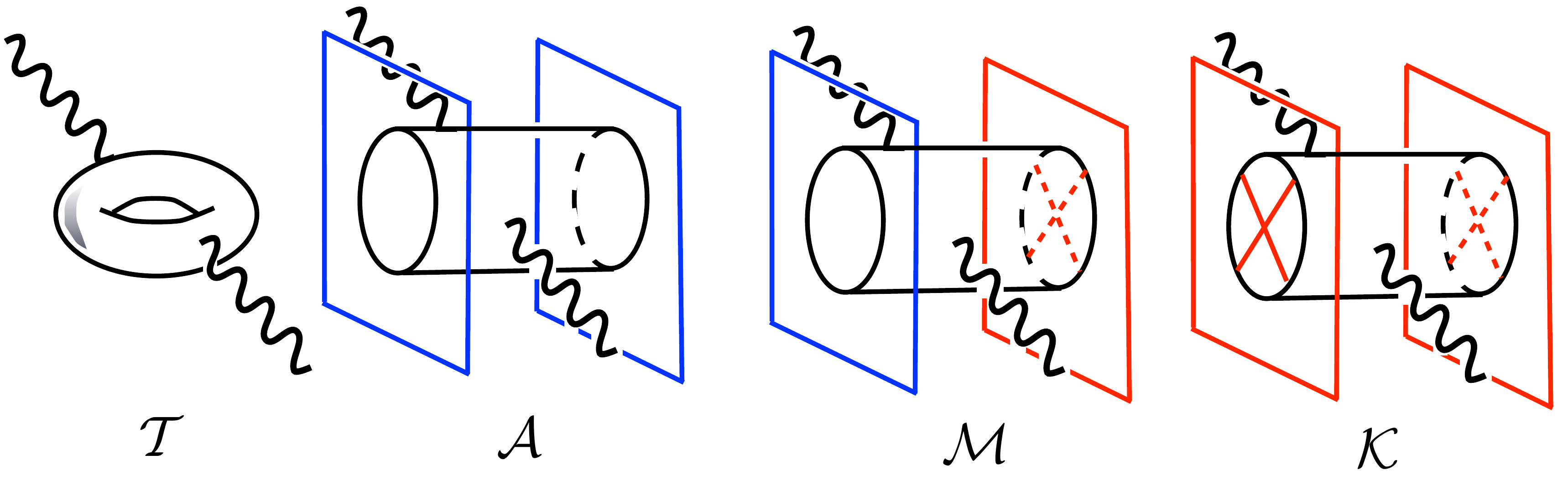}
\caption{The one-loop two-point function
on all four worldsheets of  Euler character zero.}
\label{diag}
\end{center}
\end{figure}

To obtain 
$\widetilde G^{(1)}_{T \bar T}(T)$  of the previous section, we
would like to compute the string
amplitudes of \reffig{diag}, where the external states are closed-string moduli.
Now, generic moduli are not independent,
for example a fluctuation in a D-brane scalar can cause a fluctuation
in a complex structure modulus.
Since the power of the string coupling
is higher for lowest-order open strings (disk) than
for lowest-order closed strings (sphere), this could cause a technical complication
by mixing
different orders of string perturbation theory. In  \cite{Berg:2005ja}
we dealt with this using ``K\"ahler adapted vertex operators'', that automatically capture the corrections
of interest at the desired order of perturbation theory. 
The strategy is quite  simple: rewrite the string worldsheet action
in terms of the (complex) moduli, in which 
the moduli space is actually K\"ahler, and vary the worldsheet action with respect
to these spacetime moduli. This produces combinations where the usual 
open and closed string vertex operators are used as ``building blocks''.
For the imaginary part $\tau$ of the K\"ahler modulus $T$, however,
we showed that only a single building block contributes
to its vertex operator:\footnote{The $\alpha'^{-2}$
is to cancel the string length squared contained in the closed string coupling
$g_{\rm c}$ and make the forefactor dimensionless.}
\be \label{vtauphi}
V_{\tau} = \phi\, \frac{g_c (\alpha')^{-2}}{2 \tau^{(0)}}\, V_{Z\bar{Z}}\ ,
\ee
with the vertex operator (Eq. (2.32) of \cite{Berg:2005ja})
\begin{align}
 V_{Z \bar Z}(p) = & - \frac{2}{\alpha'} \int \dd^2\nu
 \, e^{i p \, \cdot X}
 \left[i \p \bar Z + \frac{\alpha'}{2} (p \cdot \psi) \bar \Psi \right]
 \left[ i \bar \p Z + \frac{\alpha'}{2} (p \cdot \tilde \psi) \tilde \Psi \right] \nonumber \\
 & - \frac{2}{\alpha'} \int \dd^2\nu
 \, e^{i p \, \cdot X}
 \left[ i \p Z + \frac{\alpha'}{2} (p \cdot \psi) \Psi \right]
 \left[ i \bar \p \bar Z + \frac{\alpha'}{2} (p \cdot \tilde \psi) \tilde{\bar\Psi} \right] \,,
\end{align}
where
\beqn \label{zpsi}
Z = \frac{{G}^{1/4}}{(2\,  {\rm Im}\, U)^{1/2}}(X^{1} + \bar
U X^{2}) \quad , \quad
\bar Z =  \frac{{G}^{1/4}}{(2\,  {\rm Im}\, U)^{1/2}} (X^{1}
+ U X^{2})\ , \non
\Psi =\frac{{G}^{1/4}}{(2\,  {\rm  Im} \,U)^{1/2}}  (\psi^{1} + \bar U
\psi^{2}) \quad , \quad
\bar \Psi =\frac{{G}^{1/4}}{(2\,  {\rm Im} \,U)^{1/2}}  (\psi^{1}
+ U \psi^{2})
\eeqn
are the complexified coordinates of \cite{Lust:2004cx}. Here
$X^1$ and $X^2$ are the real coordinates and $\sqrt{G}$ and $U$ the volume and the complex structure of the 
torus whose K\"ahler modulus is under consideration (we will always consider the 3rd torus). 
Also, $\phi$ is the polarization of the scalar vertex operator. 
As mentioned above, the vertex operator is found from varying the worldsheet action
expressed in the relevant spacetime moduli, producing
\be \label{Vtau}
V_{\tau} = - \frac{\delta {\cal S}_{\rm ws}}{\delta \tau^{(0)}} \delta \tau^{(0)} = - \frac{1}{4 \pi} V_{Z\bar{Z}} \frac{1}{2 \tau^{(0)}} \delta \tau^{(0)}\ ,
\ee
and comparison with \eqref{vtauphi} reveals the relation
\be \label{Deltatau}
\delta \tau^{(0)} = - 4 \pi g_c (\alpha')^{-2} \phi\ ,
\ee
(cf.\ (3.7.11c) in \cite{Polchinski:1998rq}, where  the factor of $(\alpha')^{-2}$ was absorbed in $\phi$).
Note that $\delta \tau^{(0)}$ denotes the fluctuation of $\tau^{(0)}$ about its background value, and not the one-loop correction \eqref{deltaImT} which we will not discuss further in this paper. Also, the second equality in \eqref{Vtau} follows like (2.31) in \cite{Berg:2005ja}. The relation \eqref{Deltatau} will be important later on when we read off the one-loop correction to the metric from the 2-point function that we are going to calculate. 

The two-point string S-matrix for $\sigma = \A, \M, \K$ in a $T^6/\mathbb{Z}_N$ orientifold is 
\be \label{full2ptfct}
 \big\langle\!\! \big\langle V_{\tau}(p_1) V_{\tau}(p_2) \big\rangle\!\!  \big\rangle_\sigma
 = \frac{1}{8 N} \frac{iV_4}{(4 \pi^2 \alpha')^2} {\phi^2 g_c^2 (\alpha')^{-4} \over 4 (\tau^{(0)})^2} % {\rm tr}_{\rm CP}
\int_0^\infty \frac{dt}{t^3} \sum_{k=0}^{N-1} \sum_{s\, {\rm even}} Z_s \big\langle V_{Z \bar Z}(p_1) V_{Z \bar Z}(p_2) \big\rangle_\sigma^s \,,
\ee
where $V_4$ is the regularized external volume, % ${\rm tr}_{\rm CP}$ stands for the Chan-Paton trace, 
$Z_s$ stands for the partition function without the factor $(4 \pi^2 \alpha' t)^{-2}$, which we made explicit, (but including the Chan-Paton trace for annulus and M\"obius) 
and $s\equiv\{\alpha,\beta\}$ is the spin structure of the worldsheet
fermions, cf.\ \reftab{table:spinstructures} in appendix \refapp{sec:ferm}. 
Finally, $\big\langle V_{Z \bar Z}(p_1) V_{Z \bar Z}(p_2) \big\rangle_\sigma^s$ is
the conformal field theory correlator
on the worldsheet $\sigma$. For the torus one has a very similar formula, except that % there is no Chan-Paton trace, 
the integral over the worldsheet parameter involves the complex modulus of the torus, there are two independent spin structures for the left and the right movers and both the even-even and the odd-odd combinations can  in principle contribute. 

\begin{figure}
\begin{center}
\includegraphics[width=0.5\textwidth]{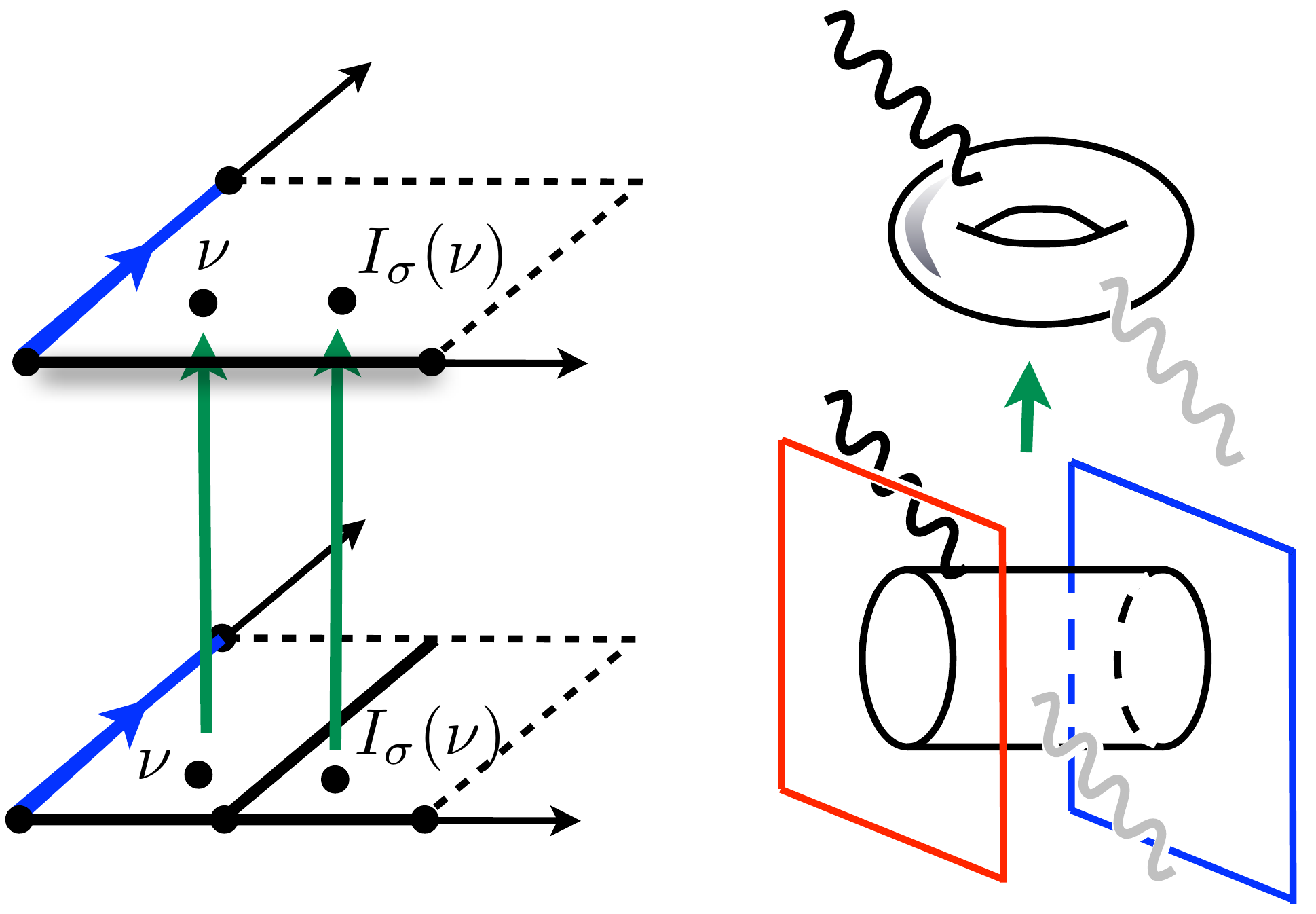}
\caption{A lifting that gives amplitudes for open and unoriented
strings from closed strings.}
\label{fig:lift}
\end{center}
\vspace{-4mm}
\end{figure}

For most of this paper we will focus on the worldsheet conformal field theory correlator and only later on reinstate the remaining ingredients in \refeq{full2ptfct}. The contraction with only bosons vanishes by supersymmetry,
cross terms with single normal-ordered operators automatically vanish, and the remaining
contributions to
the two point function are given by one piece
with four worldsheet fermions and one piece with eight worldsheet fermions:
\be
 \big\langle V_{Z \bar Z}(p_1) V_{Z \bar Z}(p_2) \big\rangle_\sigma^s  =
 \big\langle V_{Z \bar Z}(p_1) V_{Z \bar Z}(p_2) \big\rangle_\sigma^s \big|_\fourfermion +
 \big\langle V_{Z \bar Z}(p_1) V_{Z \bar Z}(p_2) \big\rangle_\sigma^s \big|_\eightfermion\ .
\ee
The latter can only contribute if there is a pole from vertex collisions, 
as we will see below. 
We begin with the four-fermion contributions:
\begin{align}
 & \big\langle V_{Z \bar Z}(p_1) V_{Z \bar Z}(p_2) \big\rangle_\sigma^s \big|_\fourfermion
 = - 2(p_1 \cdot p_2)
 \int_\sigma \dd^2 \nu_1 \int_\sigma \dd^2 \nu_2 \,
 e^{-(p_1 \cdot\, p_2) \, \langle X_1 X_2\rangle_\sigma} \nonumber \\
 & ~~~~~~~~~~~~~~~~~~ \times \Big\{
 \vev{ \bar \p Z_1 \bar \p \bar Z_2 }_\sigma
 \vev{ \Psi_2 \bar \Psi_1 }_\sigma^s
 \vev{ \psi_2 \psi_1 }_\sigma^s
 +
 \vev{ \p Z_1 \bar \p \bar Z_2 }_\sigma
 \vev{ \Psi_2 \tilde{\bar\Psi}_1 }_\sigma^s
 \vev{ \psi_2 \tilde \psi_1 }_\sigma^s \nonumber \\
 & ~~~~~~~~~~~~~~~~~~~~ +
 \vev{ \bar \p Z_1 \p \bar Z_2 }_\sigma
 \vev{ \tilde \Psi_2 \bar \Psi_1 }_\sigma^s
 \vev{ \tilde \psi_2 \psi_1 }_\sigma^s
 +
 \vev{ \p Z_1 \p \bar Z_2 }_\sigma
 \vev{ \tilde \Psi_2 \tilde{\bar\Psi}_1 }_\sigma^s
 \vev{ \tilde \psi_2 \tilde \psi_1 }_\sigma^s \Big\} \,.
\end{align}
\begin{figure}
\begin{center}
\includegraphics[width=0.5\textwidth]{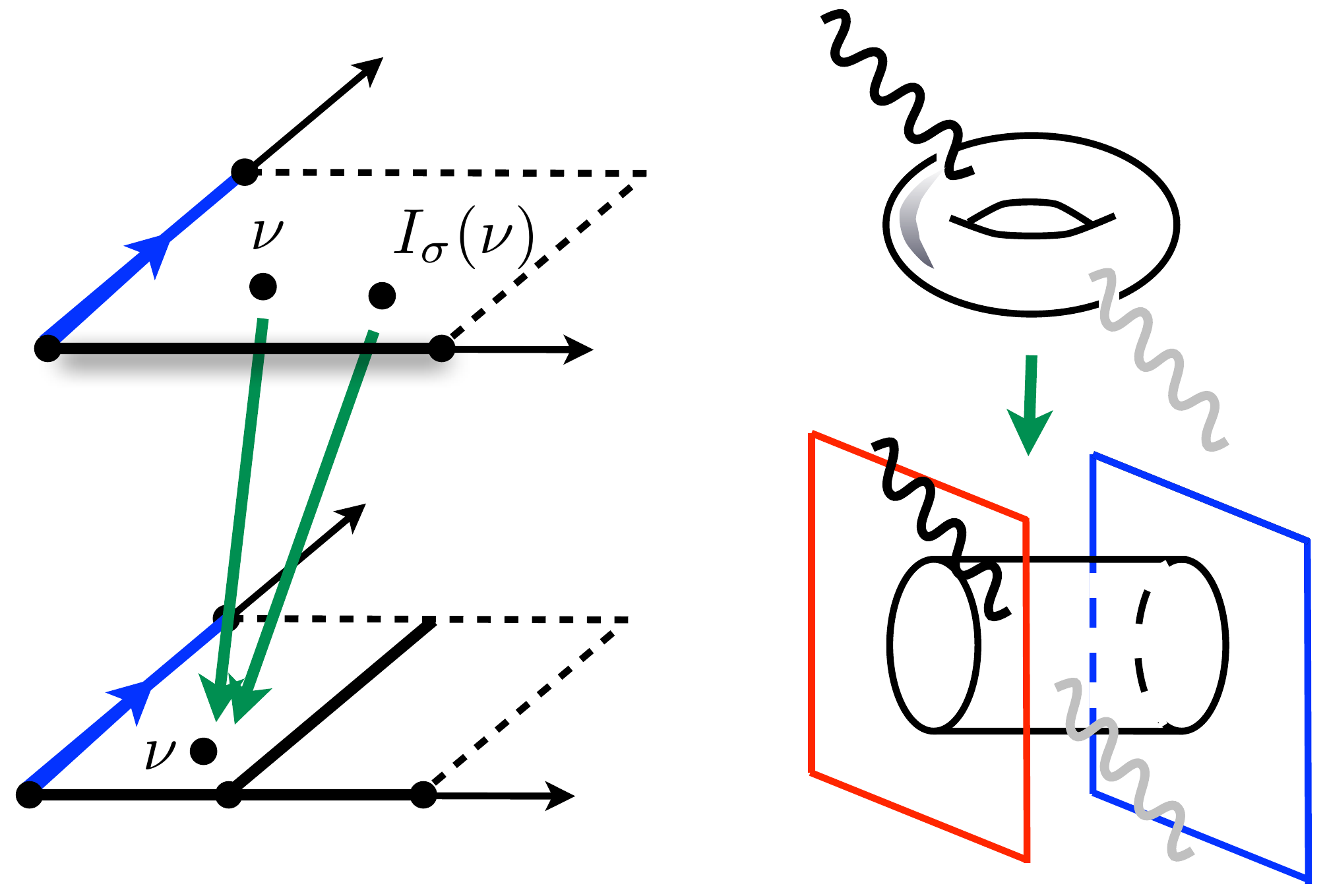}
\caption{The standard way to make the worldsheet surfaces with crosscaps
and boundaries, and their Green's functions, from that  of the worldsheet torus is by identifying under a certain involution.}
\label{involution}
\end{center}
\vspace{-4mm}
\end{figure}
Using the worldsheet correlators 
of appendices \refapp{sec:bos} and \refapp{sec:ferm} we have for the torus amplitude $\sigma = \T$:
\begin{align} \label{4fermions_torus}
 & \langle V_{Z \bar Z}(p_1) V_{Z \bar Z}(p_2) \rangle_\T^{s,\tilde s} \big|_\fourfermion
 = - 2 \, \delta \int_\T \dd^2 \nu_1 \int_\T \dd^2 \nu_2 \,
 e^{- \delta G_B^\T (\nu_1,\, \nu_2)} \nonumber \\
 & ~~~~~~~~~~ \times \Big\{
 \big[ ( \p_1 \p_2 G_B^{-\vec \gamma} ) (\nu_1,\nu_2) \big]^* \,
 G_F^{\vec \gamma} ( s ; \nu_2 , \nu_1 ) \,
 G_F ( s ; \nu_2 , \nu_1 ) \nonumber \\
 & ~~~~~~~~~~~
 + ( \p_1 \p_2 G_B^{\vec \gamma} ) (\nu_1 , \nu_2 ) \,
 \big[ G_F^{-\vec \gamma} ( \tilde s ; \nu_2 , \nu_1 ) \big]^* \,
 \big[ G_F (\tilde s ; \nu_2 , \nu_1 ) \big]^* \Big\} \,,
\end{align}
where we introduced the dimensionless momentum-squared invariant $\delta$ as\footnote{With strict momentum conservation this would vanish on-shell, but
we relax overall momentum conservation as in \cite{Minahan:1987ha}, see p. 19 of
\cite{Berg:2005ja} for more discussion.}
\be \label{deltamomentumsquare}
\delta\equiv \frac{\alpha'}{2}(p_1 \cdot p_2)
\ee
and $\vec{\gamma}$ is the twist 
(quasiperiodicity) of the worldsheet correlator, see appendices  \refapp{sec:bos} and \refapp{sec:ferm}
for details. For the annulus, M\"obius strip and Klein bottle amplitudes
($\sigma = \A, \M, \K$) we have:
\begin{align}
 & \big\langle V_{Z \bar Z}(p_1) V_{Z \bar Z}(p_2) \big\rangle_\sigma^s \big|_\fourfermion
 = - 2 \delta \int_\sigma \dd^2 \nu_1 \int_\sigma \dd^2 \nu_2 \,
 e^{- \delta G_B^\sigma(\nu_1,\, \nu_2)} \nonumber \\
 & ~~~~~~~~~~ \times \Big\{
 ( \bar \p_1 \bar \p_2 G_B^\gamma ) ( I_\sigma(\nu_1) , I_\sigma(\nu_2) ) \,
 G_F^\gamma ( s ; \nu_2 , \nu_1 ) \,
 G_F ( s ; \nu_2 , \nu_1 ) \phantom{\Big\}} \nonumber \\
 & ~~~~~~~~~~~~
 - ( \p_1 \bar \p_2 G_B^\gamma ) ( \nu_1 , I_\sigma(\nu_2) ) \,
 G_F^\gamma ( s ; \nu_2 , I_\sigma(\nu_1) ) \,
 G_F ( s ; \nu_2 , I_\sigma(\nu_1) ) \phantom{\Big\}} \nonumber \\ 
 & ~~~~~~~~~~~~
 - ( \bar \p_1 \p_2 G_B^\gamma ) ( I_\sigma(\nu_1) , \nu_2 ) \,
 G_F^\gamma ( s ; I_\sigma(\nu_2) , \nu_1 ) \,
 G_F ( s ; I_\sigma(\nu_2) , \nu_1 ) \phantom{\Big\}} \nonumber \\
 & ~~~~~~~~~~~~
 + ( \p_1 \p_2 G_B^\gamma ) ( \nu_1 , \nu_2 )
 G_F^\gamma ( s , I_\sigma(\nu_2) , I_\sigma(\nu_1) )
 G_F ( s ; I_\sigma(\nu_2) , I_\sigma(\nu_1) ) \Big\} \,, \label{4fermion}
\end{align}
%\begin{align}
% & \big\langle V_{Z \bar Z}(p_1) V_{Z \bar Z}(p_2) \big\rangle_\sigma^s \big|_\fourfermion
% = - 2 \delta \int_\sigma \dd^2 \nu_1 \int_\sigma \dd^2 \nu_2 \,
% e^{- \delta \, G_B^\sigma(\nu_1,\, \nu_2)} \nonumber \\
% & ~~~~~~~~~~ \times \Big\{
% \, \bar \p_1 \bar \p_2 \, G_B^\sigma \ss{\nicefrac{1}{2}}{\nicefrac{1}{2}+\gamma} ( \nu_1, \nu_2 )
% \, G_F^\Ts \ss{\alpha}{\beta+\gamma} (\nu_2, \nu_1 )
% \, G_F^\Ts \ss{\alpha}{\beta} (\nu_2, \nu_1 ) \nonumber \\
% & ~~~~~~~~~~~~
% \, -\p_1 \bar \p_2 \, G_B^\sigma \ss{\nicefrac{1}{2}}{\nicefrac{1}{2}+\gamma} (I_\sigma(\nu_1), \nu_2 )
% \, G_F^\Ts \ss{\alpha}{\beta+\gamma} (\nu_2, I_\sigma(\nu_1) )
% \, G_F^\Ts \ss{\alpha}{\beta} (\nu_2, I_\sigma(\nu_1) ) \nonumber \\ 
% & ~~~~~~~~~~~~
% \, -\bar \p_1 \p_2 \, G_B^\sigma \ss{\nicefrac{1}{2}}{\nicefrac{1}{2}+\gamma} ( \nu_1, I_\sigma(\nu_2) )
% \, G_F^\Ts \ss{\alpha}{\beta+\gamma} ( I_\sigma(\nu_2), \nu_1 )
% \, G_F^\Ts \ss{\alpha}{\beta} ( I_\sigma(\nu_2), \nu_1 ) \nonumber \\
% & ~~~~~~~~~~~~
% \, + \p_1 \p_2 \, G_B^\sigma \ss{\nicefrac{1}{2}}{\nicefrac{1}{2}+\gamma} ( I_\sigma(\nu_1), I_\sigma(\nu_2) )
% \, G_F^\Ts \ss{\alpha}{\beta+\gamma} ( I_\sigma(\nu_2), I_\sigma(\nu_1) )
% \, G_F^\Ts \ss{\alpha}{\beta} (I_\sigma(\nu_2), I_\sigma(\nu_1) ) \Big\} \,.
%\end{align}
where there is only a single real twist $\gamma$
(related to the orbifold twist, as in \reftab{tab:details} in \refsec{sec:addingup}) and the definition of the involutions $I_\sigma$ can be found in the appendix, cf.\ \eqref{Isigmas}, and see \reffig{involution}.
The slightly surprising choice of ordering of
arguments in \refeq{4fermion} is convenient to absorb some phases that would appear
for the more obvious ordering.
Note that although most of the correlators
refer to the covering torus, at this point the integrals
are still performed over each worldsheet surface $\sigma = \A, \M, \K$. 
We now proceed to {\it lift} the world sheet integrals to the covering torus
(see \reffig{fig:lift}), using
the results of \refapp{sec:lifting}:
\begin{align}   \label{beforespin}
 \big\langle V_{Z \bar Z}(p_1) V_{Z \bar Z}(p_2) \big\rangle_\sigma^s \big|_\fourfermion
 & = - 2 \delta \int_\Ts \dd^2 \nu_1 \int_\Ts \dd^2 \nu_2
 \, e^{- \delta G_B^\sigma(\nu_1, \,\nu_2)} \nonumber \\
 & \,~~~ \times
 (\bar \p_1 \bar \p_2 G_B^\gamma) (I_\sigma(\nu_1),I_\sigma(\nu_2))
 \, G_F^\gamma (s;\nu_2,\nu_1)
 \, G_F(s;\nu_2,\nu_1) \,.
\end{align}
As promised, the integration domains with
boundaries and crosscaps have been replaced 
by  purely torus integrals, 
and now most Green's functions
in the integrand also refer to the covering torus. The only remaining Green's functions on the surface $\sigma$ is
the $G_B^{\sigma}$ in the exponent, and since this has a forefactor that carries momenta,
we will be able to treat it separately
from the others in the low-energy limit.

In all, what we have done is
to provide a simple prescription to compute open and unoriented string (gauge theory-like) loop amplitudes
from closed string (gravity-like) loop amplitudes. This 
is backwards compared to what is done in recent work on 
field theory amplitude calculations,
where one derives gravity amplitudes from gauge theory amplitudes! 
(See for example one of the original papers \cite{Bern:2010ue},
and the more recent string theory work in   \cite{Stieberger:2014hba}.)
The lifting procedure is convenient because it allows us to systematize the calculation and
make use of known results for torus amplitudes. Note also
that the Chan-Paton states are traced over, otherwise this lifting would clearly have been impossible.

Now we go through similar steps
for the eight-fermion contributions, which read
\begin{align}
 \big\langle V_{Z \bar Z}(p_1) V_{Z \bar Z}(p_2) \big\rangle_\sigma^s \big|_\eightfermion
 & = - \left(\frac{\alpha'}{2} \right)^2(p_1 \cdot p_2)^2
 \int_\sigma \dd^2 \nu_1 \int_\sigma \dd^2 \nu_2
 \, e^{-(p_1 \cdot\, p_2) \, \langle X_1 X_2\rangle_\sigma} \nonumber \\
 &~~~\, \times  \Big\{ 
 \vev{\psi_1 \psi_2}_\sigma^s
 \vev{\tilde \psi_1 \tilde \psi_2}_\sigma^s
 - \vev{\psi_1 \tilde \psi_2}_\sigma^s
 \vev{\tilde \psi_1 \psi_2}_\sigma^s
 \Big\} \nonumber \\
 & ~~~\, \times \Big\{
 \vev{\Psi_1 \bar \Psi_2}_\sigma^s
 \vev{\tilde\Psi_2 \tilde{\bar\Psi}_1}_\sigma^s
 -
 \vev{\tilde\Psi_1 \bar \Psi_2}_\sigma^s
 \vev{\tilde \Psi_2 \bar \Psi_1}_\sigma^s \nonumber \\
 & ~~~~~ \,
 - \vev{\Psi_1 \tilde{\bar\Psi}_2}_\sigma^s
 \vev{\Psi_2 \tilde{\bar\Psi}_1}_\sigma^s
 +
 \vev{\tilde\Psi_1 \tilde{\bar\Psi}_2}_\sigma^s
 \vev{\Psi_2 \bar \Psi_1}_\sigma^s
 \Big\} \;. 
\end{align}
Again with the worldsheet correlators 
of appendices \refapp{sec:bos} and \refapp{sec:ferm}  we have for $\sigma = \T$
\begin{align}
 \big\langle V_{Z \bar Z}(p_1) V_{Z \bar Z}(p_2) \big\rangle_\T^{s, \tilde s} \big|_\eightfermion
 & = - 2 \delta^2
 \int_\T \dd^2 \nu_1 \int_\T \dd^2 \nu_2
 \, e^{- \delta G_B^\T(\nu_1,\nu_2)} \nn \\
 & \times
 G_F(s;\nu_1,\nu_2) \,
 [G_F(\tilde s;\nu_1,\nu_2)]^* \,
 G_F^{\vec \gamma}(s;\nu_1,\nu_2) \,
 [G_F^{-\vec \gamma}(\tilde s;\nu_2,\nu_1)]^* \,, 
\end{align}
and for $\sigma = \A, \M, \K$
\begin{align}
 & \big\langle V_{Z \bar Z}(p_1) V_{Z \bar Z}(p_2) \big\rangle_\sigma^s \big|_\eightfermion
 = - \delta^2 \int_\sigma \dd^2 \nu_1 \int_\sigma \dd^2 \nu_2 \,
 e^{- \delta G_B^\sigma(\nu_1,\, \nu_2)} \nonumber \\
 & ~~~~~~~~~~~ \times \Big\{
 \, G_F (s; \nu_1, \nu_2 )
 \, G_F (s; I_\sigma(\nu_1), I_\sigma(\nu_2))
  - G_F (s; \nu_1, I_\sigma(\nu_2))
 \, G_F (s; I_\sigma(\nu_1), \nu_2)
 \Big\} \nonumber \\
 & ~~~~~~~~~~~ \times \Big\{
 \, G_F^\gamma (s; \nu_1, \nu_2)
 \, G_F^\gamma (s; I_\sigma(\nu_2), I_\sigma(\nu_1)) \phantom{\Big\}} \nonumber \\
 & ~~~~~~~~~~~~~
 \, - G_F^\gamma (s; I_\sigma(\nu_1), \nu_2)
 \, G_F^\gamma (s; I_\sigma(\nu_2), \nu_1)
 \, e^{2 \pi i \gamma \delta_{\sigma \mathcal{K}}} \phantom{\Big\}} \nonumber \\
 & ~~~~~~~~~~~~~
  \, - G_F^\gamma (s; \nu_1, I_\sigma(\nu_2))
 \, G_F^\gamma (s; \nu_2, I_\sigma(\nu_1))
 \, e^{- 2 \pi i \gamma \delta_{\sigma \mathcal{K}}} \phantom{\Big\}} \nonumber \\
 & ~~~~~~~~~~~~~
 \, + G_F^\gamma (s;I_\sigma(\nu_1), I_\sigma(\nu_2))
 \, G_F (s; \nu_2, \nu_1 ) \Big\} \,. \label{8fermion}
\end{align}
Now we can use the results of \refapp{sec:lifting} again
to lift
 the world sheet integrals to the covering torus:\begin{align}
 & \big\langle V_{Z \bar Z}(p_1) V_{Z \bar Z}(p_2) \big\rangle_\sigma^s \big|_\eightfermion
 = - \delta^2 \int_\mathcal{T} \dd^2 \nu_1 \int_\mathcal{T} \dd^2 \nu_2 \,
 e^{- \delta G_B^\sigma(\nu_1,\, \nu_2)} \nonumber \\
 & ~~~~~~~~~~~ \times \Big\{
 \, G_F (s; \nu_1, \nu_2)
 \, G_F (s; I_\sigma(\nu_1), I_\sigma(\nu_2))
 \, - G_F (s; \nu_1, I_\sigma(\nu_2))
 \, G_F (s; I_\sigma(\nu_1), \nu_2)
 \Big\} \nonumber \\
 & ~~~~~~~~~~~ \times
 \, G_F^\gamma (s; \nu_1, \nu_2)
 \, G_F^\gamma (s; I_\sigma(\nu_2), I_\sigma(\nu_1)) \; . 
\label{eightferm}
\end{align}
%Another  useful form of the same expression is
%\begin{align}
%& \big\langle V_{Z \bar Z}(p_1) V_{Z \bar Z}(p_2) \big\rangle_\sigma^s \big|_\eightfermion
%= + \left( \frac{\alpha'}{2} \right)^2 (p_1 \cdot p_2)^2 \int_\mathcal{T} \dd^2 \nu_1 \int_\mathcal{T} \dd^2 \nu_2 \,
% e^{-(p_1 \cdot\, p_2) \, G_B^\sigma(\nu_1,\, \nu_2)} \\
% & ~~~~~~~~~~~~~~~~~~~~~~~~~~~~ \times \bigg\{
% \, G_F^\mathcal{T} \big( s, \nu_1, \nu_2 \big)
% \, \big[ G_F^\mathcal{T} \big(\bar s, \nu_1, \nu_2 \big) \big]^*
% \, G_F^\mathcal{T} \big( s, \gamma, \nu_1, \nu_2 \big)
% \, \big[ G_F^\mathcal{T} \big( \bar s, \gamma, \nu_1, \nu_2 \big) \big]^* \nonumber \\
% & ~~
% - (-1)^{(2\beta+1)\delta_{\sigma\mathcal{K}}} \, G_F^\mathcal{T} \big( s, \nu_1, I_\sigma(\nu_2) \big)
% \, \big[ G_F^\mathcal{T} \big(\bar s, \nu_1, I_\sigma(\nu_2) \big) \big]^* 
%  \, G_F^\mathcal{T} \big(s, \gamma, \nu_1, \nu_2 \big)
% \, \big[ G_F^\mathcal{T} \big(\bar s, \gamma, \nu_1, \nu_2 \big) \big]^*
% \bigg\} \; . \nonumber
%\end{align}
%
%%%%%%%%%%%%%%%%%%%%%%%%%%%%%%%%%%%%%%%%%%%%%%%%%%%%%%

\subsection{Four fermions}
Let us start with the four-fermion terms. We first note that the torus amplitude \eqref{4fermions_torus} does not contribute. This is due
to the fact that the two contributions from the Green's functions only depend either exclusively on the left moving spin structure or the right moving spin structure. Then, multiplying \eqref{4fermions_torus} with the partition functions and summing over both the left and the right moving spin structures produces zero. 

Thus, we are left with the four-fermion terms in \refeq{beforespin}. 
We consider
the untwisted K\"ahler modulus of the third torus. 
Then all the relevant correlators have Neumann-Neumann boundary conditions, no matter what surface we consider. Let us
concentrate on the spin structure dependent terms, i.e.\ the partition
function and the two fermion Green's functions. 
It is useful to define 
$Z_s^{\vartheta}$, the part of the  partition function
that contains the Jacobi theta functions,
which we give explicitly in \refeq{defZt}.
With this notation, we perform
the spin structure sum and obtain just a single Green's function:
\be  \label{absspin}
 \sum_s \eta_s Z_s^{\vartheta} G_F^\gamma(s;\nu_2,\nu_1) \, G_F(s;\nu_2,\nu_1)
 = G_F^\gamma(\nu_2,\nu_1)\,.
 \ee 
Here we use the following convention: if we do not indicate explicitly over which spin structure $s$
to sum, it is over even spin structures ($s=2,3,4$). 
We show \eqref{absspin} in \refapp{sec:spin_four}. The GSO sign $\eta_s$
is given in table \ref{table:spinstructures} in appendix \refapp{sec:ferm}.  What happened is that
(non-manifest) supersymmetry reduced the number of 
Green's functions in the integrand. 
We  introduced a new convention on the right hand side:
if we do not indicate the spin structure $s$ explicitly for a fermion Green's function,
it is $s=1$ (odd). 
Now we can  calculate
\begin{align}
 & \sum_s \eta_s Z_s^{\vartheta} 
 \big\langle V_{Z \bar Z}(p_1) V_{Z \bar Z}(p_2) \big\rangle_\sigma^s \big|_\fourfermion \nonumber \\
 & = - 2 \delta \int_\Ts \dd^2 \nu_1 \int_\Ts \dd^2 \nu_2
 \, e^{- \delta G_B^\sigma(\nu_1,\,\nu_2)} \,
 G_F^\gamma(\nu_2,\nu_1)
 \bar \p_2 G_F^\gamma(I_\sigma(\nu_1),I_\sigma(\nu_2)) \,, \label{ps}
\end{align}
where we used the identity
$\bar \p_1 \bar \p_2 G_B^\gamma(I_\sigma(\nu_1),I_\sigma(\nu_2)) = \bar\p_2 G_F^\gamma(I_\sigma(\nu_1),I_\sigma(\nu_2))$ to
replace a boson Green's function with a fermion Green's function. (Note that before differentiation,
the boson propagator depends on both $\nu$ and $\bar{\nu}= I_\sigma(\nu_1) - I_\sigma(\nu_2)$. We 
only display the dependence on one of them in the argument, and choose freely which one to display. The fermion propagator of course only depends on either
$\nu$ or $\bar{\nu}$. )

%%%%%%%%%%%%%%%%%%%%%%%%%%%%%%%%%%%%%%%%%%%%%%%%%%%%%%

\subsection{Evaluation  by analytic continuation}

In the limit $\delta \to 0$ the integrand in \refeq{ps} is a function of the relative coordinates $\nu \equiv \nu_2 - \nu_1$ and $\bar \nu \equiv I_\sigma(\nu_1) - I_\sigma(\nu_2)$ only. We 
change variables and trivially integrate over the ``center-of-mass'' coordinate. Setting $G_F^\gamma(\nu,0) \equiv G_F^\gamma(\nu)$ we obtain
\be
 \sum_s \eta_s Z_s^{\vartheta} 
 \big\langle V_{Z \bar Z}(p_1) V_{Z \bar Z}(p_2) \big\rangle_\sigma^s \big|_\fourfermion = - 4 \tau_2 \delta \int_{\Ts} \dd^2\nu \,
 G_F^\gamma(\nu) \bar \p G_F^\gamma(\bar\nu) + \mathscr{O}(\delta^2) \,,
 \label{fourfermionworldsheetintegral}
\ee
where the $2\tau_2$ is from the trivial integral.

\begin{figure}
\begin{center}
\includegraphics[width=0.35\textwidth]{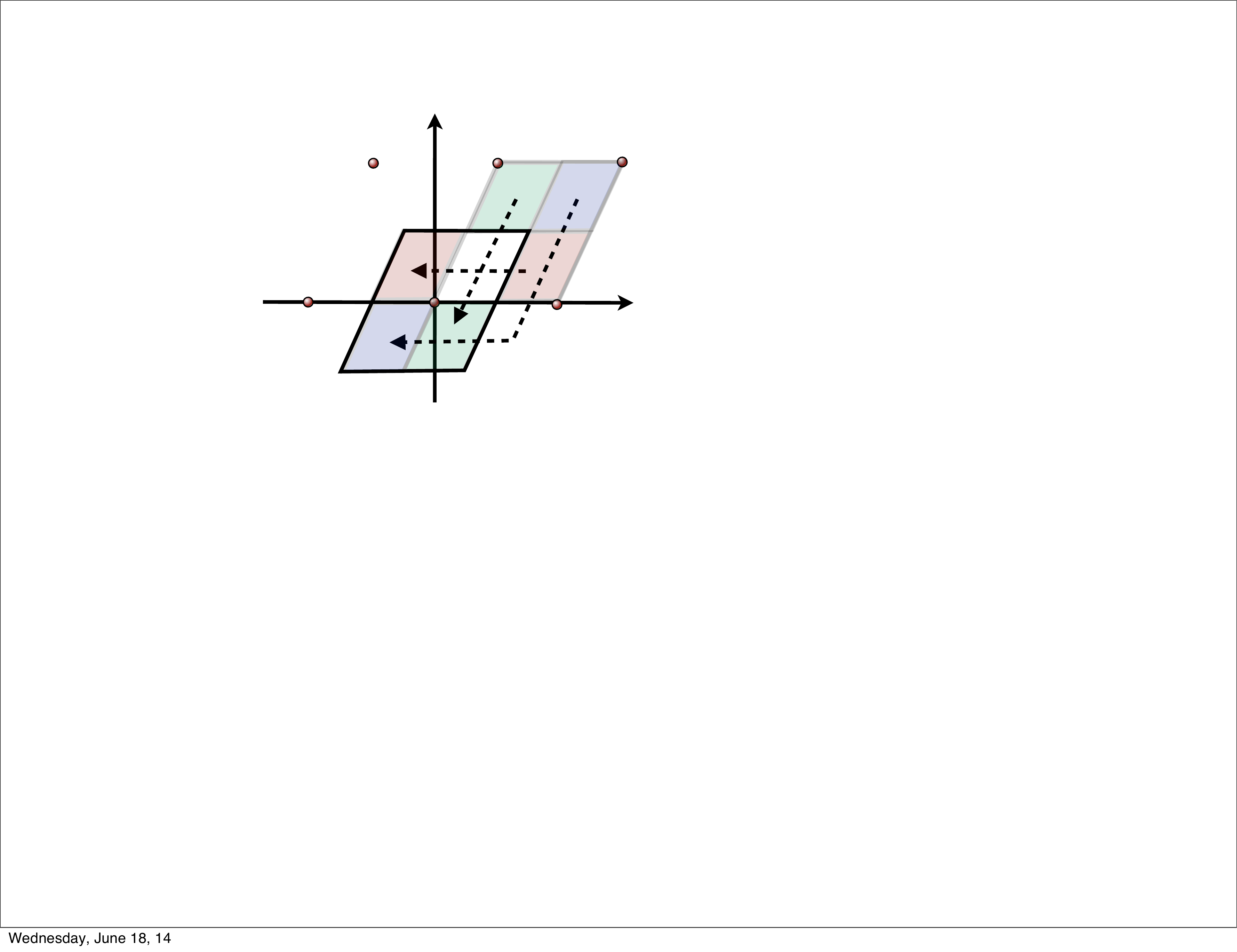}
\caption{The integrand is not invariant
under a half-integer shift, but the integral over the fundamental domain is. In this figure, each quarter-piece 
of integration domain is mapped by an {\it integer} lattice shift. }
\label{figmap}
\end{center}
\end{figure}

Naively, the integral in \refeq{fourfermionworldsheetintegral} appears divergent since the integrand is singular at the lattice points $\nu = \mathbb{Z} + \tau \mathbb{Z}$. However, when adding the contributions from the four corner points $0,1,\tau$ and $1+\tau$ we find that the divergences cancel. To see this, since the integrand in \refeq{fourfermionworldsheetintegral} is doubly periodic we can shift the integration region such that the only enclosed lattice point is the origin $\nu=0$, as in \reffig{figmap}. 
Close to $\nu = 0$ the integrand behaves as
\begin{align} \label{gfope}
 G_F^\gamma(\nu) \bar \p G_F^\gamma(\bar\nu) = - \frac{1}{\nu \bar\nu^2} - c_0 \, \frac{1}{\bar\nu^2} - c_1 \frac{\nu}{\bar\nu^2} + c_1 \frac{1}{\nu} + \text{finite} \,,
\end{align}
where $c_0$ and $c_1$ are the  first two coefficients from  the expansion $G_F^\gamma(\nu) = \frac{1}{\nu} + c_0 + c_1 \nu + \mathscr{O}(\nu^2)$. 
Because the same number of powers of $\nu$ and $\bar{\nu}$ never occur together, 
the integral of \refeq{gfope} over an infinitesimal disk centered at $\nu = 0$ must vanish:
\begin{align}
 & \lim_{\varepsilon \to 0} \int_{|\nu| \leq \varepsilon} \dd^2 \nu \, G_F^\gamma(\nu) \bar \p G_F^\gamma(\bar\nu) \nonumber \\
 & = 2 \lim_{\varepsilon \to 0} \int_0^\varepsilon \dd r \, r \int_0^{2\pi} \dd \theta \left[ - \frac{e^{i \theta}}{r^3} - c_0 \frac{e^{2 i \theta}}{r^2} - c_1 \frac{e^{3 i \theta}}{r} + c_1 \frac{e^{-i\theta}}{r} + \text{finite} \right] = 0 \,,
\end{align}
where the factor of $2$ in the second row arises from $\dd^2 \nu = 2 \, \dd(\Re \nu) \, \dd (\Im \nu)$. We changed to polar coordinates via $\nu = r e^{i \theta}$ and used $\int_0^{2 \pi} \dd \theta \, e^{i m \theta} = 0$ for all $m \in \mathbb{Z} \backslash\{0\}$.

We have argued that the integral in \refeq{fourfermionworldsheetintegral} is finite,
but finiteness is not manifest. 
To calculate the finite value explicitly we will use analytic continuation: we introduce a complex
parameter $s$ (not to be confused with the spin structure,
which is also denoted $s$) and 
define a meromorphic function $F(s)$ whose value at $s=1$ is the desired integral.  
This may seem  roundabout
for calculating a finite value, but we will find that
evaluation in some other region of the complex $s$ plane is quite simple,
and we believe that some of the intermediate steps could be of general interest.
For completeness, we rederive the result
by two alternative approaches in the appendix: partial integration 
and point splitting, in \refapp{sec:delta} and \refapp{sec:point}, respectively.  

In \refsec{sec:limitformula} we identify the Green's functions 
of bosons and fermions on the torus in terms of Eisenstein-Kronecker functions
$E_1$:
\begin{align}
 G_B^\gamma(\nu) & = E_1(\gamma,\nu) + \mathrm{const.} \,, \quad \\
 G_F^\gamma(\nu) & = - \p_\nu E_1(\gamma,\nu) \,.
\end{align}
The Eisenstein-Kronecker functions
 (see e.g.\ Ch.\ 8 of \cite{Weil}) are
 defined for any complex numbers $w$ and $z$ and for $\Re s >1$ by
\be
 E_s(w,z) = \left( \frac{\tau_2}{\pi} \right)^s {\sum_{m,n}}'\, \,
 \frac{\exp(\frac{2\pi i}{\tau_2} \Im [(w+m+n\tau)\bar z])}{|w+m+n\tau|^{2s}} \,,
\ee
where by $\sum_{m,n}'$ we intend the sum over all $m,n \in \mathbb{Z}$ such that $w + m+n\tau \neq 0$. For $\Re s > \hf$ the series converges almost everywhere, i.e. away from $z \in \mathbb{Z} + \tau \mathbb{Z}$, and in particular for $s=1$ the series converges to the boson Green's function. For other values of $s$ the Eisenstein-Kronecker function is defined by analytic continuation.

Now, we define a new function $F=F(s)$ of the complex parameter $s$, whose value at $s=1$ gives the desired worldsheet integral
\begin{align}
 F(s) := \int_\Ts \dd^2\nu \, \p E_s(\gamma,\nu) \bar\p^2 E_s(\gamma,\bar\nu) \,, \quad \label{relevantfunction}
 F(1) = \int_\Ts \dd^2\nu \, G_F^\gamma(\nu) \bar\p G_F^\gamma(\bar\nu) \,.
\end{align}
The motivation for studying $F(s)$ rather than $F(1)$ directly is that by taking $\Re s$ large enough, the integrand becomes {\it manifestly} finite everywhere in the integration region. Now, we have
\begin{align}
 F(s) = (-1)^3 \,\int \dd^2\nu \, & \left( \frac{\tau_2}{\pi} \right)^{s-1} \sum_{k,\ell}
 \frac{\gamma+k+\ell\bar\tau}{|\gamma+k+\ell\tau|^{2s}} \,
 e^{\frac{2\pi i}{\tau_2} \Im [(\gamma+k+\ell\tau)\bar \nu] } \nonumber \\
 \times & \left( \frac{\tau_2}{\pi} \right)^{s-2} \sum_{m,n} \,
 \frac{(\gamma+m+n\bar\tau)^2}{|\gamma+m+n\tau|^{2s}} \, e^{\frac{2\pi i}{\tau_2} \Im [(\gamma+m+n\tau) \nu ] } \,.
\end{align}
Using the orthogonality relation
\be
 \int \dd^2 \nu \,
 e^{\frac{2\pi i}{\tau_2} \Im[ (\gamma + k + \ell \tau) \bar\nu ]} \,
 e^{\frac{2\pi i}{\tau_2} \Im[ (\gamma + m + n \tau) \nu ]}
 = 2\tau_2 \, \delta_{\ell,-n} \, \delta_{k,m+n\delta_{\sigma\M}} \,,
\ee
which follows from $\bar\tau = - \tau + \delta_{\sigma\M}$, we arrive at
\begin{align}
 F(s) & = - 2 \pi \left( \frac{\tau_2}{\pi} \right)^{2s-2} \sum_{m,n} \frac{(\gamma+m+n\delta_{\sigma \M}-n\bar\tau)(\gamma+m+n\bar\tau)^2}{|\gamma+m+n\delta_{\sigma \M}-n\tau|^{2s}|\gamma+m+n\tau|^{2s}} \,, \\
 & = - 2 \pi \left( \frac{\tau_2}{\pi} \right)^{2s-2} \sum_{m,n} \frac{\gamma+m+n\bar\tau}{|\gamma+m+n\tau|^{2(2s-1)}} \,. \label{beforecomplex}
\end{align}
We now temporarily (\refeq{startcomplex} through \refeq{endcomplex}) treat $\gamma$ and $\bar\gamma$ as independent variables and only in the end we specify to real $\gamma$. Then we can write \refeq{beforecomplex} as
\begin{align} \label{startcomplex}
 F(s) & = 2 \pi \frac{\p}{\p\gamma} \bigg[ \frac{1}{2s-2} \left( \frac{\tau_2}{\pi} \right)^{2s-2} \sum_{m,n} \frac{1}{|\gamma+m+n\tau|^{2(2s-2)}} \bigg] \bigg|_{\gamma\,\mathrm{real}}
  = 2 \pi \frac{\p}{\p\gamma} \bigg[ \frac{E_{2s-2} (\gamma,0)}{2s-2} \bigg] \bigg|_{\gamma\,\mathrm{real}} \,.
\end{align}
The right-hand side defines the analytic continuation of $F(s)$ to the whole complex plane.

In the limit $s \to 1$ we have
\be
 \frac{E_{2s-2} (\gamma,0)}{2s-2} = \frac{E_{0} (\gamma,0)}{2s-2} + E_0'(\gamma,0) + \ldots
\ee
and 
$E_0(\gamma,0) = 0$, $E_0'(\gamma,0) = E_1(0,\gamma) $,
which is given by Kronecker's second limit formula
\be \label{endcomplex}
 E_1(0,\gamma) = - \ln \left| \frac{\vartheta_1(\gamma,\tau)}{\eta(\tau)} \right|^2 + \frac{2\pi}{\tau_2} (\Im \gamma)^2 \, ,
\ee
as we show in
 \refsec{sec:limitformula}  for completeness.
When evaluating $\p_\gamma E_1|_{\gamma\,\mathrm{real}}$ the second term  vanishes and we conclude for $F(s)$ at $s=1$:
\be
 F(1) = - 2 \pi \frac{\vartheta_1'(\gamma,\tau_\sigma)}{\vartheta_1(\gamma,\tau_\sigma)} \,.
\ee
Thus we find the four-fermion term
\be \label{4fsssum}
\sum_s \eta_s Z_s^\vartheta \vev{V_{Z \bar Z}(p_1)V_{Z \bar Z}(p_2)}_{\sigma} =  8 \pi \tau_2 \delta \, \frac{\vartheta_1'(\gamma,\tau)}{\vartheta_1(\gamma,\tau)} \, , 
\ee
which agrees with the alternative
calculations in \refapp{sec:delta} and \refapp{sec:point}.

%%%%%%%%%%%%%%%%%%%%%%%%%%%%%%%%%%%%%%%%%%%%%%%%%%%%%%%%

\subsection{Eight-fermion term: General arguments}
 \label{GenArg}
Let us now move on to the eight-fermion term \Ref{eightferm}. We first focus on $\sigma = \A,\M, \K$ and come back to the torus contribution in \refsec{sec:torus_8fermions}.
For ${\mathcal N}=4$ sectors, the eight-fermion term gave the full result
for higher-curvature corrections \cite{Pasquinucci:1997di}.
Since the eight-fermion term has four explicit momentum  factors in front, it can only contribute at order $\alpha' p^2$ if there is a pole $1 / \delta=2/(\alpha' p_1\cdot p_2)$ from vertex collisions.
For ${\mathcal N}=2$ sectors, there is no pole, so this term does not contribute at all 
to renormalization at order $\alpha' p^2$ . We will see this below, and we will also see that for ${\mathcal N}=1$ sectors there is a pole.
A more efficient way to compute
 it might be to  first change the superghost picture,
 which can reduce the number of explicit factors of momentum
 so that one does not have to extract a pole, but we will not attempt 
 that possibly instructive calculation in this paper.

 Although the present calculation will be longer than the one
 for the four-fermion term in the previous section,
 we will find that it reduces to an expression
 that is very similar to \refeq{4fsssum}. Probably
 there is a more direct way to see this,
 perhaps by worldsheet supersymmetry. 

We now return to the lifted eight-fermion term of \refeq{eightferm}:
\be \label{S8fermion}
S \equiv -\delta^2 \int_\mathcal{T} \dd^2 \nu_1 \int_\mathcal{T} \dd^2 \nu_2 \,
 e^{-\delta \, G_B^\sigma(\nu_1,\, \nu_2)} \,\, f(\nu_1,\nu_2),  \label{(9)}
\ee
where $f$ is
\begin{align}
&f(\nu_1,\nu_2)= \Big\{\, G_F \big(s; \nu_1, \nu_2\big)
 \, G_F \big(s; I_\sigma(\nu_1), I_\sigma(\nu_2)\big)
 \, - G_F \big(s; \nu_1, I_\sigma(\nu_2)\big)
 \, G_F \big(s; I_\sigma(\nu_1), \nu_2\big)
 \Big\} \nonumber \\
 & ~~~~~~~~~~~ \times
 \, G_F^\gamma \big(s;\nu_1, \nu_2\big)
 \, G_F^\gamma \big(s;I_\sigma(\nu_2), I_\sigma(\nu_1)\big) \; . 
\end{align}
One can check that $f$ as a whole is doubly periodic on the covering torus, i.e.
\be \label{doublep}
f(\nu_1, \nu_2)= f(\nu_1+1, \nu_2)=f(\nu_1, \nu_2+1)=f(\nu_1+\tau, \nu_2)=f(\nu_1, \nu_2+\tau)
\ee 
using:
\be
I_{\sigma} ( \nu+1 )=I_{\sigma}(\nu)-1, \,\, I_{\sigma}(\nu+\tau)=I_{\sigma}(\nu)- {\bar\tau}=I_{\sigma}(\nu)+\tau-\delta_{\sigma\,\mathcal{M}} \; . 
\ee
The cancellation of phases that ensures double periodicity as in \Ref{doublep}
is fairly nontrivial for the M\"obius strip, but it works.
We are interested in terms in $S$ of order $\delta$. In other words,
since $S$ has an overall factor $\delta^2$, we want to compute  
\be 
F=\left. \frac{S}{\delta}\right|_{\delta\rightarrow 0}=\left\{ - \delta \int_\mathcal{T} \dd^2 \nu_2 \int_\mathcal{T} \dd^2 \nu_1 \,
 e^{-\delta \, G_B^\sigma(\nu_1,\, \nu_2)} \,\, f(\nu_1,\nu_2) \right\}_ {\delta \rightarrow 0}. \label{FSdelta}
\ee
In the limit $\delta \rightarrow 0$ the only contribution to the integral on the right hand side arises from vertex operator collisions, i.e.\ from 
the integration regions with $\nu_1 \rightarrow \nu_2$ or  $\nu_1 \rightarrow I_\sigma (\nu_2)$. 
This can lead to an additional factor $1/\delta$, which then cancels 
the explicit factor of $\delta$ and leads to a finite result. In principle, a similar factor of $1/\delta$ could also arise from integrating $F$ 
over the world-sheet modulus $t$, but we will argue in a moment that this does not occur. Thus, the range of integration in 
\eqref{FSdelta} can effectively be restricted to the fundamental domain for one of the vertex operator positions (let us choose $\nu_2$) and
disks of infinitesimal radius for $\nu_1$ around $\nu_2$ and $I_\sigma (\nu_2)$. Therefore, schematically we need to calculate (and sum) integrals 
of the form
\be \label{Fdef}
F= \lim_{\varepsilon \to 0} \left\{ - \delta \int_\mathcal{T} \dd^2 \nu_2 \int_{D_\varepsilon} \dd^2 \nu_1 \,
 e^{-\delta \, G_B^\sigma(\nu_1,\, \nu_2)} \,\, f(\nu_1,\nu_2) \right\}_ {\delta \rightarrow 0},
\ee 
where $D_\varepsilon$ is a disk of radius $\varepsilon$ around $\nu_2$ or $I_\sigma (\nu_2)$ (whose dependence on $\nu_2$ we keep implicit).
We can now expand $f$ around poles and isolate the terms which survive the $\delta \to 0$ limit. At this point it is useful to separate $f$ as
\be
f=f_1-f_2,
\ee
where
\be
f_1&=& G_F \big(s; \nu_1, \nu_2\big)
 \, G_F \big(s; I_\sigma(\nu_1), I_\sigma(\nu_2)\big)
\, G_F^\gamma \big(s; \nu_1, \nu_2\big)
 \, G_F^\gamma \big(s; I_\sigma(\nu_2), I_\sigma(\nu_1)\big)  \label{f1}\ ,\\
f_2&=&  G_F \big(s; \nu_1, I_\sigma(\nu_2)\big)
 \, G_F \big(s; I_\sigma(\nu_1), \nu_2\big)
\, G_F^\gamma \big(s; \nu_1, \nu_2\big)
 \, G_F^\gamma \big(s; I_\sigma(\nu_2), I_\sigma(\nu_1)\big) \label{f2} \; . 
\ee   
Here $f_1$ has a pole at $\nu_1=\nu_2$, while in addition to this pole $f_2$ has one more pole at $\nu_1=I_\sigma(\nu_2)$ 
unless $\nu_2=I_\sigma(\nu_2)$ modulo torus periodicity. This region is a fixed line under involution and it is the boundary of annulus and M\"obius strip worldsheets.

\begin{figure}
\begin{center}
\includegraphics[width=0.4\textwidth]{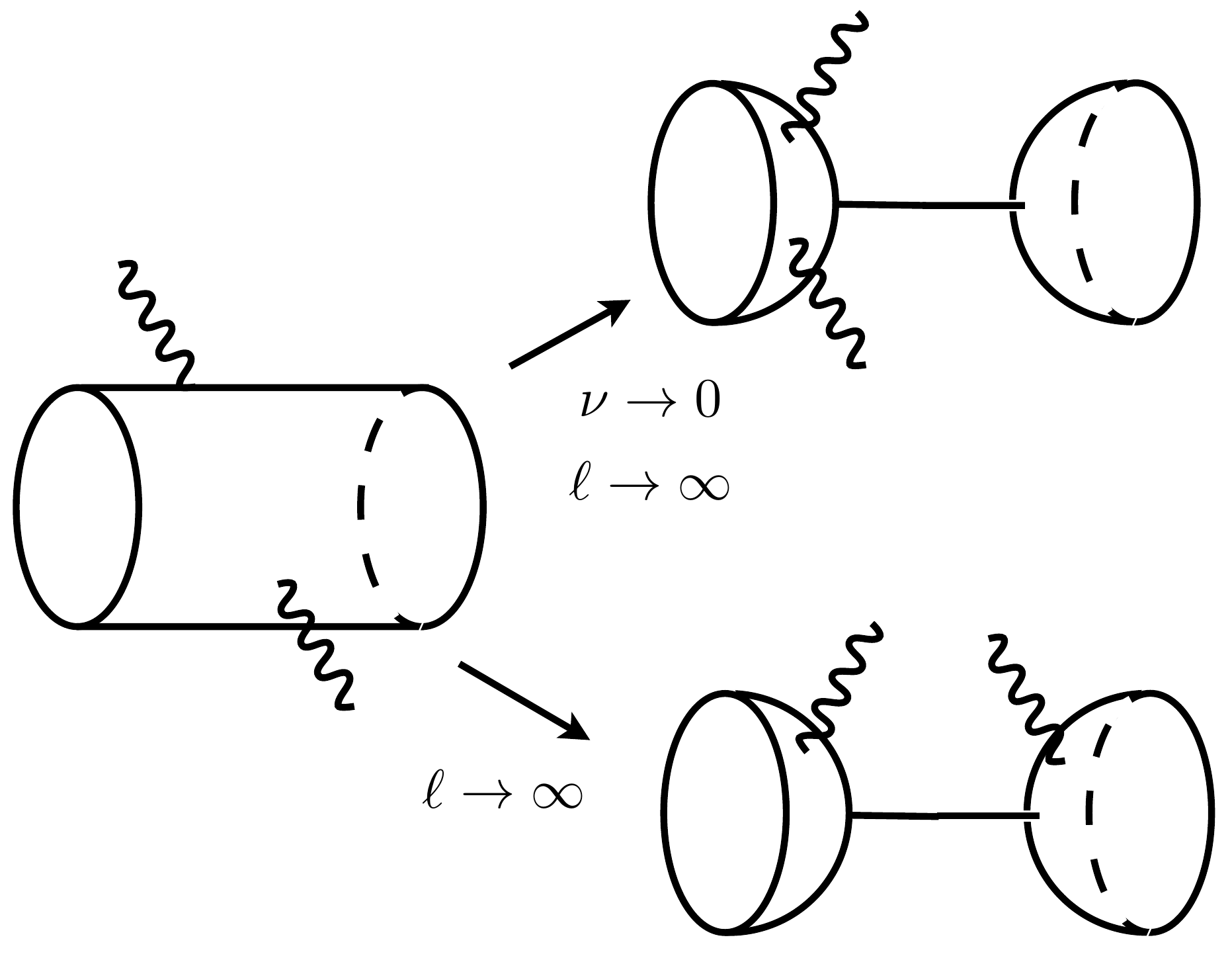}
\caption{One possible factorization of the one-loop two-point function
on the annulus.}
\label{factor}
\end{center}
\vspace{-5mm}
\end{figure}

Before we continue with the calculation, a comment about factorization. As stated above, the eight-fermion terms are of order $\delta^2$ to start with (recall $\delta \sim \alpha' p^2$). In order to contribute to the spacetime kinetic term we are interested in, the integral of \eqref{S8fermion} over the worldsheet modulus $t$ would have to provide a factor of $\delta^{-1}$. As just mentioned this could in principle either arise from a vertex collision, i.e.\ from a pole in the integral over $\nu \equiv \nu_1 - \nu_2$ (or $\nu \equiv \nu_1 - I_\sigma(\nu_2)$), or from the UV region of the $t$ integral (i.e. from $t \rightarrow 0$), as in eq.\ (A.6) of \cite{Antoniadis:2002cs}. We will see in the following that there are indeed contributions from vertex collisions. However, we do not expect any additional contributions from the $t$ integral. The reason can be understood from \reffig{factor}. 
For $\nu \neq 0$ and $t \rightarrow 0$ (or equivalently $\ell \rightarrow \infty$), the amplitude factorizes as in the lower part of that figure. There are also similar factorization limits of the M\"obius and Klein bottle amplitudes that add to the annulus of \reffig{factor}. For the amplitude not to vanish in this factorization limit, 
there would need to exist a disk coupling (or projective
plane coupling, for the unoriented worldsheets) between the modulus $T$ and some massless closed string state $C$
(which could be $T$ itself). In order for this coupling to lead to the $\delta^2$ factor in the numerator of the amplitude, the coupling should arise at second order in derivatives, with the $\delta^{-1}$ factor resulting from the exchange of $C$. Thus, 
for the amplitude not to vanish in this factorization limit, there should be a coupling proportional to $\partial_\mu T \partial^\mu C$  at the level of the disk or projective plane. However, it is known that
there is no contribution to the kinetic term of the closed string moduli at disk level \cite{Lust:2004cx}, and similarly
we do not expect any contribution from the projective plane. Thus, in the following we will 
single out vertex collision as
the only possible source for a $\delta^{-1}$ pole in the eight-fermion terms for this amplitude.\footnote{A similar argument applied to the four-fermion terms shows the absence of a mass term for the modulus $T$.}

The expert reader may have realized
that in the well-known calculation of mass terms for anomalous $U(1)$ gauge fields \cite{Antoniadis:2002cs}, the analogous argument must fail. There the factorization limit of  (the analog of the lower panel of) 
\reffig{factor} {\it does} lead to a non-vanishing contribution at order $\delta^0$ to the 2-point function of the gauge fields, signaling a mass term for the naively anomalous $U(1)$ gauge fields. In their case, the amplitude is of order $\delta$ to start with, but a $\delta^{-1}$ pole arises from the UV-region of the $t$ integral due to mixing between the anomalous $U(1)$ gauge field $A^\mu$ and the RR axion $a$ at disk level, i.e.\ a  coupling $A^\mu\partial_\mu a $. This involves only one derivative, in accordance with the factor $\delta$ in the numerator of the 2-point function of gauge fields. The $\delta^{-1}$ pole results from the exchange of the massless RR axion, in a process similar to the lower part of \reffig{factor} but
with the vertex operators of the vectors inserted at the boundaries. 
So there
is no contradiction. 

Finally, for completeness we mention that 
\cite{Antoniadis:2002cs} also have scale-dependent wave function renormalization terms like $\delta \log \delta$. Generic complete factorization
of diagrams such as those in \reffig{factor}
would also include field theory
limits of the tree-level string diagrams on each side, 
for example the closed string vertices could a priori collide 
on a sphere factorized away from the 
worldsheet, see e.g. \cite{Hashimoto:1996bf}, but note that
our spacetime momenta are much less generic than in that paper: our polarizations
are never parallel to any of the momenta.
Finally we mention \cite{Dudas:1999gz}
for an example of a somewhat analogous discussion for a four-point one-loop open string diagram.

%%%%%%%%%%%%%%%%%%%%%%%%%%%%%%%%%%%%%%%%%%%%%

\subsection{First piece of eight-fermion term}
\label{first_piece}
Consider $f_1$ from \refeq{Fdef}. This depends only on $\nu=\nu_1- \nu_2$,
so the pole comes from direct vertex collision (i.e. not collision with an image).
The relevant local disk $D_\varepsilon$ is $|\nu|<\varepsilon$,
and we have
\be
f_1= G_F \big(s; \nu \big)
 \, G_F \big(s; - \bar{\nu}\big)
\, G_F^\gamma \big(s; \nu\big)
 \, G_F^\gamma \big(s; \bar{\nu}\big)
 = -G_F \big(s; \nu \big)
 \, G_F \big(s; \bar{\nu}\big)
\, G_F^\gamma \big(s; \nu\big)
 \, G_F^\gamma \big(s; \bar{\nu}\big). 
\ee
In the second equality we assumed even spin structure.
A non-vanishing contribution can only arise from a term
precisely at order $|\nu|^{-(2-2\delta)}$. To isolate this term,
let us write generic terms of the untwisted correlator as
$b_{m,n}\nu^m\bar{\nu}^n$ and generic terms of the twisted correlator
as ${c}_{m,n}\nu^m\bar{\nu}^n$, then 
\bea
 && - \delta \, e^{-\delta G_B}\, f_1  \stackrel{|\nu|\to0}{\longrightarrow} \nonumber \\
&& \delta \cdot |\nu|^{2 \delta} \left({1 \over \nu}  + b_{1,0} \nu + \ldots \right)
\left({1 \over \bar{\nu}}  + b_{1,0}\bar{\nu} + \ldots \right)
\left({1 \over \nu} + c_{0,0} + c_{1,0}\nu + \ldots \right)
\left({1 \over \bar{\nu}} + c_{0,0} + c_{1,0} \bar{\nu} + \ldots \right)\ ,
  \nonumber 
\eea
where we used that $b_{0,0}=0$ for the untwisted correlator. 
The most divergent term will 
be cancelled by spin structure summation.
This also shows that for ${\mathcal N}=2$,
when all correlators are untwisted and hence also $c_{0,0}=0$, the relevant
pole is absent and the eight-fermion term cannot contribute.
For the next term we obtain (see \refapp{sec:delta} for the computation of $c_{0,0}$)
\be
 \delta \cdot (c_{0,0})^2 {1 \over |\nu|^{2-2\delta}} = \left( {\vartheta_{s}'(\gamma_3) \over \vartheta_{s}(\gamma_3)} \right)^{\!2} 
 {\delta \over |\nu|^{2-2\delta}}\; .
\ee
Note that there are no $\tau$-dependent pieces from the untwisted correlators  here since they 
only supplied the $1/|\nu|^2$ (whose residue 1 is determined by the OPE). 

Thus, integration over the disk for $\nu$ gives 
\be
\lim_{\delta\to 0} \int_{|\nu|<\varepsilon} \dd^2 \nu \, (c_{0,0})^2 {\delta \over |\nu|^{2-2\delta}} =
\lim_{\delta\to 0} 4 \pi \int_0^\varepsilon \dd |\nu| \, (c_{0,0})^2 {\delta \over |\nu|^{1-2\delta}} =  2 \pi (c_{0,0})^2.
\ee
So the final result for this piece is
\be
 \lim_{\varepsilon \to 0} \left\{ - \delta \int_\mathcal{T} \dd^2 \nu_2 \int_{D_\varepsilon} \dd^2 \nu_1 \,
 e^{-\delta \, G_B^\sigma(\nu_1,\, \nu_2)} \,\, f_1(\nu_1,\nu_2) \right\}_ {\delta\rightarrow 0} &=& 2 \pi \left( {\vartheta_{s}'(\gamma_3) \over \vartheta_{s}(\gamma_3)} \right)^{\!2} \int_\mathcal{T} \dd^2 \nu_2\ .
\ee
%There is another term at order $|\nu|^0$:
%\be
%\delta \cdot |\nu|^{2 \delta} \big[ (b_{1,0})^2 + (c_{1,0})^2 + b_{1,0} c_{1,0} + b_{1,0} c_{1,0} \big]
%\ee
%This could contribute a pole from the $t$ integral, as we will see below.

%%%%%%%%%%%%%%%%%%%%%%%%%%%%%%%%%%%%%%%%%%%%%

 \subsubsection{Spin structure sum}

We would like to compute 
\be \label{sss}
\sum_{{\rm even} } \eta_s\, Z_s^{\vartheta}\, \left( {\vartheta_{s}'(\gamma_3) \over \vartheta_{s}(\gamma_3)} \right)^{\!2} .
\ee   
In \refapp{sec:spin} we obtain
\be
\sum_{{\rm even} } \eta_s\, Z_s^{\vartheta}\, \left( {\vartheta_{s}'(\gamma_3) \over \vartheta_{s}(\gamma_3)} \right)^{\!2} &=& 
- \frac{\vartheta_1' (\gamma_3)}{\vartheta_1 (\gamma_3)}+ \frac{\vartheta_1' (g_1+h_1\tau+\gamma_1)}{\vartheta_1(g_1+h_1\tau+\gamma_1)} 
+\frac{\vartheta_1' (-g_1-h_1\tau+\gamma_2)}{\vartheta_1(-g_1-h_1\tau+\gamma_2)}+ 2\frac{\vartheta_1'(2 \gamma_3)}{\vartheta_1(2 \gamma_3)} \non
&&=  - 3\frac{\thbwp{1/2}{1/2+\gamma_3} (0)}{\thbw{1/2}{1/2+\gamma_3} (0)}+ 
\frac{\thbwp{1/2+h_1}{1/2+g_1+\gamma_1}(0)}{\thbw{1/2+h_1}{1/2+g_1+\gamma_1}(0)} 
+\frac{\thbwp{1/2-h_1}{1/2-g_1+\gamma_2}(0)}{\thbw{1/2-h_1}{1/2-g_1+\gamma_2}(0)}\ , \label{spinstructure_f1}
\ee
see \reftab{tab:details} for $g_i$ and $h_i$. In the last equality we used $2 \gamma_3=n-\gamma_3$ (with some integer $n$) for $\mathbb{Z}_6'$, which is our main example, so that
\be
\frac{\thbwp{1/2}{1/2+2 \gamma_3}(0)}{\thbw{1/2}{1/2+2 \gamma_3}(0)} = 
-\frac{\thbwp{1/2}{1/2+\gamma_3}(0)}{\thbw{1/2}{1/2+\gamma_3}(0)}\ .
\ee  
Note that when moving variables in the argument into the characteristics for the second and third term in the first line 
of \eqref{spinstructure_f1} they generate additional constants proportional to $h$, but they add up to zero in this case. 
Now we turn to the $f_2$ part.

%%%%%%%%%%%%%%%%%%%%%%%%%%%%%%%%%%%%%%%%%%%%%

\subsection{Second piece of eight-fermion term} 

Unlike the first piece $f_1$, the second piece $f_2$ of the eight-fermion term has two poles: $\nu_1=\nu_2$ coming from twisted correlators and $\nu_1=I_\sigma(\nu_2)$ from untwisted ones. We will compute them separately. Note that the two poles coincide if $\nu_2=I_\sigma(\nu_2)$ (these are the fixed points under the involution of $\nu_2$, i.e.\ the open string boundaries). One might be worried about overcounting, but  the integration region $\nu_2=I_\sigma(\nu_2)$ does not contribute, since it is a region of measure zero. Thus, we may in fact include it in both terms for convenience. We will come back to this at the end of this subsection. 

%%%%%%%%%%%%%%%%%%%%%%%%%%%%%%%%%%%%%%%%%%%%%

\subsubsection{Pole $\nu_1=\nu_2$ from twisted correlators}
\label{second_piece_nu1_nu2}

Let us write $\nu=\nu_1-\nu_2$ so the
pole of twisted correlators is at $\nu=0$.
Then 
\be
f_2 \, \stackrel{|\nu|\to0}{\longrightarrow} \, \frac{G_F \big(s; \nu_2, I_\sigma(\nu_2)\big)
 \, G_F \big(s; I_\sigma(\nu_2), \nu_2\big)}{|\nu|^2} 
 = - \frac{G_F\big(s; \tilde{\nu}_2 \big)^2}{|\nu|^2},
\ee
where $\tilde{\nu}_2 \equiv  I_\sigma(\nu_2)-\nu_2$ and in the equality we assumed even spin structure for $s$. Therefore the $\nu$-integral for small $\varepsilon$ gives
\be
 \left\{-\delta \int_{\nu\leq\varepsilon} \dd^2 \nu R_\delta\, f_2\right\}_{\delta=0} =
G_F\big(s; \tilde{\nu}_2 \big)^2  \left\{ \delta \int_{\nu\leq\varepsilon} \dd^2 \nu
\frac{ |\nu|^{2 \delta}}{|\nu|^2}\right\}_{\delta=0} = 2 \pi \, G_F \big(s; \tilde{\nu}_2 \big)^2\ .
\ee
Now we need to compute the $\nu_2$ integral together with the spin structure summation.
We use the well-known identity for the square of the fermion propagator (e.g.\  (3.6) in \cite{Stieberger:2002wk})
\be \label{GFsquared}
(G_F(s; \tilde{\nu}_2))^2=\partial_{\nu}^2 \ln \vartheta_s(0) + g(\tilde{\nu}_2) \; , 
\ee
where $g(\tilde{\nu}_2)$ depends on $\tilde \nu_2$, but is independent of the spin structure. So it cannot survive spin structure summation and can be discarded. %(Indeed, note that $g(\tilde{\nu}_2)$ has a pole at $\nu_2=I_\sigma(\nu_2)$,  the  involution fixed point set. When approaching $|\nu|\to 0$, $f_2$ goes like $1/|\nu|^4$, which cancels by spin structure summation.) 
On the other hand the first term on the right hand side depends only on the spin structure, not on $\nu_2$. This means that the integral becomes trivial after spin structure summation:
\be
\sum_{{\rm even}} \eta_s\, Z_s^{\vartheta}  \int_\mathcal{T} \dd^2 \nu_2 \, G_F(s; \tilde{\nu}_2)^2=
\int_\mathcal{T} \dd^2 \nu_2 \, \sum_{{\rm even} } \eta_s\, Z_s^{\vartheta}\, G_F(s; \tilde{\nu}_2)^2=
\left(\sum_{{\rm even} } \eta_s\, Z_s^{\vartheta} \,\partial_{\nu}^2 \ln \vartheta_s(0)\right) \int_\mathcal{T} \dd^2 \nu_2\ .
\ee    
The spin structure sum in the bracket is
 computed in the appendix in \refeq{eightsumresult}.
Thus we obtain
\be
\sum_s \eta_s\, Z_s^{\vartheta}\,\lim_{\varepsilon \to 0} \left\{  \delta \int_\mathcal{T} \dd^2 \nu_2 \int_{|\nu|<\varepsilon} \dd^2 \nu_1 \,
 e^{-\delta \, G_B^\sigma(\nu_1,\, \nu_2)} \,\, f_2(\nu_1,\nu_2) \right\}_ {\delta\rightarrow 0} =
- 2 \pi   \sum_{i=1}^3 \frac{\thbwp{1/2+h_i}{1/2+g_i+\gamma_i}(0)}{\t{1/2+h_i}{1/2+g_i+\gamma_i}(0)} \int_\mathcal{T} \dd^2 \nu_2.
\ee

%%%%%%%%%%%%%%%%%%%%%%%%%%%%%%%%%%%%%%%%%%%%%

\subsubsection{Pole $\nu_1=I_\sigma(\nu_2)$ from untwisted correlators}
\label{second_piece_nu1_Inu2} 

Now we consider the pole  $\nu_1=I_\sigma(\nu_2)$ from untwisted correlators.
There is a minor complication for the Klein bottle amplitude, where 
points that are mapped outside the original torus by the involution
should in principle be mapped back. 
This introduces a step function for the involution (see \eqref{polekleinbottle_step} in appendix \refapp{stepfunctionappendix}),
but  due to the fact that the integrand as a whole is doubly periodic on the torus,
this actually leads to the same result as without the step function, 
as we show in  \refapp{stepfunctionappendix}. So,
 in order not to overload the notation, we neglect the step function 
 in the following calculations.

By introducing  $\nu\equiv\nu_1-I_\sigma(\nu_2)$ and $\tilde{\nu}_2\equiv I_\sigma(\nu_2)-\nu_2$, one can rewrite the arguments of the correlators as follows: 
\be 
\nu_1-I_\sigma(\nu_2)=\nu,\,\,\, 
I_\sigma(\nu_1)-\nu_2=-\bar{\nu}+\tau \delta_{\sigma \mathcal{K}},\,\,\, 
\nu_1-\nu_2= \nu+\tilde{\nu}_2, \,\,\,
I_\sigma(\nu_2)-I_\sigma(\nu_1)= \bar{\nu}+\tilde{\nu}_2 - \tau \delta_{\sigma \mathcal{K}}\ .
\ee
Then near the pole, we have (cf.\ \eqref{GFnu1tau})
\be
&&f_2 \, \stackrel{|\nu|\to0}{\longrightarrow} \, - e^{2 \pi i \gamma \delta_{\sigma \mathcal{K}}} \,\, \frac{G_F^\gamma \big(s; \tilde{\nu}_2 \big)^2}{|\nu|^2}, \\
&&G_B^{\sigma}(\nu_1,\nu_2) = G_B^{\sigma}(\nu_1,I_\sigma(\nu_2))
\stackrel{|\nu|\to0}{\longrightarrow} -\ln |\nu|^2
\ee
and the $\nu$-integral over the disk $|\nu_1-I_\sigma(\nu_2)|\leq\varepsilon$ for small $\varepsilon$ gives
\be \label{spinstructtwo}
\left\{-\delta \int_{\nu\leq\varepsilon} \dd^2 \nu\, e^{-\delta\, G_B^\sigma}\, f_2\right\}_{\delta=0} =
e^{2 \pi i \gamma \delta_{\sigma \mathcal{K}}}\,G_F^\gamma \big(s; \tilde{\nu}_2 \big)^2  \left\{ \delta \int_{\nu\leq\varepsilon} \dd^2 \nu
\frac{ |\nu|^{2 \delta}}{|\nu|^2}\right\}_{\delta=0} =  2 \pi \,e^{2 \pi i \gamma \delta_{\sigma \mathcal{K}}}\, G_F^\gamma \big(s; \tilde{\nu}_2 \big)^2\ . \non  
\ee
Now let us perform the spin structure summation including the partition function before $\nu_2$ integration.
%
%%%%%%%%%%%%%%%%%%%%%%%%%%%%%%%%%%%%%%%%%%%%%%
%
%\subsubsection*{Spin structure sum}
%We would like to compute 
%\be
%\label{spinstructtwo}
%\int_{\mathcal{T}} \dd^2 \nu_2\,\,\sum_{{\rm even} } \eta_s\, Z_s^{\vartheta}\, G_\gamma(s;\tilde{\nu}_2)^2.
%\ee   
This differs from \Ref{absspin} in that {\it both} correlators are twisted.
In \refapp{sec:spin_eight} we give the details, leading to (cf.\ eq.\ \eqref{spinsumeightfinal})
\be \label{spin_sum_G}
\sum_{{\rm even} } \eta_s\, Z_s^{\vartheta}\, (G_F^\gamma(s;\tilde{\nu}_2))^2\,\, =\,\, 2\, G_F^{2\gamma}(\tilde{\nu}_2)\ .
\ee
%
%%%%%%%%%%%%%%%%%%%%%%%%%%%%%%%%%%%%%%%%%%%%%%
%
%\subsubsection*{Integral over $\nu_2$}
Now let us consider the $\nu_2$ integration, namely we have to compute
\be
\int_{\mathcal{T}} \dd^2 \nu_2\,\ 4 \pi\,e^{2 \pi i \gamma \delta_{\sigma \mathcal{K}}}\,G_F^{2\gamma}(\tilde{\nu}_2) \; . 
\ee
At first sight one might think that this integral
is ambiguous, since the integrand is a twisted correlator,  i.e. 
the value of the integral could depend on the choice of fundamental region.
But the original integrand was doubly periodic, so that cannot be.
Indeed the twist does not affect the $\nu\rightarrow \nu+\tau$
direction due to the combination $\tilde{\nu}_2=1-2\,{\rm Re} \, \nu_2 + \frac{\tau}{2}\,\delta_{\sigma\ck}$ appearing in the argument. Also, it has no twist in the upper characteristic and is therefore periodic  in the horizontal direction $\nu\rightarrow \nu+1$. 
%Indeed, using that
%for annulus, M\"obius and Klein bottle one has $\tilde{\nu}_2=1-2\,{\rm Re} \, \nu_2 + \frac{\tau}{2}\,\delta_{\sigma\ck}$:
%\be
%&&G_F^{2\gamma}(\tilde{\nu}_2)=
%G_F^{2\gamma}(1-2\,{\rm Re} \, \nu_2+ \frac{\tau}{2}\,\delta_{\sigma\ck})=
%G_F^{2\gamma}(-2\,{\rm Re} \, \nu_2+ \frac{\tau}{2}\,\delta_{\sigma\ck}) \non \\
%&& G_F^{2\gamma}(-2\,{\rm Re} \, (\nu_2+1)+ \frac{\tau}{2}\,\delta_{\sigma\ck})= 
%G_F^{2\gamma}(-2\,{\rm Re} \, \nu_2-2+ \frac{\tau}{2}\,\delta_{\sigma\ck})= \non 
%&&=G_F^{2\gamma}(-2\,{\rm Re} \, \nu_2+ \frac{\tau}{2}\,\delta_{\sigma\ck})\ .
%\ee

For the measure, we can write $\dd^2\, \nu_2=2 \, \dd( {\rm Im} \, \nu_2) \dd\, ( {\rm Re}\, \nu_2)$. For the M\"obius strip it is possible to map the tilted covering torus  to a rectangular torus. This can be done by cutting out the  triangle in the   M\"obius fundamental region 
that lies to the right of the line $\nu_2=1$, and shifting this triangle by one horizontal lattice shift to the left. 
The resulting region is square, as that for the annulus and the Klein bottle, and the integral over this region is the same as the integral over the original covering torus. This is because the correlator is periodic  in the horizontal direction, as noted above.
Thus we have
\be
\int_{\mathcal{T}} \dd^2 \nu_2\,4 \pi e^{2 \pi i \gamma \delta_{\sigma \mathcal{K}}}\,G_F^{2\gamma}(\tilde{\nu}_2)&=&
8 \pi \int_0^{\tau_2} \dd\, {\rm Im} \, \nu_2 \int_0^1 \dd\, {\rm Re} \, \nu_2\,e^{2 \pi i \gamma \delta_{\sigma \mathcal{K}}}\, G_F^{2\gamma}(-2\,{\rm Re} \, \nu_2+ \frac{\tau}{2} \,\delta_{\sigma\ck}) \non
&=& 8 \pi \,\tau_2\, e^{2 \pi i \gamma \delta_{\sigma \mathcal{K}}}\,\int_0^1 \dd\, x\,\ G_F^{2\gamma}(-2x+ \frac{\tau}{2}\,\delta_{\sigma\ck})\ .
\ee
Using the following well-known representation for $G_F$ (see e.g.\ the appendix
of \cite{Berg:2011ij}):
\be
G_F^y(z)=\frac{\vartheta_1(y+z)\vartheta_1'(0)}{\vartheta_1(y)\vartheta_1(z)}=\pi \cot(\pi y)+\pi\cot(\pi z)+
4 \pi \sum_{m=1}^\infty\sum_{n=1}^\infty q^{mn}\sin(2 \pi m y+2 \pi n z),
\ee 
we end up with
\be
8 \pi^2 \,\tau_2\, e^{2 \pi i \gamma \delta_{\sigma \mathcal{K}}}\,\int_0^1 \dd\, x\,\Big[ \cot(2 \pi \gamma)+ \cot\left( \pi(-2x+\frac{\tau}{2}\,\delta_{\sigma\ck})\right)\Big] \; . 
\ee
The other pieces integrate to zero, as happened in  \cite{Berg:2011ij}. 
Using
\be
\int_0^1\dd x \cot\left(\pi(- 2x +\frac{\tau}{2} \delta_{\sigma\ck}) \right) = - i\,\delta_{\sigma\ck} \,{\rm sgn}\left({\rm Im} \tau\right), 
\ee
we get
\be
\int_{\mathcal{T}} \dd^2 \nu_2\,\ 4 \pi\,e^{2 \pi i \gamma \delta_{\sigma \mathcal{K}}}\,G_F^{2\gamma}(\tilde{\nu}_2)
&=& 8 \pi^2 \,\tau_2\, e^{2 \pi i \gamma \delta_{\sigma \mathcal{K}}}\big[\cot(2 \pi \gamma)- i \,\delta_{\sigma\ck}\big] \\
&= &8 \pi^2 \,\tau_2\,\frac{\cot(2 \pi \gamma)}{\cos(2 \pi \gamma \,\delta_{\sigma\ck})}
\ee
and thus for  all surfaces
\be \label{nu2integral_untwisted}
 \sum_s \eta_s\, Z_s^{\vartheta}\,\lim_{\varepsilon \to 0} \left\{ \delta \int_\mathcal{T} \dd^2 \nu_2 \int_{|\nu_1-I_\sigma(\nu_2)|\leq\varepsilon} \dd^2 \nu_1 \,
 e^{-\delta \, G_B^\sigma(\nu_1,\, \nu_2)} \,\, f_2(\nu_1,\nu_2) \right\}_ {\delta=0}  = - 8 \pi^2 \,\tau_2\,\frac{\cot(2 \pi \gamma)}{\cos(2 \pi \gamma \,\delta_{\sigma\ck})}.  
\ee
Let us come back to the issue raised at the beginning of this subsection, concerning the potential overcounting of the boundary contributions. In \refsec{second_piece_nu1_nu2} we calculated the contributions from poles at $\nu_1 = \nu_2$. The contribution from the pole is still integrated over $\nu_2$, including the boundary region. However, as discussed below \refeq{GFsquared}, the integrand after spin structure summation does not depend on $\nu_2$ at all. Since the boundary contribution corresponds to a region of measure zero, it vanishes. 

Similarly, in \refsec{second_piece_nu1_Inu2} we integrated ${\rm Re} (\nu_2) \equiv x$ from $0$ to $1$ which includes the boundary regions for $\ca$ and $\cm$. However, as in \refsec{second_piece_nu1_nu2}, these regions turn out not to contribute at all. This is because the (in principle divergent) contributions from $x=0, x=1/2$ and $x=1$ cancel each other by periodicity of $\cot(-2 \pi x)$ and for the integral of $\cot(2 \pi \gamma)$, which is independent of $x$, the boundary contribution vanishes again because the boundary is of measure zero. 

%%%%%%%%%%%%%%%%%%%%%%%%%%%%%%%%%%%%%%%%%%%%%

\subsection{Adding up eight-fermion: $\sigma=\A,\M,\K$}
It is useful to add up contributions from $f_1$ and $f_2$ before $t$-integration:
\be
F&=& \lim_{\varepsilon \to 0} \left\{ - \delta \int_\mathcal{T} \dd^2 \nu_2 \int_{D_\varepsilon} \dd^2 \nu_1 \,
 e^{-\delta \, G_B^\sigma(\nu_1,\, \nu_2)} \,\, f(\nu_1,\nu_2) \right\}_ {\delta=0} \\
 &=& \lim_{\varepsilon \to 0} \left\{ - \delta \int_\mathcal{T} \dd^2 \nu_2 \int_{D_\varepsilon} \dd^2 \nu_1 \,
 e^{-\delta \, G_B^\sigma(\nu_1,\, \nu_2)} \,\, \left[f_1(\nu_1,\nu_2)-f_2(\nu_1,\nu_2) \right]\right\}_{\delta=0} \; .
 \ee
From the results obtained above we have
\be 
\sum_{\rm even} \eta_s Z_s^{\vartheta}\, F & =& -16 \pi \, \tau_2\,
 \frac{\thbwp{1/2}{1/2+\gamma_3} (0)}{\thbw{1/2}{1/2+\gamma_3} (0)}
- 8 \pi^2\, \tau_2\,\frac{\cot(2\pi \gamma)}{\cos(2 \pi \gamma\, \delta_{\sigma\ck})} \\
&=& -8 \pi \,\tau_2\,
\left[2\, \frac{\thbwp{1/2}{1/2+\gamma_3} (0)}{\thbw{1/2}{1/2+\gamma_3} (0)}
+ \frac{\pi\,\cot(2\pi \gamma)}{\cos(2 \pi \gamma\, \delta_{\sigma\ck})} \right] \ .
 \label{8f}
\ee
Note that the factor $\tau_2$ comes from $\int_{\mathcal T} \dd^2 \nu_2=2 \tau_2 $. This was also the case for 
the 4-fermion integral. So in both cases we arrive at the following $t$ integral:
\be
I=\int_{0}^{\infty} \frac{dt}{t^2} \, \frac{\tht_1'(\gamma,\tau)}{\tht_1(\gamma,\tau)}\ .
\ee

%%%%%%%%%%%%%%%%%%%%%%%%%%%%%%%%%%%%%%%%%%%%%

\subsection{Eight-fermion term: torus}
\label{sec:torus_8fermions}
For the torus the spin structure $s$ for left-movers
and the spin structure $\tilde{s}$ for right-movers are a priori independent.
The sum over even-even spin structures gives
\begin{align}
 & \sum_{s, \tilde s} \eta_s \tilde{\eta}_{\tilde{s}} \, Z_s^\vartheta \tilde{Z}_{\tilde{s}}^\vartheta \,
 G_F(s;\nu_1,\nu_2) \,
 \big[ G_F(\tilde s;\nu_1,\nu_2) \big]^* \,
 G_F^{\vec \gamma}(s; \nu_1,\nu_2) \,
 \big[ G_F^{-\vec \gamma}(\tilde s; \nu_2,\nu_1) \big]^* \\
 & = -
 G_F^{\vec \gamma}(\nu_1,\nu_2) \,
 \big[ G_F^{\vec \gamma}(\nu_1,\nu_2) \big]^*\,,
\end{align}
where we used $G_F^{-\vec \gamma}(\tilde s; \nu_2,\nu_1) = - G_F^{\vec \gamma}(\tilde s; \nu_1,\nu_2)$ before performing the spin structure sum.
Note that the spin structure summation is the same
as in the analogous calculation for open-string external states, cf.\ \cite{Berg:2011ij}, but now for left-movers and right-movers separately. 
Focusing on the $|\nu|^{-2+2 \delta}$ term as above, we have, using the same expansion
around $\nu=0$ as above,
\bea
\delta^2\cdot   |\nu|^{2 \delta} 
\left({1 \over \nu}  + c_{0,0} + c_{1,0}\nu  + \ldots \right)
\left({1 \over \bar{\nu}}  +  \tilde{c}_{0,0} + \tilde{c}_{1,0}\nu  + \ldots \right)
= \delta^2\cdot   |\nu|^{2 \delta} 
{1 \over \nu}  
{1 \over \bar{\nu}}  +\ldots\ , \label{torusGF2}
\eea
i.e.\ there is indeed a contribution to linear order in $\delta$, as the $\nu$-integral of \eqref{torusGF2} will give a factor of $\delta^{-1}$. 
%What happened?
%Because the spin structure summation lowered the strength of the singularity
%(as a consequence of worldsheet supersymmetry),
%there is no pole contribution. 
%One could guess that the constant
%result from the torus will be cancelled by one from the Klein bottle.
This even-even 8-fermion term should reproduce the known one-loop correction from the torus
that is proportional to the Euler number, as calculated in a different way
in \cite{Antoniadis:1997eg} for Type II strings on a general Calabi-Yau manifold.
The torus amplitude
is inherited from the $\cn=2$ parent theory
(i.e.\ before orientifolding), and it would be interesting to
 check agreement with \cite{Antoniadis:1997eg} in more detail. 
Finally, the odd-odd amplitude does not contribute to 
the 2-point function of the K\"ahler modulus, as there are not enough external worldsheet fermions $\psi^{\mu}$ to cancel the zero modes.
This is consistent with the calculation in \cite{Antoniadis:1997eg}. 
The odd-odd 8-fermion term is  relevant when calculating the 
one-loop correction to the Einstein-Hilbert term via scattering of two (external) gravitons, cf.\  \cite{Kohlprath:2003pu}.

%%%%%%%%%%%%%%%%%%%%%%%%%%%%%%%%%%%%%%%%%%%%%%

\subsection{Adding up orbifold $k$ sectors}
\label{sec:addingup}

Let us summarize what we have obtained. For the 4-fermion term (see \refeq{4fsssum}) we have
\be \label{4F}
\sum_s \eta_s Z_s^{\vartheta} 
\big\langle V_{Z \bar Z}(p_1) V_{Z \bar Z}(p_2) \big\rangle_\sigma^s \big|_{\substack{\text{Four}~~~~\, \\ \text{Fermions}}}
=\delta \cdot 8\pi\, \tau_2\, \frac{\tht_1'(\gamma_3,\tau)}{\tht_1(\gamma_3,\tau)} 
\ee
and for the 8-fermion term (see \refeq{8f})
\be \label{8F}
\sum_s \eta_s Z_s^{\vartheta} 
\big\langle V_{Z \bar Z}(p_1) V_{Z \bar Z}(p_2) \big\rangle_\sigma^s \big|_{\substack{\text{Eight}~~~~\, \\ \text{Fermions}}}
=\delta \cdot (-8 \pi) \,\tau_2\,
\left[2\, \frac{\thbwp{1/2}{1/2+\gamma_3} (0)}{\thbw{1/2}{1/2+\gamma_3} (0)}
+ \frac{\pi\,\cot(2\pi \gamma_3)}{\cos(2 \pi \gamma_3\, \delta_{\sigma\ck})} \right] . 
\ee
Thus altogether
\be
&&\sum_s \eta_s Z_s^{\vartheta} 
\left[\big\langle V_{Z \bar Z}(p_1) V_{Z \bar Z}(p_2) \big\rangle_\sigma^s \big|_\fourfermion
+\big\langle V_{Z \bar Z}(p_1) V_{Z \bar Z}(p_2) \big\rangle_\sigma^s \big|_\eightfermion \right] \non
&&= \delta \cdot (-8 \pi) \,\tau_2\,
\left[\frac{\thbwp{1/2}{1/2+\gamma_3} (0)}{\thbw{1/2}{1/2+\gamma_3} (0)}
+ \frac{\pi\,\cot(2\pi \gamma_3)}{\cos(2 \pi \gamma_3\, \delta_{\sigma\ck})} \right] \non
&&\equiv \delta \cdot C_\sigma^{(k)},
\ee
where we defined the $t$-dependent function
\be
C_\sigma^{(k)} \equiv-8 \pi \,\tau_2\,
\left[\frac{\thbwp{1/2}{1/2+\gamma_3} (0)}{\thbw{1/2}{1/2+\gamma_3} (0)}
+ \frac{\pi\,\cot(2\pi \gamma_3)}{\cos(2 \pi \gamma_3\, \delta_{\sigma\ck})} \right]. \label{Ck}
\ee
Here $\gamma_3= k\, v_3$ for annulus and M\"obius, whereas $\gamma_3= 2 k\, v_3$ for Klein bottle.
It is convenient to add up orbifold sectors for each surface before 
performing the integral over worldsheet moduli ($t$ or $\ell$). 
To be explicit, we focus on a particular orientifold,
the ${\mathcal N}=1$ sectors of the $\mathbb{Z}_6'$ orientifold, though our results
should generalize straightforwardly to other examples. In the notation of \cite{Berg:2005ja}
(section 4 and appendix A.1) we compute the following combination: 
\bea
&&\sum_{k=1,2,4,5} \ck_{\Theta^3}^{(k)} + \sum_{k=2,4} \left[ (\mbox{tr} \gamma_9^k) (\mbox{tr} \gamma_5^k) \ca_{95}^{(k)} \right] + \sum_{k=1,5} (\mbox{tr} \gamma_9^{2k}) \cm_{9}^{(k)} + \sum_{k=2,4} (\mbox{tr} \gamma_5^{2k}) \cm_{5}^{(k)} \nonumber \\
&&+ \sum_{k=1,5}  \left[ (\mbox{tr} \gamma_9^k) (\mbox{tr} \gamma_5^k) \ca_{95}^{(k)} \right]  + \sum_{k=1,5} \left[(\mbox{tr} \gamma_9^k)^2 \ca_{99}^{(k)}  + (\mbox{tr} \gamma_5^k)^2 \ca_{55}^{(k)}\right].
\eea
Note that these are  not partition (zero-point) functions but two-point functions, i.e.
 $\langle V_{Z\bar{Z}}
V_{Z\bar{Z}} \rangle_{\sigma}$ for surfaces indicated above. There should be no confusion 
as we will always use $\Z$ for the partition function
here in the main text. 
Also, we separated CP factors explicitly from the amplitudes. Due to the tadpole condition $\mbox{tr}(\gamma_9^k)=0=\mbox{tr}(\gamma_5^k)$ for $k=1, 5$, the second line in the above expression vanishes. 
For ${\mathcal N}=1$ sectors, the internal partition function of any surface $\sigma$ can be put in the following form (cf.\ appendix A.1 of \cite{Berg:2005ja} and also \cite{Aldazabal:1998mr} and appendix \refapp{sec:tadpoles} below):
\be
\Z^{\rm int}_{\sigma,k}\zba{\alpha}{\beta} = \eta_{\alpha \beta} \, \tilde{\chi}_{\sigma}\, (-2 \sin (\pi \gamma_3)) \frac{\thbw{\alpha}{\beta+\gamma_3}{(0,\tau)}}{\thbw{\frac{1}{2}}{\frac{1}{2}+\gamma_3}{(0,\tau)}} \prod_{i=1}^2  \frac{f(\gamma_i) \, \thbw{\alpha + h_i}{\beta+\gamma_i +g_i}{(0,\tau)}}{\thbw{\frac{1}{2}+h_i}{\frac{1}{2}+\gamma_i+g_i}{(0,\tau)}}\ .
\ee
The details are given in \reftab{tab:details}.
\begin{table}[t]
\renewcommand{\arraystretch}{1.2}
 \[
\begin{array}{ccccccccc}
\toprule
\sigma & k & \tilde{\chi} & \gamma_i & f(\gamma_i) & h_1 & h_2 & g_1 & g_2 \\
\midrule
\ck & 1,2,4,5 &
 \tilde{\chi}(\Theta^3,\Theta^k)&2 k v_i &1 
& \frac{1}{2} &  -\frac{1}{2}  & 0 & 0 \\
\ca_{95} & 2,4 &
2 &  k v_i &1 &  \frac{1}{2} &  -\frac{1}{2}  & 0 & 0 \\
\cm_{9} &1,5
& -1  & k v_i &-2 \sin(\pi \gamma_i) & 0 &0 &0 & 0 \\ 
 \cm_{5} & 2,4 &
 -1 &k v_i & 2 \cos(\pi \gamma_i)& 0 & 0 &\frac{1}{2}& -\frac{1}{2} \\
\bottomrule
\end{array}
\]
\caption{Constants associated with partition functions.}
\label{tab:details}
\end{table}
We can immediately write down
\be
\ca_{95}^{(k)}&=&-4 \sin(\pi\,k\,v_3) \,C_\ca^{(k)}\ , \\
\cm_9^{(k)}&=& 8\sin(\pi\,k\,v_3) \sin(\pi\,k\,v_1) \sin(\pi\,k\,v_2) \,C_\cm^{(k)}\ , \\
\cm_5^{(k)}&=& 8\,\sin(\pi\,k\,v_3) \cos(\pi\,k\,v_1) \cos(\pi\,k\,v_2)\, C_\cm^{(k)}\ ,\\
\ck_{\Theta^3}^{(k)}&=& -2\, \tilde{\chi}(\Theta^3,\Theta^k)\, \sin(2 \pi\,k\,v_3)\, C_\ck^{(k)}\ ,
\ee 
where we have suppressed an overall surface- and $k$-independent constant factor coming from
the  universal {\it external} partition function, that we will reinstate at the end of the calculation, cf.\ \eqref{full2ptfct}. 
Here $v_i=\frac{1}{6} (1,-3,2)$. For $\mathbb{Z}_6'$ with $v_3=1/3$ we observe that
\be
C_\sigma^{(3 N \pm k)}=\pm C_\sigma^{(k)}, \label{C_odd}
\ee
which will be very useful in the following.
We will also need the following Chan-Paton traces
that follow from tadpole cancellation 
(see \refapp{sec:tadpoles}):  $\mbox{tr} \gamma_{9}^k =\mbox{tr} \gamma_{5}^k =-8$ (for $k=2$), 
$\mbox{tr} \gamma_{9}^k =\mbox{tr} \gamma_{5}^k =8$ (for $k=4$) and $ \gamma_{9}^6 = \gamma_{5}^6 =-1$. 

Now we perform the orbifold sector sum. Let us start with the D9-D5 annulus:
\be
\sum_{k=2,4} (\tr \gamma_9^k) (\tr \gamma_5^k) \ca_{95}^{(k)}  = 64 (\ca_{95}^{(2)}+\ca_{95}^{(4)})=  512 \sin( \pi v_3)\, C_\ca^{(1)}\ .
\ee
For the D9-O9 M\"obius strip:
\be
\sum_{k=1,5} (\tr \gamma_9^{2k}) \cm_{(9)}^{(k)}= - 8(\cm_9^{(1)}+\cm_9^{(5)})=-128 \sin(\pi v_1)\sin(\pi v_2) \sin(\pi v_3) \, C_\cm^{(1)} , 
\ee
and for the D5-O5 M\"obius strip:
\be
\sum_{k=2,4} (\tr \gamma_5^{2k}) \cm_{5}^{(k)} = 8 (\cm_5^{(2)}+\cm_5^{(4)}) =  -128  \sin(\pi v_1)\sin(\pi v_2) \sin(\pi v_3) \, C_\cm^{(1)}.
\ee
In the last equality we used $\cos(2\pi v_1)=\sin(\pi v_1)$ and $\cos(2\pi v_2)=\sin(\pi v_2)$. 

For the Klein bottle ($\tilde{\chi}(\Theta^3,\Theta^k)=4$ for $k=1,2,4,5$, see (A.4) of \cite{Font:1989aj}\footnote{Note that there is a mistake in eq.\ (A.6) of \cite{Berg:2005ja}.}) using \refeq{C_odd}
\be
\sum_{k=1,2,4,5} \ck_{\Theta^3}^{(k)}= -8\, \sum_{k=1,2,4,5}  \sin(2 \pi\,k\,v_3)\, C_\sigma^{(k)} =-32 \sin( \pi v_3)\, C_\ck^{(1)}\ .
\ee
Thus altogether we have
\be \label{sum_bf_tint}
&&\hspace{-1cm}\sum_{k=1,2,4,5} \ck_{\Theta^3}^{(k)} + \sum_{k=2,4} \left[ (\mbox{tr} \gamma_9^k) (\mbox{tr} \gamma_5^k) \ca_{95}^{(k)} \right] + \sum_{k=1,5} (\mbox{tr} \gamma_9^{2k}) \cm_{9}^{(k)} + \sum_{k=2,4} (\mbox{tr} \gamma_5^{2k}) \cm_{5}^{(k)} \nonumber \\
&&= 32 \sin( \pi v_3) \left[ -C_\ck^{(1)}+ 16 \, C_\ca^{(1)}  -8 \sin(\pi v_1)\sin(\pi v_2) \, C_\cm^{(1)}\right] \non
&&= 32 \sin( \pi v_3) \left[ -C_\ck^{(1)}+ 16 \, C_\ca^{(1)}  +4\, C_\cm^{(1)}\right]  \label{k-sum} \;. 
\ee
To finish the calculation of the one-loop two-point function
of closed strings, we should perform the $t$ integral. 

%%%%%%%%%%%%%%%%%%%%%%%%%%%%%%%%%%%%%%

\subsection{Computing the $t$ integral}
\label{tint}

We want to calculate the $t$ integral of $C_\sigma^{(1)}(t)$, defined in \refeq{Ck}.
The non-trivial integral we need to evaluate takes the form
\be \label{singularmodintegral}
 I_\sigma(\gamma) = \int_0^\infty \frac{\dd t}{t^2} \,
 \frac{\vartheta_1'(\gamma,\tau_\sigma)} {\vartheta_1(\gamma,\tau_\sigma)} \,,
\ee
where $\tau_\sigma=\tau_\sigma(t)$ is given in \refeq{onelooptau}. 
Here we present the results, which are obtained using
the generalized Mellin transform discussed in \refapp{sec:gmt}. 
Moreover, in \refapp{sec:direct} a more direct version of the same calculation is given, that is
justified by producing the same results as using the generalized Mellin transform.

The integral in \refeq{singularmodintegral} is not well-defined since the integrand is singular in the limit $t \to 0$:
\be
 \frac{\vartheta_1'(\gamma,\tau_\sigma)}{\vartheta_1(\gamma,\tau_\sigma)}
 = \frac{a_\sigma}{t} + \text{rapid decay} \,, \quad
 a_\sigma(\gamma) = \begin{cases}
 2\pi(1- 2\gamma) & \text{for } \A \\
 \pi(1- 4\gamma) & \text{for } \M \\
 \frac{\pi}{2}(1- 2\gamma) & \text{for } \K 
 \end{cases} \,,
\ee
where we have assumed $0 < \gamma < 1$ for $\A,\K$ and $0 < \gamma < \thf$ for $\M$. We introduce a small $t$ cutoff and separate the integral into one divergent  and one regular part by subtracting off the asymptotic behavior of the integrand:
\begin{align} \label{regularizedmodintegral}
 \int_{\frac{1}{e_\sigma\Lambda}}^\infty \frac{\dd t}{t^2} \,
 \frac{\vartheta_1'(\gamma,\tau_\sigma)}{\vartheta_1(\gamma,\tau_\sigma)}
 & = \frac{1}{2} a_\sigma e_\sigma^2 \Lambda^2 + \int_0^\infty \frac{\dd t}{t^2} \left[
 \frac{\vartheta_1'(\gamma,\tau_\sigma)}{\vartheta_1(\gamma,\tau_\sigma)}
 - \frac{a_\sigma}{t} \right] \,, \quad
 e_\sigma = 
 \begin{cases}
 1 & \text{for } \A \\
 4 & \text{for } \M,\K
 \end{cases} \,,
\end{align}
where we have taken the limit $\Lambda \to \infty$ in the second term of the right-hand side. Note that the $t$ integrals of $\A$ on the one hand and $\M$ and $\K$ on the other hand have different cutoffs in order to ensure that the cutoff in the closed string channel (i.e.\ on $\ell$) is the same for all world sheets, cf.\ \cite{Angelantonj:2002ct}. The finite part of \refeq{regularizedmodintegral} is identified with a so-called generalized Mellin transform and it is analyzed in detail in \refapp{sec:gmt}. Using \refeqs{annulusintegerresult}{mobiussintegerresult} with $n=1$ and taking into account $\tau_\sigma = \tau_\sigma(t)$ in \refeq{onelooptau} we find (still assuming $0 < \gamma < 1$ for $\A,\K$ and $0 < \gamma < \thf$ for $\M$)
\begin{align}
I_\ca= \int_{\frac{1}{\Lambda}}^\infty \frac{\dd t}{t^2} \,
 \frac{\vartheta_1'(\gamma,\frac{it}{2})}{\vartheta_1(\gamma,\frac{it}{2})}
 & = \pi(1-2\gamma) \Lambda^2 + \frac{\pi}{24} \left[ \psi'(\gamma) - \psi'(1-\gamma) \right] \,, \\
I_\cm= \int_{\frac{1}{4\Lambda}}^\infty \frac{\dd t}{t^2} \,
 \frac{\vartheta_1'(\gamma,\hf+\frac{it}{2})}{\vartheta_1(\gamma,\hf+\frac{it}{2})}
 & = 8\pi(1-4\gamma) \Lambda^2 + \frac{\pi}{12} \left[ \psi'(\gamma) - \psi'(1-\gamma)- \thf \psi'(\thf+\gamma) + \thf \psi'(\thf-\gamma) \right] \,, \\ 
 I_\ck=\int_{\frac{1}{4\Lambda}}^\infty \frac{\dd t}{t^2} \,
 \frac{\vartheta_1'(\gamma,2it)}{\vartheta_1(\gamma,2it)}
 & = 4\pi(1-2\gamma) \Lambda^2 + \frac{\pi}{6} \left[ \psi'(\gamma) - \psi'(1-\gamma) \right] \,,
\end{align}
for $\A, \M$ and $\K$, respectively. Here $\psi'(x)$ denotes the trigamma function, i.e. the derivative of the digamma function $\psi(x) = {\Gamma'(x)}/{\Gamma(x)}$.

%%%%%%%%%%%%%%%%%%%%%%%%%%%%%%%%%%%%%%%%%%%%%

\subsection{Cancellation of UV divergences}

Using the above results for $t$ integrals, we will
now combine them as in \refeq{sum_bf_tint}. 
We will check the cancellation of UV divergences between diagrams. 
There are two kinds of divergence, linear and quadratic in the cutoff $\Lambda$,
that must cancel separately.  
The linear divergences come from the $\cot(2 \pi \gamma_3)$ in $C_\sigma^{(1)}$ in \refeq{sum_bf_tint}, 
after $t$ integration. 
The quadratic divergences are evident in the $t$ integrals.
Let 
\be \label{sum_bf_tint2}
&& \hspace{-1cm} \int_0^\infty\frac{dt}{t^3}\left[\sum_{k=2,4} (\tr \gamma_9^k) (\tr \gamma_5^k) \ca_{95}^{(k)}
+ \sum_{k=1,5} (\tr \gamma_9^{2k}) \cm_{9}^{(k)}
+ \sum_{k=2,4} (\tr \gamma_5^{2k}) \cm_{5}^{(k)}
+ \sum_{k=1,2,4,5} \ck_{\Theta^3}^{(k)} \right]\nonumber \\
&&
 = 16 a \int_{\frac{1}{\Lambda}}^{\infty} \frac{dt}{t^3} \,C_\ca^{(1)}
 + 4 b \int_{\frac{1}{4\Lambda}}^{\infty} \frac{dt}{t^3} \, C_\cm^{(1)}
 - c \int_{\frac{1}{4 \Lambda}}^{\infty} \frac{dt}{t^3}\, C_\ck^{(1)} \non
&& = 8a \int_{\frac{1}{\Lambda}}^{\infty} \frac{dt}{t^2} \, \frac{C_\ca^{(1)}}{\tau_2}
+ 2 b \int_{\frac{1}{4\Lambda}}^{\infty} \frac{dt}{t^2} \, \frac{C_\cm^{(1)}}{\tau_2}
 - 2c \int_{\frac{1}{4 \Lambda}}^{\infty} \frac{dt}{t^2}\, \frac{C_\ck^{(1)}}{\tau_2} \label{C_tint}\ , \label{Ct}
\ee
where $c=32 \sin( \pi v_3) $,
and we introduced coefficients 
$a$, $b$ that keep track of the different contributions arising from the different worldsheets,
where  $a=b=c$ corresponds to the actual calculation at hand. 
Up to an overall factor, the 
$\cot(2 \pi \gamma_3)$ terms in $C_\sigma^{(k)}$ give
\be
 \Lambda\, \left(- \frac{a}{2} -  \frac{b}{2}+c\right) \; , 
\ee 
using $\gamma_3=2 v_3 $ for Klein bottle and $\int_{1/ \Lambda}^\infty {dt/ t^2}= \int_0^\Lambda d
\ell$.
On the other hand, using the results of \refsec{tint},  
the theta parts in $C_\sigma^{(1)}$  give for \refeq{C_tint}  up to an overall factor
\be
 8\, a \, I_\ca+ 2\, b \, I_\cm-2\, c I_\ck   &\propto&  \Big[8\, a\,(1-2 v_3)+16\, b \,(1-4 v_3)-8\, c (1-4 v_3)\Big]\, \Lambda^2  +\text{finite constants} \non
&=& \Lambda^2\,  \frac{8(a - 2\, b+c)}{3}+\text{finite constants}\ .
\ee
We see that for all linear and quadratic UV divergences to cancel we need:
\be
 - \frac{a}{2} -\frac{b}{2}+c &=&0 \qquad \text{from} \; \Lambda\ , \\
a - 2\,b+c&=&0  \qquad  \text{from} \; \Lambda^2\ .
\ee
Thus, all UV divergences cancel for $a=b=c$, as required.\footnote{We are not aware of other calculations of divergences in
open and unoriented diagrams where two
different powers of the cutoff appear.
%To our knowledge this is the first occasion 
%where the tadpole cancellation in a string amplitude involves two different powers of the cutoff for the worldsheet parameter. 
On the technical level this is due to the contribution of the eight-fermion term (which is absent in other related calculations in the literature,
such as partition functions, open string two-point functions, and the ${\mathcal N}=2$
contribution to a closed string two-point function). 
It would be interesting to have better physical understanding of the appearance of the different powers.}

%%%%%%%%%%%%%%%%%%%%%%%%%%%%%%%%%%%%%%%%%%%%%%
\subsection{Finite result}

%\begin{align}
% & \left. \left[ 4I_\A\left(\ts{\frac{1}{3}}\right) + I_\K \left(\ts{\frac{2}{3}}\right) + I_\M\left(\ts{\frac{1}{3}}\right) \right] \right|_\text{finite} \\
% & = \ts{\frac{\pi}{6}} \left[ \psi' \left(\ts{\frac{1}{3}}\right) - \psi'\left(\ts{\frac{2}{3}}\right) \right]
% + \ts{\frac{\pi}{12}} \left[ \psi'\left(\ts{\frac{1}{3}}\right) - \psi'\left(\ts{\frac{2}{3}}\right) + \thf \psi'\left(\ts{\frac{5}{6}}\right) - \thf \psi'\left(\ts{\frac{1}{6}}\right) \right]
% - \ts{\frac{\pi}{6}} \left[ \psi'\left(\ts{\frac{2}{3}}\right) - \psi'\left(\ts{\frac{1}{3}}\right) \right] \\
% & = \ts{\frac{5\pi}{4}} \left[ \psi'\left(\ts{\frac{1}{3}}\right) - \frac{2\pi^2}{3} \right] \\
% & = 5 \pi \sin \left( \ts{\frac{\pi}{3}} \right) \Cl_2 \left( \ts{\frac{\pi}{3}} \right)
%\end{align}
We would now like to compute the finite terms of \refeq{sum_bf_tint2}. To this end we first consider
\begin{align}
 & 4I_\A\left(\frac{1}{3}\right) + I_\M\left(\frac{1}{3}\right) - I_\K \left(\frac{2}{3}\right) \nonumber \\
 & = \frac{\pi}{3} \left[ \psi' \left(\frac{1}{3}\right) - \psi'\left(\frac{2}{3}\right) \right]
 + \frac{\pi}{12} \left[ \psi'\left(\frac{1}{3}\right) - \psi'\left(\frac{2}{3}\right) - \hf \psi'\left(\frac{5}{6}\right) + \hf \psi'\left(\frac{1}{6}\right) \right] \,, \label{finiteIs}
\end{align}
where $I_\sigma$ was defined in \refeq{singularmodintegral} and we noticed that $I_\K(\frac{2}{3}) = - 4 I_\A(\frac{1}{3})$. The different terms can be expressed in terms of $\psi'(\frac{1}{3})$ using the reflection and duplication formulas for the trigamma function.\footnote{In particular, we have
$\psi'\left(\frac{2}{3}\right) = - \psi'\left(\frac{1}{3}\right) + \frac{4\pi^2}{3}$, 
$\psi'\left(\frac{5}{6}\right) = - 5\psi'\left(\frac{1}{3}\right) + \frac{16\pi^2}{3}$ and
$\psi'\left(\frac{1}{6}\right) =  5 \psi'\left(\frac{1}{3}\right) -\frac{4\pi^2}{3}$.}
The value of $\psi'(\frac{1}{3})$ is itself conveniently expressed in terms of the second Clausen function (cf. \refapp{appendixclausen})
\be \label{clausenrelation}
 \psi'\left(\frac{1}{3}\right) = 4\sin \left(\frac{\pi}{3}\right) \Cl_2\left(\frac{\pi}{3}\right) + \frac{2\pi^2}{3} \,,
\ee
see for instance eq.\ (12) in \cite{Coffey:2005jd}. Taking the above into account we get
\be
 4I_\A\left(\frac{1}{3}\right) + I_\M\left(\frac{1}{3}\right) - I_\K \left(\frac{2}{3}\right)
 = 5 \pi \sin \left( \frac{\pi}{3} \right) \Cl_2 \left( \frac{\pi}{3} \right) \,.
\ee
Taking into account the overall factors given in \eqref{full2ptfct} (for $N=6$), reinstating the factors of \eqref{sum_bf_tint2} that are suppressed in \eqref{finiteIs} and making the substitution (cf.\ \eqref{Deltatau})
\be
V_4 \rightarrow \dd^4 x \sqrt{-g}\quad , \quad i(p_1 \cdot p_2) g_c^2 (\alpha')^{-4} \phi^2 \rightarrow -\frac{1}{16 \pi^2}\cdot {1 \over 2}\cdot\partial_\mu \tau^{(0)} \partial^\mu \tau^{(0)}\ ,
\ee
where the $1/2$ is for identical fields,
we finally obtain 
\be
&& \hspace{-1.5cm}\frac{1}{48} \frac{iV_4}{(4 \pi^2 \alpha')^2}  {\phi^2 g_c^2 (\alpha')^{-4} \over 4 (\tau^{(0)})^2} \cdot \delta \cdot (-8 \pi) \cdot 32 \sin \left( \frac{\pi}{3} \right) \cdot 2
 \left[ 4I_\A\left(\frac{1}{3}\right) + I_\M\left(\frac{1}{3}\right) - I_\K \left(\frac{2}{3}\right) \right]\nonumber \\[2mm]
&\rightarrow&  \dd^4 x\,  \sqrt{-g}\, (\alpha')^{-1} \left(\frac{5}{2^9 \pi^4 } \Cl_2\left( \frac{\pi}{3} \right) \right){1 \over (\tau^{(0)})^2}\partial_\mu \tau^{(0)} \partial^\mu \tau^{(0)}\ .
\ee
Comparing with \eqref{kinetic_term} (where there is an explicit factor of $\kappa_4^{-2} = (2 \pi \sqrt{\alpha'})^6 \kappa_{10}^{-2}=(\pi \alpha')^{-1}$ in front), we can read off the $\N=1$ sector one-loop contribution to the metric of $\tau^{(0)}$ (in the string frame)
\be \label{finalanswer}
\widetilde G^{(1)}_{\N = 1}\, \partial_\mu \tau^{(0)} \partial^\mu \tau^{(0)} =  \left(\frac{5}{2^9 \pi^3 } \Cl_2\left( \frac{\pi}{3} \right) \right){1 \over (\tau^{(0)})^2}\partial_\mu \tau^{(0)} \partial^\mu \tau^{(0)}\ .
\ee
The number in parenthesis is about 0.0003197.
In the Einstein frame this expression is multiplied with $e^{2 \Phi_4}$ due to Weyl rescaling, cf.\ eq.\ \eqref{finalmetric}. This gives the expected suppression in the string coupling compared to the tree level term \eqref{Gzero}, as appropriate for a one-loop contribution. 
Note that, even though the string amplitude produced a finite real constant, the final result \eqref{finalanswer} is in fact moduli dependent, due to the normalization of the vertex operators \eqref{Vtau} and the Weyl-rescaling.
Note also that in the $\mathbb{Z}_6'$ orientifold (in contrast to for example $\mathbb{Z}_3$) there are also $\N=2$ sectors, whose contribution was already calculated in \cite{Berg:2005ja}. The total $\widetilde G^{(1)}$ 
is the sum of those contributions and the $\N=1$ sector contribution \eqref{finalanswer} that we computed  in this paper. 

\newpage
\section{Conclusions and Outlook}

We have computed one-loop contributions to the K\"ahler metric
of closed string moduli in minimal supersymmetry in four dimensions. 
This was a necessary step
towards completing the moduli dependent 
one-loop string effective
action of minimally supersymmetric toroidal orientifolds. As discussed in section \refsec{sec:sea}, to complete 
the one-loop metric of the K\"ahler modulus under consideration in this paper, one would also 
need to know the correction to the Einstein-Hilbert term in the $\mathbb{Z}'_6$ orientifold (which is work in progress) as well as the corrected
definitions of the K\"ahler variables. Moreover, we only discussed the contribution of $\N = 1$ sectors in this paper, 
to which one 
would add known contributions from $\N=2$ sectors, cf.\ \cite{Berg:2005ja}.

In this paper we developed a few necessary techniques to calculate one-loop terms
in the string effective action with minimal supersymmetry in four dimensions. We would like to highlight 
three aspects of this development. We derived the boson and fermion correlators for twisted fields, that contain 
new phases for the Klein bottle, cf.\ \eqref{dZdZbar}-\eqref{dbarZdbarZbar} and \eqref{PsiPsibar}-\eqref{tildePsitildePsibar}. 
We clarified the relation between the (un)twisted correlators and generalized Eisenstein series, cf.\ \eqref{greenB}-\eqref{greenspin2}. Finally, we introduced the method of generalized Mellin transforms for calculation of integrals over the open string 
world-sheet parameter $t$, cf.\ appendix \refapp{sec:gmt}. We hope that these methods will find further applications 
in orientifold loop calculations. 

There are many areas of techniques that need development
to make further progress here, but luckily similar techniques
are well established in mathematics. 
We have seen that the integrals
require some understanding of the theory 
of nonholomorpic automorphic forms. 
A complementary direction is the recent substantial progress on the use of automorphic forms
in related closed
string calculations with more supersymmetry
(e.g.\ \cite{Angelantonj:2011br,Angelantonj:2012gw,Angelantonj:2013eja}),
where it is noted that various recent generalizations
of classical results in mathematics 
(somewhat reminiscient of  the generalizations we discuss in appendix \ref{sec:gmt}) 
are very useful for application to these kinds of string calculations. It would be very useful
to connect those  techniques to ours. 

As noted earlier, there has been
impressive
development in field theory amplitude calculations in recent years, 
and some of the driving forces have been
the string theory KLT relations at tree level and the
field theory BCJ relations, i.e.\ ``gravity is the square of gauge theory''
\cite{Bern:2010ue}.
How and
whether KLT-type relations can be systematized for string loops is not well
understood at the moment. This becomes particularly nontrivial when massive
string states enter the game, as they do here, unlike what often happens in
extended supersymmetry, as emphasized in the introduction. It is also of
special interest in orientifolds where open string amplitudes are not
obviously just "square roots" of closed string amplitudes, as emphasized in
\cite{Lust:2004cx} for string tree-level. But with lifting to the covering
torus, perhaps one could find a square-root structure at least for individual
terms of the amplitude.

For phenomenology,  moduli stabilization is needed and one is typically not too
interested in orbifold limits of Calabi-Yau manifolds per se, but rather the
smooth (blown-up) points of Calabi-Yau moduli space. With this in mind, it
would be very interesting to attempt to generalize some of these calculations
to blown-up orbifolds and orientifolds. 

In general, there are many phenomenological applications
that require understanding the effective action at least at the one-loop level. For example, there is an ongoing 
discussion
in the literature of how to incorporate so-called ``anomaly mediation'' in string theory, i.e.\ 
calculating one-loop contributions to the effective action
of models that have been arranged to solve
phenomenological problems with flavor-changing neutral currents at tree-level
(see e.g.\ \cite{deAlwis:2010sw}). Ultimately
the safest way to perform these computations
would be in string theory --- see e.g.\ \cite{Conlon:2010qy} for comments on this.

Another obvious avenue of interest would be to use true flux
compactifications in curved space as background, rather than toroidal
orientifolds that are almost everywhere flat. A good start would be AdS
compactifications. String loop calculations in AdS quickly become technically
challenging,  for example in the pure spinor formalism, see for instance \cite{Linch:2008rw}. 
There is also interesting recent work in this direction using
integrability (see for example \cite{Bianchi:2014rfa}). With
the confluence of new methods, perhaps this kind of calculation will become
more feasible in the future. 

%%%%%%%%%%%%%%%%%%%%%%%%%%%%%%%%%%%%%%%%%%%%%%%%%%%%%%

\section{Acknowledgments}
We thank Carlo Angelantonj, Costas Bachas, Paolo Di Vecchia, J\"urgen Fuchs, James Gray, Thomas Grimm, Bo Sundborg
and Don Zagier for helpful discussions and Mark Goodsell and Stephan Stieberger for helpful comments on the draft.
J.K. thanks Fernando Quevedo for useful discussion and the hospitality during the visits to ICTP, where a substantial proportion of his research work was done. M.B. would like to thank the University of North Carolina at Chapel Hill for hospitality during the completion of this work. He would
also like to acknowledge financial support from the Wenner-Gren Foundations. 
The work of M.H. is supported by the Excellence Cluster ``The Origin and the Structure of the Universe'' in Munich and by the German Research Foundation (DFG) within the Emmy-Noether-Program (grant number: HA 3448/3-1).
The work of S.S. was supported by the DFG Transregional Collaborative Research Centre TRR 33 ÒThe Dark UniverseÓ.

%%%%%%%%%%%%%%%%%%%%%%%%%%%%%%%%%%%%%%%%%%%%%%%%%%%%%%

\appendix

\section{Useful formulas}
The $\eta$ and $\tht$ functions are
\be 
\eta(\tau) &=& q^{1/24} \prod_{n=1}^\infty \left( 1-q^n \right) \ , \non
\tht\ba{\vec \alpha}{\vec\beta}(\vec \nu,G) &=& \sum_{\vec n\in \mathbb{Z}^N} 
 e^{i\pi(\vec n+\vec \alpha)^{\rm T} G (\vec n+\vec \alpha)} 
e^{2\pi i(\vec \nu+\vec \beta)^{\rm T}(\vec n+\vec\alpha)}  \ .
\label{thetamatrix}
\ee
Modular transformation:
\be \label{modtransmoment}
\tht\ba{\vec \alpha}{\vec 0}(0,it G^{-1}) &=& \sqrt{G} \, t^{-N/2}\,  
 \tht\ba{\vec 0}{\vec 0 }(\vec \alpha,it^{-1} G) \ .
\ee
Modular transformation for annulus and Klein bottle:
\be \label{modtransannulus}
%\thba{\alpha}{\beta}(\nu,\tau)
%&=& e^{\pi i \alpha(\alpha+1)}
%\thba{\alpha}{\beta-\alpha-1/2}(\nu,\tau+1) \label{T} \ , \non
\eta(\tau) & = & (-i \tau)^{-1/2} \eta(-1/\tau)\ , \non
\thba{\alpha}{\beta}(\nu,\tau)
&=& (-i \tau)^{-1/2} e^{2\pi i \alpha \beta - \pi i \nu^2/\tau}
\thba{-\beta}{\alpha}(\nu/\tau,-1/\tau) \label{S} \ . 
\ee
Modular transformation for 
the M\"obius strip:\footnote{Note that there is a typo in formula (122) of \cite{Berg:2004ek}.}
\be \label{modtransmoebius}
\eta(\tau) & = & e^{-\frac{\pi i}{6}} (1 - 2 \tau)^{-1/2} \eta \left( \frac{\tau}{1 - 2 \tau} \right)\ ,\non
\thba{\alpha}{\beta}(\nu,\tau)
&=& (1-2\tau)^{-1/2} 
e^{-2\pi i \beta^2} e^{-\pi i\nu^2/(\tau-1/2)}
\thba{\alpha+2\beta}{\beta}\Big(\frac{\nu}{1-2\tau},
\frac{\tau}{1-2\tau}\Big) \; .
\ee
Shifts in characteristics:
\be \label{shift}
\thba{\alpha+1}{\beta}(\nu,\tau) &=& \thba{\alpha}{\beta}(\nu,\tau)\ , \non
\thba{\alpha}{\beta+1}(\nu,\tau) &=& e^{2 \pi i \alpha} \thba{\alpha}{\beta}(\nu,\tau)\ .
\ee 
$\nu$-periodicity formula:
\be  \label{period}
\thba{\alpha}{\beta}(\nu+a\tau + b,\tau)
&=&
e^{-2\pi i ab} e^{-\pi i a^2\tau}e^{-2\pi i a(\nu+\beta)}
\thba{\alpha+a}{\beta+b}(\nu,\tau)\ . 
\ee
For $-1/2 < \alpha < 1/2$ (and arbitrary $\beta$) one has the $q\rightarrow 0$ limits (with $q=e^{2 \pi i \tau}$)
\be \label{limits}
\lim_{q \rightarrow 0} \thba{\alpha}{\beta} (\nu, \tau) &=& e^{2 \pi i \alpha (\nu + \beta)} q^{\alpha^2/2} \ , \non
\lim_{q \rightarrow 0} \thba{1/2}{\beta} (\nu, \tau) &=& 2 \cos[\pi (\nu + \beta)] q^{1/8}\ ,  \non
\lim_{q \rightarrow 0} G_F \zba{\alpha}{\beta} (\nu, \tau) &=& \pi \frac{e^{2 \pi i \alpha \nu}}{\sin(\pi \nu)} \ ,  \non
\lim_{q \rightarrow 0} G_F \zba{1/2}{\beta} (\nu, \tau) &=& \frac{\cos[\pi (\nu + \beta)]}{\cos(\pi \beta)} \frac{\pi }{\sin(\pi \nu)} \ ,  \non
\lim_{q \rightarrow 0} \eta(\tau) &=& q^{1/24}\ .
\ee

A single transformation $S$ and the sequence of transformations $S T^2 S$ give respectively
\begin{align}
 \frac{\vartheta_1'(\gamma,it)}{\vartheta_1(\gamma,it)}
 & = - \frac{2 \pi \gamma}{t}
 - \frac{i}{t} \frac{\vartheta_1'(-\frac{i\gamma}{t},\frac{i}{t})}{\vartheta_1(-\frac{i\gamma}{t},\frac{i}{t})} \,,
 \label{modulartransformation1} \\
  \frac{\vartheta_1'(\gamma,\hf + it)}{\vartheta_1(\gamma,\hf + it)}
 & = - \frac{2 \pi \gamma}{t}
 - \frac{i}{2t} \frac{\vartheta_1'(-\frac{i\gamma}{2t},-\frac{1}{2}+ \frac{i}{4t})}{\vartheta_1(-\frac{i\gamma}{2t},-\frac{1}{2}+ \frac{i}{4t})} \,.
 \label{modulartransformation2}
\end{align}

Exercise 12 on p.\ 489 of \cite{WW} (or \cite{AbSt})
\be
 \p_\nu \ln \vartheta_1(\nu,\tau)
 = \pi \cot \pi \nu + 4 \pi \sum_{n=1}^\infty \frac{q^n}{1-q^n} \sin 2 \pi n \nu \,, \quad
 |{\Im \nu}| < \Im \tau \,.
 \label{wwseries}
\ee

%%%%%%%%%%%%%%%%%%%%%%%%%%%%%%%%%%%%%%%%%%%%%%

\section{The geometry of one-loop surfaces}
\label{sec:geometry}

\begin{figure}
\begin{center}
\subfigure[Annulus]{\includegraphics[scale=0.96]{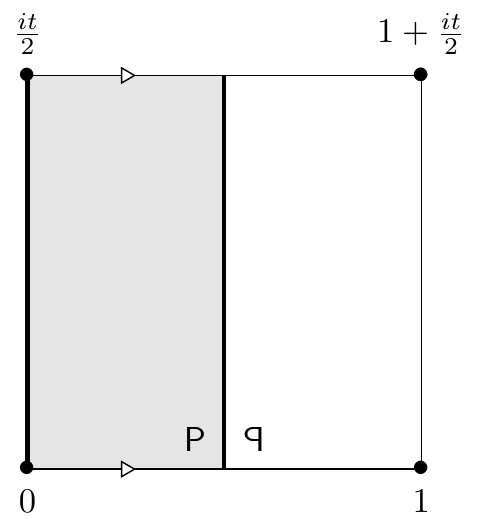}\label{annulusws}}
\subfigure[M\"obius strip]{\includegraphics[scale=0.96]{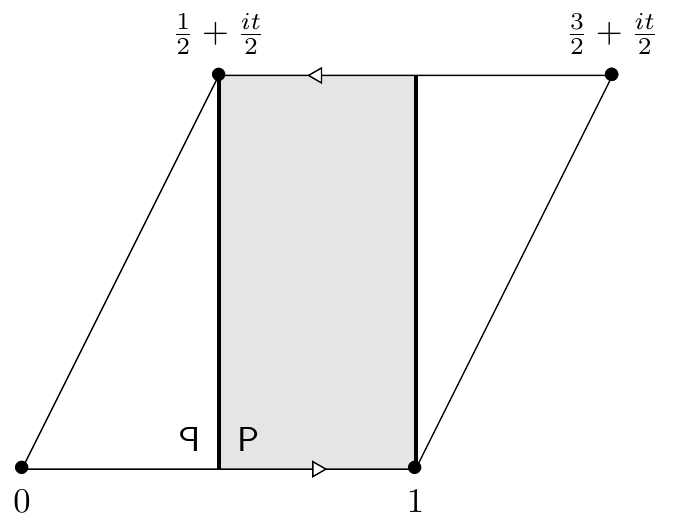}\label{mobiusws}}
\subfigure[Klein bottle]{\includegraphics[scale=0.96]{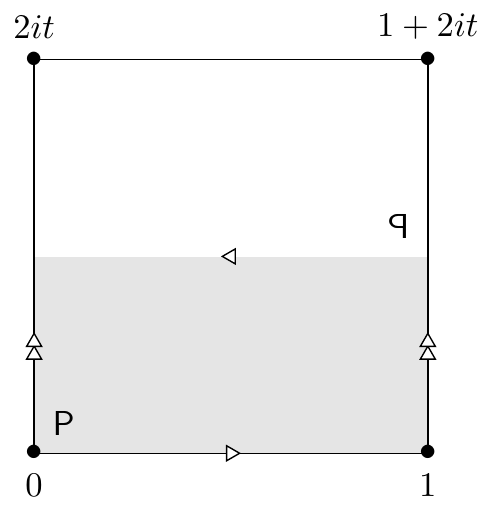}\label{kleinws}}
\caption{In \ref{annulusws}, \ref{mobiusws} and \ref{kleinws} we illustrate how the worldsheets $\A,\M$ and $\K$ (depicted in gray) are constructed from their respective covering tori (bounded by thin black lines). The action of $\nu \to I_\sigma (\nu)$ on the point $\sf{P}$ 
is given as illustration. The thick black lines denote fixed lines of the involution $I_\sigma$ and correspond to boundaries of the worldsheet.}
\label{fig:ws}
\end{center}
\end{figure}

The torus $\T$, with modulus $\tau = \tau_1 + i \tau_2$, is obtained by identifying points $\nu \in \mathbb{C}$ under the translations
\be \label{torusshifts}
 \nu \to \nu + 1 \,, \quad \nu \to \nu + \tau \,.
\ee
As a fundamental domain we can take the
parallelogram spanned by the points $0,1,\tau,1+\tau$.
With the natural line element $\dd s^2 = |\dd \nu|^2$ on $\mathbb{C}$,
the area of the parallelogram is $\tau_2$.
The annulus $\mathcal A$, M\"obius strip $\mathcal M$
and Klein bottle $\mathcal K$, collectively denoted by $\sigma = \A,\M,\K$,
can be obtained from tori $\T_\sigma$, with moduli
\be \label{onelooptau}
 \tau_\A = \frac{i t}{2} \,, ~~~~~~
 \tau_\M = \frac{1}{2} + \frac{i t}{2} \,, ~~~~~~
 \tau_\K = 2 i t \,,
\ee
by dividing out by the involutions
\be \label{Isigmas}
 I_\A(\nu) = 1 - \bar \nu \,, ~~~~~~
 I_\M(\nu) = 1 - \bar \nu \,, ~~~~~~
 I_\K(\nu) = 1 - \bar \nu + \frac{\tau_\K}{2} \,,
\ee
cf.\ the appendix of \cite{Antoniadis:1996vw}. The construction is illustrated in \reffig{fig:ws}. This allows us to construct the worldsheet propagators on $\sigma = \A, \M, \K$ from those on $\T_\sigma$ using the method of images as explained in \refsec{sec:bos} and \refsec{sec:ferm}.

%%%%%%%%%%%%%%%%%%%%%%%%%%%%%%%%%%%%%%%%%%%%%%

\subsection{Lifting to the covering torus}
\label{sec:lifting}

The covering torus $\T_\sigma$ can be recovered from the surface $\sigma$
and its image $I_\sigma(\sigma)$ as $\T_\sigma = \sigma \cup I_\sigma(\sigma)$.
Using the following equation backwards \cite{Antoniadis:1996vw}
\be
 \int_{\T_\sigma} \dd^2 \nu \, f(\nu)
 = \bigg[ \int_{\sigma} \dd^2\nu + \int_{I_\sigma(\sigma)} \dd^2\nu \bigg] \, f(\nu) 
 = \int_{\sigma} \dd^2\nu \, \Big[ f(\nu) + f(I_\sigma(\nu)) \Big] \,,
\ee
integrals involving $\nu$ and $I_\sigma(\nu)$ can be ``lifted'' from $\sigma$ to $\T_\sigma$.
For example in \refsec{sec:two-point} we make use of
\be \label{lift}
 \int_{\sigma} \dd^2 \nu \, \Big[ \p f(\nu) - \bar\p f( I_\sigma(\nu)) \Big]
 = \int_{\T_\sigma} \dd^2 \nu \, \p f(\nu) \,,
\ee
since $\p/\p\bar\nu = - \p/\p I_\sigma(\nu)$.

%%%%%%%%%%%%%%%%%%%%%%%%%%%%%%%%%%%%%%%%%%%%%%%%%%%%%%

\section{Boson correlation functions}
\label{sec:bos}

Some of the material in this and the next appendix is contained in the literature (\cite{Antoniadis:1996vw,Berg:2005ja} and references therein), some of it
is new.

\subsection{Scalar correlation functions on $\sigma=\T,\A,\M,\K$}

On the torus, the scalar correlation function is given by
\be \label{periodicboson}
 \vev{X(\nu_1)X(\nu_2)}_\T = \frac{\alpha'}{2} G_B(\nu_1,\nu_2) \,, \quad
 G_B(\nu_1,\nu_2) = - \ln \left| \frac{\vartheta_1(\nu_{12})}{\vartheta_1'(0)} \right|^2  + \frac{2\pi}{\tau_2} (\Im \nu_{12})^2 \,,
\ee
where $\nu_{12} = \nu_1-\nu_2$. Since each surface $\sigma = \A, \M, \K$ is obtained from the appropriate covering torus $\T_\sigma$, the correlation functions are simply obtained by symmetrizing under the action of $I_\sigma$ as
\begin{align}
 \vev{X(\nu_1)X(\nu_2)}_{\sigma} \label{methodofimagesscalar}
 = \thf \big[ \vev{X(\nu_1)X(\nu_2)}_{\T_\sigma}
 + \vev{X(I_\sigma(\nu_1))X(\nu_2)}_{\T_\sigma} \\
 + \, \vev{X(\nu_1)X(I_\sigma(\nu_2))}_{\T_\sigma}
 + \vev{X(I_\sigma(\nu_1))X(I_\sigma(\nu_2))}_{\T_\sigma} \big] \,. \nonumber
\end{align}

%%%%%%%%%%%%%%%%%%%%%%%%%%%%%%%%%%%%%%%%%%%%%%%%%%%%%%%%%

\subsection{(Doubly) twisted current correlation functions on $\T$}

On the torus, the (doubly) twisted current correlation functions are given by\footnote{For short expressions we use a bar to indicate complex conjugation but for longer expressions we prefer to use a star instead.}
\be
 \vev{\p Z(\nu_1) \p \bar Z(\nu_2)}_\T = \frac{\alpha'}{2} (\p_1 \p_2 G_B^{\vec\gamma}) (\nu_1,\nu_2) \,, \quad 
 \vev{\bar\p Z(\bar\nu_1) \bar\p \bar Z(\bar\nu_2)}_\T = \frac{\alpha'}{2} [(\p_1\p_2 G_B^{-\vec{\gamma}})(\nu_1,\nu_2)]^* \,,
\ee
where $\vec \gamma = (\delta,\gamma)$ (do not confuse this $\delta$ with the momentum-squared invariant defined in \eqref{deltamomentumsquare}) and
\be \label{GB-GF}
 (\p_1 \p_2 G_B^{\vec\gamma})(\nu_1,\nu_2) = \p_1 G_F^{\vec\gamma}(\nu_1,\nu_2) = - \p_2 G_F^{\vec\gamma}(\nu_1,\nu_2) \,, \quad
 G_F^{\vec\gamma}(\nu_1,\nu_2) = \frac{\vartheta \ss{1/2+\delta}{1/2+\gamma}(\nu_{12}) \vartheta_1'(0)}{\vartheta\ss{1/2+\delta}{1/2+\gamma}(0) \vartheta_1(\nu_{12})} \,,
\ee
with the transformation properties
\begin{align}
 G_F^{\vec\gamma}(\nu_1+1,\nu_2) & = e^{+2 \pi i \delta} G_F^{\vec\gamma}(\nu_1,\nu_2) \,, \\
 G_F^{\vec\gamma}(\nu_1+\tau,\nu_2) & = e^{-2 \pi i \gamma} G_F^{\vec\gamma}(\nu_1,\nu_2) \,. \label{GFnu1tau}
\end{align}

%%%%%%%%%%%%%%%%%%%%%%%%%%%%%%%%%%%%%%%%%%%%%%%%%%%%%%%%%

\subsection{Twisted current correlation functions on $\sigma=\A,\M,\K$}

\label{sec:relieving}

In the following we use the method of images to determine the twisted current correlation function
\be
 \vev{\J(\nu_1) \bar{\J}^T(\nu_2)}_\sigma \,, \quad
 \J (\nu) = \sqrt{\frac{2}{\alpha'}} \begin{pmatrix} \p Z(\nu) \\ \bar \p Z(\bar \nu) \end{pmatrix} \,, \quad
 \bar{\J} (\nu) = \sqrt{\frac{2}{\alpha'}} \begin{pmatrix} \p \bar Z(\nu) \\ \bar \p \bar Z(\bar \nu) \end{pmatrix} \,.
\ee
On $\sigma=\A,\M,\K$ we only have twist along the cycle $\nu \to \nu+\tau$ thus the relevant correlation function is the one with $\delta=0$. Using
\be \label{twistedoddfermion}
 [G_F^{\gamma}(\nu_1,\nu_2)]^* = -G_F^{-\gamma}(I_\sigma(\nu_1),I_\sigma(\nu_2)) \,, \quad
 G_F^\gamma(\nu_1,\nu_2) = \frac{\vartheta_1(\nu_{12}+\gamma) \vartheta_1'(0)}{\vartheta_1(\gamma)\vartheta(\nu_{12})}
\ee
the twisted current correlation function on the covering torus $\T_\sigma$ of $\sigma = \A,\M,\K$ takes the form
\be \label{twistedcorrelatortorus}
 \vev{ \J (\nu_1) \bar{\J}^T (\nu_2)}_{\T_\sigma}
 = \begin{pmatrix}
 (\p_1 \p_2 G_B^\gamma) (\nu_1,\nu_2) & 0 \\
 0 & (\bar \p_1 \bar \p_2 G_B^\gamma)(I_\sigma(\nu_1),I_\sigma(\nu_2)) \end{pmatrix} \,,
\ee
where
\begin{align}
 (\p_1 \p_2 G_B^\gamma)(\nu_1,\nu_2)
 & = \p_1 G_F^\gamma(\nu_1,\nu_2)
 = -  \p_2 G_F^\gamma(\nu_1,\nu_2) \,, \\
 (\bar \p_1 \bar \p_2 G_B^\gamma)(I_\sigma(\nu_1),I_\sigma(\nu_2))
 & = - \bar\p_1 G_F^\gamma(I_\sigma(\nu_1),I_\sigma(\nu_2))
 = \bar \p_2 G_F^\gamma(I_\sigma(\nu_1),I_\sigma(\nu_2)) \,.
\end{align}
Note that for the twisted correlation function on the torus, eq.\ \eqref{twistedcorrelatortorus}, there are no off-diagonal terms in contrast to the corresponding untwisted one. This is due to the absence of zero modes. 

Now, the twisted current correlation function on $\sigma = \A,\M,\K$ is obtained by symmetrizing under the action of $I_\sigma$
\begin{align} \label{JcorrelatorAMK}
 \vev{ \J (\nu_1) \bar{\J}^T(\nu_2)}_\sigma
 = \thf \big[ \vev{ \J (\nu_1) \bar{\J}^T(\nu_2)}_{\T_\sigma}
 + e^{+\pi i \gamma \delta_{\sigma \K}} \mathbf{P} \vev{ \J (I_\sigma(\nu_1)) \bar{\J}^T(\nu_2)}_{\T_\sigma} \\
 + \, e^{- \pi i \gamma \delta_{\sigma \K}} \vev{ \J (\nu_1) \bar{\J}^T(I_\sigma(\nu_2))}_{\T_\sigma} \mathbf{P}^T
 + \mathbf{P} \vev{ \J (I_\sigma(\nu_1)) \bar{\J}^T(I_\sigma(\nu_2))}_{\T_\sigma} \mathbf{P}^T \big] \,. \nonumber
\end{align}
Here $\vev{ \J (I_\sigma(\nu_1)) \bar{\J}^T(\nu_2)}_{\T_\sigma}$, for instance, has to be understood as \eqref{twistedcorrelatortorus} with arguments $I_\sigma(\nu_1)$ and $\nu_2$. Compared to the scalar correlation function the twisted current correlation function offers two novelties: First, since $\J$ and $\bar\J$ transform as vectors under the involution we must include factors of $\mathbf{P}$ in the symmetrization, where $\mathbf{P}$ denotes the vector representation of $I_\sigma$, i.e. $I_\sigma(\bs{\p}) = \mathbf{P} \bs{\p}$ for $\bs{\p} = (\p, \bar \p)^T$ and $\mathbf{P} = -\sigma^1$.
Second, since $\J$ and $\bar\J$ are twisted and $\I_\A^2(\nu) = \I_\M^2(\nu) = \nu$ but $\I_\K^2(\nu) = \nu + \tau$ we must include phases $e^{\pm \pi i \gamma \delta_{\sigma \K}}$ in the symmetrization such that appropriate phases are produced when applying the involution twice. More precisely, these phases ensure that $\vev{ \J (I_\K^2 (\nu_1)) \bar{\J}^T(\nu_2)}_\K = \vev{ \J (\nu_1 + \tau ) \bar{\J}^T(\nu_2)}_\K$ comes out to be equal to $e^{-2 \pi i \gamma} \vev{ \J (\nu_1) \bar{\J}^T(\nu_2)}_\K$, as it should. Taking the above into account we conclude for $\sigma = \A,\M,\K$
\be
 \vev{ \J (\nu_1) \bar{\J}^T (\nu_2)}_\sigma
 = \begin{pmatrix}
 (\p_1 \p_2 G_B^\gamma)(\nu_1,\nu_2)
 & e^{-\pi i \gamma \delta_{\sigma \K}} (\p_1 \bar \p_2 G_B^\gamma)(\nu_1,I_\sigma(\nu_2)) \\
 e^{+\pi i \gamma \delta_{\sigma \K}} (\bar \p_1 \p_2 G_B^\gamma)(I_\sigma(\nu_1),\nu_2)
 & (\bar \p_1 \bar \p_2 G_B^\gamma)(I_\sigma(\nu_1),I_\sigma(\nu_2))
 \end{pmatrix} \,.
\ee
In terms of the components of $\J$ and $\bar\J$ we have for $\sigma=\A,\M,\K$
\begin{align}
 \vev{\p Z(\nu_1) \p \bar Z(\nu_2)}_\sigma
 & = \frac{\alpha'}{2} (\p_1\p_2 G_B^\gamma) (\nu_1,\nu_2) \,, \label{dZdZbar} \\
 \vev{\p Z(\nu_1) \bar \p \bar Z(\bar \nu_2)}_\sigma
 & = \frac{\alpha'}{2} (\p_1 \bar \p_2 G_B^\gamma) (\nu_1,I_\sigma(\nu_2)) \, e^{-\pi i \gamma \delta_{\sigma \K}} \,, \\
 \vev{\bar\p Z(\bar \nu_1) \p \bar Z(\nu_2)}_\sigma
 & = \frac{\alpha'}{2} (\bar\p_1 \p_2 G_B^\gamma) (I_\sigma(\nu_1),\nu_2) \, e^{+\pi i \gamma \delta_{\sigma \K}} \,, \\
 \vev{\bar \p Z(\bar \nu_1) \bar \p \bar Z(\bar \nu_2)}_\sigma
 & = \frac{\alpha'}{2} (\bar \p_1 \bar \p_2 G_B^\gamma) (I_\sigma(\nu_1),I_\sigma(\nu_2)) \label{dbarZdbarZbar} \,,
\end{align}
where
\begin{align}
 (\p_1\p_2 G_B^\gamma) (\nu_1,\nu_2)
 & = \p_1 G_F^\gamma(\nu_1,\nu_2)
 = - \p_2 G_F^\gamma(\nu_1,\nu_2) \,, \label{current1} \\
 (\p_1 \bar \p_2 G_B^\gamma) (\nu_1,I_\sigma(\nu_2))
 & = - \p_1 G_F^\gamma(\nu_1,I_\sigma(\nu_2))
 = - \bar \p_2 G_F^\gamma(\nu_1,I_\sigma(\nu_2)) \,, \label{current2} \\
 (\bar\p_1 \p_2 G_B^\gamma) (I_\sigma(\nu_1),\nu_2)
 & = \bar\p_1 G_F^\gamma (I_\sigma(\nu_1),\nu_2)
 = \p_2 G_F^\gamma (I_\sigma(\nu_1),\nu_2) \,, \label{current3} \\
 (\bar \p_1 \bar \p_2 G_B^\gamma) (I_\sigma(\nu_1),I_\sigma(\nu_2))
 & = - \bar \p_1 G_F^\gamma(I_\sigma(\nu_1),I_\sigma(\nu_2))
 = \bar \p_2 G_F^\gamma(I_\sigma(\nu_1),I_\sigma(\nu_2)) \,, \label{current4}
\end{align}
and
\be
 G_F^\gamma(\nu_1,\nu_2) = \frac{\vartheta_1(\nu_{12}+\gamma) \vartheta_1'(0)}{\vartheta_1(\gamma)\vartheta_1(\nu_{12})} \,.
\ee

%%%%%%%%%%%%%%%%%%%%%%%%%%%%%%%%%%%%%%%%%%%%%%%%%%%%%%%%%

\section{Fermion propagators}
\label{sec:ferm}

%%%%%%%%%%%%%%%%%%%%%%%%%%%%%%%%%%%%%%%%%%%%%%%%%%%%%%

\subsection{(Doubly) twisted and untwisted fermion correlation functions on $\T$} \label{doublytwistedtorus}

On the torus, the correlation functions of the untwisted holomorphic and anti-holomorphic fermions, with spin structures $s$ and $\tilde s$ respectively, are given by
\be
 \vev{\psi(\nu_1)\psi(\nu_2)}_\T^s  = G_F(s; \nu_1, \nu_2) \,, \quad 
 \vev{\tilde\psi(\bar\nu_1)\tilde\psi(\bar\nu_2)}_\T^{\tilde s} = [G_F(\tilde s;\nu_1,\nu_2)]^* \,,
\ee
where (using the labels $s$ and $(\alpha,\beta)$ interchangeably, according to \reftab{table:spinstructures})
\begin{align}
 G_F(s; \nu_1, \nu_2)
 = \frac{\vartheta_s(\nu_{12})\vartheta'_1(0)}{\vartheta_s(0)\vartheta_1(\nu_{12})}
 = \frac{\vartheta\ss{\alpha}{\beta}(\nu_{12})\vartheta'_1(0)}{\vartheta\ss{\alpha}{\beta}(0)\vartheta_1(\nu_{12})} \,, \quad s=2,3,4 \,.
\end{align}
The untwisted fermion correlation functions satisfy
\begin{align}
 G_F(s;\nu_1 + 1,\nu_2)
 & = - e^{+ 2 \pi i \alpha}
 G_F(s;\nu_1,\nu_2) \,,
 \label{transformationproperties1} \\
 G_F(s;\nu_1 + \tau,\nu_2)
 & = - e^{- 2 \pi i \beta}
 G_F(s;\nu_1,\nu_2) \,.
 \label{transformationproperties2}
\end{align}
On the torus, the correlation functions of the twisted holomorphic and twisted anti-holomorphic fermions, with twist $\vec \gamma = (\delta,\gamma)$, and spin structures $s$ and $\tilde s$ respectively, are given by
\begin{align}
 \vev{\Psi(\nu_1)\bar\Psi(\nu_2)}_\T^s
 = G_F^{\vec \gamma}(s; \nu_1, \nu_2) \,, \quad 
 \vev{\tilde\Psi(\bar\nu_1)\tilde{\bar\Psi}(\bar\nu_2)}_\T^{\tilde s} = [G_F^{-\vec{\gamma}}(\tilde s;\nu_1,\nu_2)]^* \,,
\end{align}
where
\begin{align}
 G_F^{\vec{\gamma}} (s; \nu_1, \nu_2) =
 \frac{\vartheta\ss{\alpha+\delta}{\beta+\gamma}(\nu_{12})\vartheta'_1(0)}{\vartheta\ss{\alpha+\delta}{\beta+\gamma}(0)\vartheta_1(\nu_{12})} \,, \quad s=2,3,4 \,.
\end{align}
The twisted fermion correlation functions satisfy
\begin{align}
 G_F^{\vec \gamma}(s;\nu_1 + 1,\nu_2)
 & = -e^{+2\pi i (\alpha+\delta)}
 G_F^{\vec \gamma}(s;\nu_1,\nu_2) \,,
 \label{transformationproperties3} \\
 G_F^{\vec \gamma}(s;\nu_1 + \tau,\nu_2)
 & = - e^{-2\pi i (\beta+\gamma)}
 G_F^{\vec \gamma}(s;\nu_1,\nu_2) \,.
 \label{transformationproperties4}
\end{align}

\begin{table}[t]
\renewcommand{\arraystretch}{1.3}
\begin{center}
\begin{tabular}{ccccc}
\toprule
$s$ & 1 & 2 & 3 & 4 \\
\midrule
$\ss{\alpha}{\beta}$ & $\ss{1/2}{1/2}$ & $\ss{1/2}{0}$ & $\ss{0}{0}$ & $\ss{0}{1/2}$ \\
$\eta_s$ & $-1$ & $-1$ & $+1$ & $-1$ \\
\bottomrule
\end{tabular}
\end{center}
\vspace{-4mm}
\caption{Spin structures can be expressed in $(\alpha,\beta)$ or $s$.}
\label{table:spinstructures}
\end{table}%

%%%%%%%%%%%%%%%%%%%%%%%%%%%%%%%%%%%%%%%%%%%%%%%%%%%%%%

\subsection{Twisted and untwisted fermion propagators on $\sigma = \A, \M, \K$}

In the following we use the method of images to determine the twisted and untwisted fermion correlation functions on $\sigma = \A, \M, \K$. Let us first consider the twisted correlation function, since the untwisted is simply obtained by putting $\gamma=0$:
\be
 \vev{\bs{\Psi}(\nu_1)\bar{\bs{\Psi}}^T(\nu_1)}_\sigma^s \,, \quad
 \bs{\Psi}(\nu) = \begin{pmatrix} \Psi(\nu) \\ \tilde\Psi(\bar\nu) \end{pmatrix} \,, \quad
 \bar{\bs{\Psi}}(\nu) = \begin{pmatrix} \bar\Psi(\nu) \\ \tilde{\bar{\Psi}}(\bar\nu) \end{pmatrix} \,.
\ee
First, the boundary conditions on $\sigma = \A, \M, \K$ relate the holomorphic and anti-holomorphic spin structures $s$ and $\tilde s$. On the annulus and Klein bottle the two are the same, i.e. $s = \tilde s$ on $\sigma=\A,\K$.  However on the M\"obius strip, $\sigma = \M$, we have $\tilde 1 = 1$, $\tilde 2 = 2$, $\tilde 3 = 4$ and $\tilde 4 = 3$. Taking this into account we have for the relevant correlation function
\be \label{complex}
 [G_F^{\gamma}(s; \nu_1, \nu_2)]^* = - G_F^{-\gamma}(\tilde s; I_\sigma(\nu_1), I_\sigma(\nu_2)) \,, \quad
 G_F^{\gamma} (s; \nu_1, \nu_2) = \frac{\vartheta_s(\nu_{12}+\gamma)\vartheta'_1(0)}{\vartheta_s(\gamma)\vartheta_1(\nu_{12})} \,, \quad s = 2,3,4 \,.
\ee
Thus, for the propagator on the covering torus of each surface we have
\begin{align}
 \vev{\bs{\Psi}(\nu_1) \bar{\bs{\Psi}}^T(\nu_2)}_{\T_\sigma}
 & = \begin{pmatrix}
 G_F^\gamma (s;\nu_1, \nu_2) & 0 \\
 0 & - G_F^\gamma (s; I_\sigma(\nu_1),I_\sigma(\nu_2))
 \end{pmatrix} \,.
\end{align}
The correlation function on $\sigma=\A,\M,\K$ is then obtained by symmetrizing under the involution $I_\sigma$
\begin{align} \label{symmetrized}
 \vev{\bs{\Psi}(\nu_1) \bar{\bs{\Psi}}^T(\nu_2)}_\sigma
 = \thf \big[
  \vev{ \bs{\Psi}(\nu_1) \bar{\bs{\Psi}}^T(\nu_2) }_\T
 + e^{+ \pi i \gamma \delta_{\sigma \mathcal{K}}} \bs{\gamma}_2 \, \vev{ \bs{\Psi}(I_\sigma(\nu_1)) \bar{\bs{\Psi}}^T(\nu_2)}_\T \\
 + \, e^{- \pi i \gamma \delta_{\sigma \mathcal{K}}} \vev{ \bs{\Psi}(\nu_1) \bar{\bs{\Psi}}^T(I_\sigma(\nu_2))  }_\T \bs{\gamma}_2^T
 + \, \bs{\gamma}_2 \vev{ \bs{\Psi}(I_\sigma(\nu_1)) \bar{\bs{\Psi}}^T(I_\sigma(\nu_2)) }_\T \bs{\gamma}_2^T \big] \,. \nonumber
\end{align}
Here the gamma matrix $\bs{\gamma}_2$, with $\bs{\gamma}_2^{-1} \bs{\gamma}^b \bs{\gamma}_2 =  \bs{\gamma}^a \mathbf{P}_a^{~b}$, forms the spinorial representation of $I_\sigma$, that takes into account the transformation properties of $\bs{\Psi}$ and $\bar{\bs{\Psi}}$ under the involution. ($\mathbf{P}$ is the same matrix as before, just in real coordinates now instead of complex; thus, it is given by $-\sigma^3$ instead of $-\sigma^1$). For the annulus and the M\"obius strip, the sign of the Dirac algebra on the covering torus is determined by the requirement $(\bs{\gamma}_2)^2 = \bs{1}$, since $I_\A^2(\nu) = I_\M^2(\nu) = \nu$. Thus a consistent choice for $\sigma=\A,\M$ is $\gammamatrix_1 = \sigma_1$, $\gammamatrix_2 = \sigma_2$, with $\{ \gammamatrix_a , \gammamatrix_b \} = 2 \delta_{ab}$.
For the Klein bottle, the sign of the Dirac algebra on the covering torus is determined by the requirement $(\gammamatrix_2)^2 = - e^{-2 \pi i\beta} \bs{1}$, since $I_\K^2(\nu) = \nu + \tau$. Thus, for $\sigma=\K$ and $s=4$, a consistent choice is $\gammamatrix_1 = \sigma_1$, $ \gammamatrix_2 = \sigma_2$, with $\{ \gammamatrix_a , \gammamatrix_b \} = 2 \delta_{ab}$ and for $\sigma=\K$ and $s=2,3$, a consistent choice is $\gammamatrix_1 = -i \sigma_2$, $\gammamatrix_2 = i \sigma_1$, with $\{ \gammamatrix_a, \gammamatrix_b \} = - 2 \delta_{ab}$\,. With the above choices we find for $\sigma = \A,\M,\K$ and $s=2,3,4$
\be
 \vev{\bs{\Psi}(\nu_1,\bar\nu_1)\bar{\bs{\Psi}}^T(\nu_2,\bar\nu_2)}_{\sigma}^s = 
 \begin{pmatrix}
 G_F^{\gamma}(s;\nu_1,\nu_2) &
 i e^{- \pi i \gamma \delta_{\sigma\K}} G^\gamma_F(s;\nu_1,I_\sigma(\nu_2)) \\
 i e^{+ \pi i \gamma \delta_{\sigma\K}} G^\gamma_F(s;I_\sigma(\nu_1),\nu_2) &
 - G_F^\gamma ( s;I_\sigma(\nu_1), I_\sigma(\nu_2) )
 \end{pmatrix} \,.
\ee
Thus, the twisted fermion correlation functions are
\begin{align}
 \vev{\Psi(\nu_1) \bar \Psi(\nu_2)}_\sigma^s
 & = G_F^\gamma(s;\nu_1,\nu_2) \,, \label{PsiPsibar} \\
 \vev{\tilde \Psi (\bar\nu_1) \bar \Psi(\nu_2)}_\sigma^s
 & = i e^{+\pi i \gamma \delta_{\sigma\mathcal{K}}} \, G_F^\gamma(s;I_\sigma(\nu_1),\nu_2) \,, \\
 \vev{\Psi(\nu_1) \tilde{\bar\Psi}(\bar\nu_2)}_\sigma^s
 & = i e^{- \pi i \gamma \delta_{\sigma\mathcal{K}}} \, G_F^\gamma(s;\nu_1,I_\sigma(\nu_2)) \,, \\
 \vev{\tilde \Psi (\bar\nu_1) \tilde{\bar\Psi}(\bar\nu_2)}_\sigma^s
 & = - G_F^\gamma(s;I_\sigma(\nu_1),I_\sigma(\nu_2)) \label{tildePsitildePsibar} \,,
\end{align}
where
\begin{align}
 G_F^{\gamma} (s; \nu_1, \nu_2) =
 \frac{\vartheta_s(\nu_{12}+\gamma)\vartheta'_1(0)}
 {\vartheta_s(\gamma)\vartheta_1(\nu_{12})} \,, \quad
 s=2,3,4 \,.
\end{align}
The untwisted correlation functions are simply obtained by putting $\gamma = 0$
\begin{align}
 \vev{\psi(\nu_1)\psi(\nu_2)}_\sigma^s
 & = G_F( s; \nu_1,\nu_2) \,, \\
 \vev{\tilde\psi(\bar\nu_1)\psi(\nu_2)}_\sigma^s
 & = i G_F(s;I_\sigma(\nu_1),\nu_2) \,, \\
 \vev{\psi(\nu_1)\tilde \psi(\bar\nu_2)}_\sigma^s
 & = i G_F(s;\nu_1,I_\sigma(\nu_2)) \,, \\
 \vev{\tilde\psi(\bar\nu_1)\tilde\psi(\bar\nu_2)}_\sigma^s
 & = - G_F(s;I_\sigma(\nu_1),I_\sigma(\nu_2)) \,,
\end{align}
where
\begin{align}
 G_F(s; \nu_1, \nu_2) = \frac{\vartheta_s(\nu_{12})\vartheta'_1(0)}{\vartheta_s(0)\vartheta_1(\nu_{12})} \,, \quad s=2,3,4 \,.
\end{align}

%%%%%%%%%%%%%%%%%%%%%%%%%%%%%%%%%%%%%%%%%%%%%%%%%%%

\section{The generalized Eisenstein series}

Let $w,z$ be two arbitrary complex numbers and define the following generalized Eisenstein series,
or Eisenstein-Kronecker function:
\be \label{epsteinzetafunction}
 E_s(w,z)
 = \left( \frac{\tau_2}{\pi} \right)^s {\sum_{m,n}}'
 \frac{e^{\frac{2\pi i}{\tau_2} \Im [(w+m+n\tau)\bar z] }}{|w+m+n\tau|^{2s}} \,, \quad
 \Re s >1 \,,
\ee
where the prime on the sum indicates that
 the sum runs over all $m,n \in \mathbb{Z}$ such that $w + m+n\tau \neq 0$. 
 (Note that this $s$ is not the spin structure. Since they never occur in the same place,
 hopefully there will be no confusion.)
 In \refsec{sec:proofofreflectionformula} we give the analytic continuation of \refeq{epsteinzetafunction} to the whole complex $s$-plane: 
 \begin{itemize}
\item For $z \not \in \mathbb{Z} + \tau \mathbb{Z}$ the analytic continuation defines an entire function in the whole complex $s$-plane.
\item For $z \in \mathbb{Z} + \tau \mathbb{Z}$ the analytic continuation defines a meromorphic function in the whole complex $s$-plane, with a single pole at $s=1$ and residue $e^{\frac{2\pi i}{\tau_2} \Im w\bar z}$. 
\end{itemize}
In both cases the analytic continuations satisfy the reflection formula
\be \label{reflectionformula}
 \Gamma(s) E_s(w,z) = e^{\frac{2 \pi i}{\tau_2} \Im w\bar z} \, \Gamma(1-s) E_{1-s}(z,w) \,.
\ee
The reflection formula can be used to prove the more general reflection formula
\be
 \Gamma(s+\ts{\frac{k}{2}}) E_s^{(k)}(w,z)
 = e^{\frac{2 \pi i}{\tau_2} \Im w \bar z} \, \Gamma(1-s+\ts{\frac{k}{2}}) E_{1-s}^{(k)}(z,w) \,,
\ee
where $E_s^{(k)}$ is essentially the $k^\text{th}$ derivative of $E_{s+\frac{k}{2}}$ with respect to $z$
\be
 E_s^{(k)}(w,z) = \left( \frac{\tau_2}{\pi} \right)^{s+\frac{k}{2}} {\sum_{m,n}}'
 \frac{(\bar w + m + n \bar \tau)^k e^{\frac{2\pi i}{\tau_2} \Im [(w+m+n\tau)\bar z] }}{|w+m+n\tau|^{2s+k}} \,, \quad \Re s>1 \,.
\ee

The first reflection formula can also be used to prove the following useful limit formula
\be
 E_0(w,z) = 0 \,, \quad
 E'_0(w,z) = e^{\frac{2 \pi i}{\tau_2} \Im w \bar z} \, E_1(z,w) \,, \quad
 w \not \in \mathbb{Z} + \tau \mathbb{Z} \,,
\ee
where $E'_s=\frac{\dd}{\dd s}E_s$. We can for example use this limit formula to calculate functional determinants. In \refsec{sec:limitformula} we identify $E_1(w,z)$ with the twisted scalar Green's function. Using \refeq{E10} in \refsec{sec:limitformula} gives
\be
 e^{-E'_0(\beta-\hf + \tau(\alpha-\hf),0)} = e^{-2\pi(\alpha-\hf)^2\tau_2} \left| \frac{\vartheta_1(\beta-\hf+(\alpha-\hf)\tau)}{\eta(\tau)} \right|^2 = \left| \frac{\vartheta\ss{\alpha}{\beta}(0,\tau)}{\eta(\tau)} \right|^2 \,,
\ee
where we invoked the definition of $\vartheta\ss{\alpha}{\beta}$ in terms of $\vartheta_1$ in the last equality.

%%%%%%%%%%%%%%%%%%%%%%%%%%%%%%%%%%%%%%%%%%%%%%%%%%%%%%

\subsection{Relation to Green's functions}

\label{sec:limitformula}

To relate the generalized Eisenstein series to various Green's functions it is useful to introduce real coordinates $u,v$ and $x,y$ for $w$ and $z$ through
\begin{alignat}{3}
 w & = u + \tau v : & \quad\quad
 u & = \Re w - \frac{\tau_1}{\tau_2} \Im w \,, & \quad
 v & = \frac{1}{\tau_2} \Im w \,, \\
  z & = x + \tau y : & \quad\quad
 x & = \Re z - \frac{\tau_1}{\tau_2} \Im z \,, & \quad
 y & = \frac{1}{\tau_2} \Im z \,.
\end{alignat}
In terms of the real coordinates \refeq{epsteinzetafunction} takes the form
\be \label{epsteinzetafunctionreal}
 E_s(w,z)
 = \left( \frac{\tau_2}{\pi} \right)^s {\sum_{m,n}}'
 \frac{e^{2 \pi i (n+v)x}e^{-2 \pi i (m+u)y}}{|m+u+(n+v)\tau|^{2s}} \,.
\ee

From \refeq{epsteinzetafunction} it is clear that
\be \label{deltafunctionsource}
 \p_z \p_{\bar z} E_1(w,z) =
 \begin{dcases}
  -2 \pi \delta^2(z) + \frac{\pi}{\tau_2} & w \in \mathbb{Z} + \tau \mathbb{Z} \\
  - 2 \pi e^{\frac{2\pi i}{\tau_2} \Im w \bar z} \, \delta^2(z) & w \not \in \mathbb{Z} + \tau \mathbb{Z}
 \end{dcases} \,,
\ee
where $\delta^2(z)$ denotes the periodic delta function
\be
 \delta^2(z)
 = \frac{1}{2 \tau_2} \delta(x) \delta(y)
 = \frac{1}{2\tau_2} \sum_{m} e^{2 \pi i m x} \sum_{n} e^{2 \pi i n y} \,.
\ee
At first glance one might think that 
the phase in the second equation of \refeq{deltafunctionsource} is unimportant
since it evaluates to 1 at $z=0$, but because $\delta^2(z)$ is the periodic delta
function, the phase can be nontrivial elsewhere.

From \refeq{epsteinzetafunctionreal} it is clear that
\begin{align}
 E_s(w+1,z) & = E_s(w,z) \,, \label{trans1} \\
 E_s(w+\tau,z) & = E_s(w,z) \,, \label{trans2} \\
 E_s(w,z+1) & = e^{+2 \pi i v} E_s(w,z) \label{trans3} \,, \\
 E_s(w,z+\tau) & = e^{-2 \pi i u} E_s(w,z) \,. \label{trans4}
\end{align}

\refEq{deltafunctionsource} and \refeqs{trans3}{trans4} establishes $E_1(0,z)$ ($E_1(w,z)$) as the periodic (twisted) scalar Green's function on the torus. E.g. for $w=0$ we have the familiar result
\be \label{E10}
 E_1(0,z) = - \ln \left| \frac{\vartheta_1(z,\tau)}{\eta(\tau)} \right|^2 + \frac{2\pi}{\tau_2} (\Im z)^2 \,.
\ee

Thus we conclude
\begin{align}
 G_B(\nu) & = E_1(0,\nu) + \ln (4 \pi^2 |\eta|^4) \,, \label{greenB} \\
 G_B^\gamma(\nu) & = E_1(\gamma,\nu) \,, \\
 G_F(\nu) & = - \p_\nu E_1(0,\nu)
 = \ts{\frac{\pi}{\tau_2}} E_{\frac{1}{2}}^{(1)}(0,\nu) \,, \\
 G_F^\gamma(\nu) & = - \p_\nu E_1(\gamma,\nu)
 = \ts{\frac{\pi}{\tau_2}} E_{\frac{1}{2}}^{(1)}(\gamma,\nu) \,, \\
 G_F(s;\nu) & = - \p_\nu E_1(\beta-\thf + (\alpha-\thf)\tau,\nu)
 = \ts{\frac{\pi}{\tau_2}} E_{\frac{1}{2}}^{(1)}(\beta-\thf + (\alpha-\thf)\tau,\nu) \,, \label{greenspin1} \\
 G_F^\gamma(s;\nu) & = - \p_\nu E_1(\beta-\thf+\gamma+(\alpha-\thf)\tau,\nu)
 = \ts{\frac{\pi}{\tau_2}} E_{\frac{1}{2}}^{(1)}(\beta-\thf + \gamma + (\alpha-\thf)\tau,\nu) \,,\label{greenspin2}
\end{align}
where $\nu$ is the worldsheet coordinate,
$\gamma \in \mathbb{R} \backslash \mathbb{Z}$ denotes the twist
and $s$ labels the spin structures $(\alpha,\beta)$, according to \reftab{table:spinstructures}.
 The $G_F$ without the twist $\gamma$
 is the odd spin structure ($(\alpha,\beta)=(1/2,1/2)$). Note
 that it has a linear nonholomorphic term due to the zero mode in this spin structure. 
 (cf.\ eq.\ (3.266) in \cite{D'Hoker:1988ta}). The zero mode is absent 
 for arbitrarily small twist. In \refeq{greenspin1} and \eqref{greenspin2} we intend
 even spin structures $s=2,3,4$. 

%%%%%%%%%%%%%%%%%%%%%%%%%%%%%%%%%%%%%%%%%%%%%

\subsection{Proof of the reflection formula \refeq{reflectionformula}}

\label{sec:proofofreflectionformula}

If we define the following two-component vectors and matrices
\be \label{matrixnotation}
 \mathbf{m} = \begin{pmatrix} m \\ n \end{pmatrix} \,, \quad
 \mathbf{w} = \begin{pmatrix} u \\ v \end{pmatrix} \,, \quad
 \mathbf{z} = \begin{pmatrix} x \\ y \end{pmatrix} \,, \quad
 \mathbf{G} = \frac{1}{\tau_2} \begin{pmatrix} 1 & \tau_1 \\ \tau_1 & |\tau|^2 \end{pmatrix} \,, \quad
 \bs{\Omega} = \begin{pmatrix} 0 & - 1 \\ 1 & 0 \end{pmatrix} \,,
\ee
we can write $\refeq{epsteinzetafunction}$ as
\be \label{epsteinzetafunctionvector}
 E_s(\mathbf{w},\mathbf{z}) = \frac{1}{\pi^s} {\sum_\mathbf{m}}'
 \frac{e^{2 \pi i (\mathbf{m}+\mathbf{w})^T \bs{\Omega} \mathbf{z}}}{\big[(\mathbf{m}+\mathbf{w})^T \mathbf{G} (\mathbf{m}+\mathbf{w})\big]^s} \,,
\ee
where by $\sum_\mathbf{m}'$ we denote the sum over all $\mathbf{m} \in \mathbb{Z}^2$ such that $\mathbf{m}+\mathbf{w} \neq \bs{0}$.

Now, assuming $\Re s>1$, we have
\begin{align}
 \Gamma(s) E_s(\mathbf{w},\mathbf{z})
 & = \int_0^\infty \dd t \, t^{s-1} {\sum_\mathbf{m}}' e^{- \pi t (\mathbf{m}+\mathbf{w})^T \mathbf{G} (\mathbf{m} + \mathbf{w}) + 2 \pi i (\mathbf{m}+\mathbf{w})^T \bs{\Omega} \mathbf{z}} \label{mellintransformofepstein1} \\
 & = \int_0^\infty \dd t \, t^{s-1} {\sum_\mathbf{m}}' f(t,\mathbf{m},\mathbf{w},\mathbf{z}) \,, \label{mellintransformofepstein2}
\end{align}
where we have defined
\be
 f(t,\mathbf{m},\mathbf{w},\mathbf{z})
 = e^{- \pi t (\mathbf{m}+\mathbf{w})^T \mathbf{G} (\mathbf{m} + \mathbf{w}) + 2 \pi i (\mathbf{m}+\mathbf{w})^T \bs{\Omega} \mathbf{z}} \,.
\ee
Let us now analyze the $[0,1]$ integral in \refeq{mellintransformofepstein2}. If $\mathbf{w} \in \mathbb{Z}^2$ we add the term $\mathbf{m} = -\mathbf{w}$ to the sum
\be
 \int_0^1 \dd t \, t^{s-1} {\sum_{\mathbf{m}}}' f(t,\mathbf{m},\mathbf{w},\mathbf{z})
 = - \frac{\varrho(\mathbf{w})}{s} + \int_0^1 \dd t \, t^{s-1} \sum_{\mathbf{m}} f(t,\mathbf{m},\mathbf{w},\mathbf{z}) \,,
\ee
where $\varrho(\mathbf{w}) = 1$ if $\mathbf{w} \in \mathbb{Z}^2$ and $\varrho(\mathbf{w}) = 0$ if $\mathbf{w} \not\in \mathbb{Z}^2$.

We now make a Poisson resummation ($S$ modular transformation)
\be
 \sum_\mathbf{m} f(t,\mathbf{m},\mathbf{w},\mathbf{z})
 = e^{2 \pi i \mathbf{w}^T \mathbf{\Omega} \mathbf{z}} \sum_\mathbf{m} t^{-1} f(t^{-1},\mathbf{m},\mathbf{z},\mathbf{w}) \,,
\ee
which allows us to write the $[0,1]$ integral as
\begin{align}
 \int_0^1 \dd t \, t^{s-1} {\sum_{\mathbf{m}}}' f(t,\mathbf{m},\mathbf{w},\mathbf{z})
 & = - \frac{\varrho(\mathbf{w})}{s} + e^{2 \pi i \mathbf{w}^T \bs{\Omega} \mathbf{z}} \int_0^1 \dd t \, t^{s-2} \sum_{\mathbf{m}} f(t^{-1},\mathbf{m},\mathbf{z},\mathbf{w}) \\
 & = - \frac{\varrho(\mathbf{w})}{s} + e^{2 \pi i \mathbf{w}^T \bs{\Omega} \mathbf{z}} \left[ \frac{\varrho(\mathbf{z})}{s-1} + \int_1^\infty \dd t \, t^{-s} {\sum_{\mathbf{m}}}' f(t,\mathbf{m},\mathbf{z},\mathbf{w}) \right] \,,
\end{align}
where in the last line by $\sum_\mathbf{m}'$ we denote the sum over all $\mathbf{m} \in \mathbb{Z}$ such that $\mathbf{m}+\mathbf{z} \neq \bs{0}$.

Thus we conclude
\begin{align}
 \Gamma(s)E_s(\mathbf{w},\mathbf{z})
 & = - \frac{\varrho(\mathbf{w})}{s} +
 \int_1^\infty \dd t \, t^{s-1} {\sum_{\mathbf{m}}}' f(t,\mathbf{m},\mathbf{w},\mathbf{z}) \nonumber \\
 & + \, e^{2 \pi i \mathbf{w}^T \bs{\Omega} \mathbf{z}} \left[ \frac{\varrho(\mathbf{z})}{s-1} + \int_1^\infty \dd t \, t^{-s} \,\, {\sum_{\mathbf{m}}}' f(t,\mathbf{m},\mathbf{z},\mathbf{w}) \right] \,,\label{epsteinanalyticcontunation}
\end{align}
where by $\sum_\mathbf{m}'$ we denote the sum over all $\mathbf{m} \in \mathbb{Z}$ such that $\mathbf{m}+\mathbf{w} \neq \bs{0}$ ($\mathbf{m}+\mathbf{z} \neq \bs{0}$) in the first (second) line. Eq. \eqref{epsteinanalyticcontunation} gives the desired
analytic continuation of \refeq{epsteinzetafunctionvector} to the whole complex $s$-plane. The sums and integrals define entire functions of $s$, and the only poles are those of
\be
 \frac{1}{\Gamma(s)} \left[ - \frac{\varrho(\mathbf{w})}{s} + \frac{\varrho(\mathbf{z})}{s-1} e^{2 \pi i \mathbf{w}^T \bs{\Omega} \mathbf{z}} \right] \,.
\ee
In fact, the singular behavior of $1/s$ at $s=0$ is cancelled by that of $1/\Gamma(s)$, so $s=0$ is a smooth point. Thus the only pole of $E_s(\mathbf{w},\mathbf{z})$ is $s=1$ and it is only present for $\mathbf{z} \in \mathbb{Z} + \tau \mathbb{Z}$ with residue $e^{2 \pi i \mathbf{w}^T \bs{\Omega} \mathbf{z}}$.

Finally we observe from \refeq{epsteinanalyticcontunation}
\be
 \Gamma(s)E_s(\mathbf{w},\mathbf{z}) =
 e^{2 \pi i \mathbf{w}^T \bs{\Omega} \mathbf{z}} \, \Gamma(1-s)E_{1-s}(\mathbf{z},\mathbf{w}) \,,
\ee
or in complex notation
\be
 \Gamma(s)E_s(w,z) =
 e^{\frac{2 \pi i}{\tau_2}  \Im w \bar z} \, \Gamma(1-s)E_{1-s}(z,w) \,.
\ee
This is the reflection formula we wanted to prove.

%%%%%%%%%%%%%%%%%%%%%%%%%%%%%%%%%%%%%%%%%%%%%

\subsection{Relation to Epstein zeta functions}

The Epstein zeta function in $n$ dimensions is defined as \cite{siegel}
\be
 \zeta(s,\mathbf{u},\mathbf{v},\mathbf{Q},\mathbf{P}_g)
 = {\sum_{\mathbf{m}}}' \,
 \frac{\mathbf{P}_g[\mathbf{m}+\mathbf{v}] e^{2 \pi i \mathbf{m}^T \mathbf{u}}}{\mathbf{Q}[\mathbf{m}+\mathbf{v}]^{s+\frac{g}{2}}} \,.
\ee
Here $\mathbf{u},\mathbf{v} \in \mathbb{R}^n$ and $\mathbf{m} \in \mathbb{Z}^n$
with $\mathbf{m}+\mathbf{v} \neq \mathbf{0}$. By $\mathbf{Q}[\mathbf{x}] = \mathbf{x}^T \mathbf{Q} \mathbf{x}$, with $\mathbf{x} \in \mathbb{R}^n$,
we denote a positive definite quadratic form
and by $\mathbf{P}_g[\mathbf{x}]$ we denote a
spherical function (from German Kugelfunktion) of degree $g$
with respect to $\mathbf{Q}$, i.e. $\mathbf{P}_g[\mathbf{x}]$ denotes
a homogenous polynomial of degree $g$ solving Laplace's equation
$(\mathbf{Q}^{-1})_{ab} \p_a \p_b \mathbf{P}[\mathbf{x}] = 0$.
All such functions can be written as
$\mathbf{P}_g[\mathbf{x}] = \sum_i \big( \mathbf{x}^T \mathbf{Q} \, \bs{w}_i \big)^g$,
where $\{ \bs{w}_i \in \mathbb{C}^n \}$ comprise a finite collection of isotropic vectors
of $\mathbf{Q}$, i.e. vectors $\bs{w} \in \mathbb{C}^n$ with $\mathbf{Q}[\bs{w}] = \bs{w}^T \mathbf{Q} \bs{w} = 0$.

The Epstein zeta function satisfies the following reflection formula \cite{siegel}
\begin{align}  \label{refl}
 \pi^{-s} \, \Gamma(s + \ts{\frac{g}{2}}) \, \zeta(s,\mathbf{u},\mathbf{v},\mathbf{Q},\mathbf{P}_g)
 = \frac{e^{- 2 \pi i \mathbf{u}^T \mathbf{v}}}{i^g \sqrt{\det \mathbf{Q}}} \,
 \pi^{-(\frac{n}{2}-s)} \,
 \Gamma( \ts{\frac{n}{2}} - s + \ts{\frac{g}{2}}) \,
 \zeta(\ts{\frac{n}{2}} - s,\mathbf{v},-\mathbf{u},\mathbf{Q}^{-1},\mathbf{P}^{-1}_g) \,,
\end{align}
where $\mathbf{P}_g^{-1}[\mathbf{x}] \equiv \mathbf{P}_g[\mathbf{Q}^{-1}\mathbf{x}]$ defines a spherical function of degree $g$ with respect to $\mathbf{Q}^{-1}$.

On the two-torus, i.e. $n=2$, and with the choice of quadratic form $\mathbf{Q} = \mathbf{G}$, with $\mathbf{G}$ as in \refeq{matrixnotation}, there are two isotropic vectors $\bs{w}_1 = (-\bar\tau,1)$ and $\bs{w}_2 = (-\tau,1)$, and we can identify the generalized Eisenstein series of the above with Epstein zeta functions as
\begin{align}
 E_s(\mathbf{w},\mathbf{z})
 & = \pi^{-s} e^{2 \pi i \mathbf{w}^T \mathbf{\Omega} \mathbf{z}} \zeta(s,\bs{\Omega}\mathbf{z},\mathbf{w},\mathbf{G},1) \\
 E_s^{(k)}(\mathbf{w},\mathbf{z})
 & = i^{-k} \pi^{-s-\frac{k}{2}} e^{2 \pi i \mathbf{w}^T \mathbf{\Omega} \mathbf{z}} \zeta(s,\bs{\Omega}\mathbf{z},\mathbf{w},\mathbf{G},\mathbf{P}_k) \,,
\end{align}
where $\mathbf{P}_k[\mathbf{x}] = (\mathbf{x}^T\mathbf{G}\bs{w}_1)^k = i^k (x_1 + x_2\bar \tau)^k$ for $\mathbf{x} = (x_1,x_2)$, with $\bs{w}_1 =(-\bar\tau,1)$ and $\mathbf{G},\bs{\Omega}$ as in \refeq{matrixnotation}.

%%%%%%%%%%%%%%%%%%%%%%%%%%%%%%%%%%%%%%%%%%%%%%%%%%%%%%%%%%%%%%%%%%%%%%%%

\section{Twisted five-theta identity}
\label{sec:fivetheta}

We want to generalize the five-theta identity \cite{Atick:1986rs}\footnote{Note that in some ways this is simpler than the four-theta identity! Note also that the sum in \eqref{5Riemann} is over all spin structures, even and odd; on the other hand, in \eqref{spinstructuresum}, in view of the factor $\t{\alpha}{\beta}(0)$ we can either sum over all spin structures or just over the even ones. Finally, note that due to the same factor and, thus, due to the absence of the odd spin structure in the sum, we could equally well use $\tilde \eta_{\alpha \beta}$ in \eqref{spinstructuresum}.} 
\begin{align}
\label{5Riemann}
 & \sum_{\alpha, \beta} \tilde{\eta}_{\alpha \beta} {  \thbw{\alpha}{\beta}(z_1)  \thbw{\alpha}{\beta}(z_2)  \thbw{\alpha}{\beta}(z_3)  \thbw{\alpha}{\beta}(z_4)  \thbw{\alpha}{\beta}(z_5) \over  \thbw{\alpha}{\beta}(\sum z_i) } 
=  -2 \ 
{\vartheta_1 (z_1') \vartheta_1 (z_2') \vartheta_1 (z_3')  \vartheta_1 (z_4') \vartheta_1 (z_5') \over  \vartheta_1({\scriptsize{ 1 \over 2}}\sum z_i' )}\ ,
\end{align}
where $\tilde \eta_{00}=\tilde \eta_{1/2,1/2}=1$ and $\tilde \eta_{0,1/2}=\tilde \eta_{1/2,0}= -1$,
with
\bea
&& z_1'=z_1+z_2+z_3+z_4\ , \nonumber \\
&& z_2'=z_2+z_3+z_4+z_5\ , \nonumber\\
&& z_3'=z_1+z_3+z_4+z_5\ ,  \nonumber\\
&& z_4'=z_1+z_2+z_4+z_5\ , \nonumber\\
&& z_5'=z_1+z_2+z_3+z_5 \nonumber
\eea
 to shifted characteristics, a ``twisted'' five-theta identity. 
 One can certainly rederive this
 using similar arguments as \cite{Atick:1986rs}. 
 However, since the arguments $z_i$ are general, 
 the identity is  already equivalent to a twisted one, using
\be  \label{period2}
\thba{\alpha}{\beta}(\nu+a\tau + b,\tau) 
&=&
e^{-2\pi i ab} e^{-\pi i a^2\tau}e^{-2\pi i a(\nu+\beta)}
\thba{\alpha+a}{\beta+b}(\nu,\tau)
\ee
(which is e.g.\ (C.13) of \cite{Kiritsis:2007zza} in our conventions).  Using this, the left hand side of \Ref{5Riemann} (with appropriately shifted arguments) is
\bea 
{\rm lhs} &=& C \,  \sum_{\alpha, \beta} \tilde{\eta}_{\alpha \beta} { \thbw{\alpha}{\beta}(z_1) \thbw{\alpha+h_1}{\beta+g_1}(z_2)\thbw{\alpha+h_2}{\beta+g_2}(z_3)\thbw{\alpha+h_3}{\beta+h_3}(z_4) \thbw{\alpha-h_3}{\beta-g_3}(z_5) \over \thbw{\alpha-h_3}{\beta-g_3}(\sum z_i) } 
\eea
with the forefactor
\be
C
   &=& e^{-2\pi i (h_1 g_1+h_2g_2+h_3 g_3)}
e^{-\pi i (h_1^2+h_2^2+h_3^2)\tau} 
e^{-2\pi i (h_1 z_2 + h_2z_3 + h_3 z_4 -h_3 z_5+h_3 \sum z_i)}
\nonumber 
\ee
using $h_1+h_2+h_3=0$.
The important thing is that the dependence on the spin structure label $\beta$ cancelled
in this phase (the spin structure label $\alpha$ never appeared at all in \Ref{period}). If this had not happened
we could not have hoped to obtain a simple generalization.
Similarly for the right hand side, we have an analogous 
forefactor 
\be
C' &=& e^{-2\pi i (h_1 g_1+h_2g_2+h_3 g_3)}
e^{-\pi i (h_1^2+h_2^2+h_3^2)\tau} 
e^{-2\pi i (-h_3 z_2' + h_2z_3' + h_1 z_4' -2h_3 z_5' +h_3\sum z_i' )}e^{-2\pi i (h_1+h_2-h_3)\cdot 1/2}  
\nonumber 
\ee
but in fact
\be  \label{phiphi}
{C' \over C} =  e^{2\pi i h_3}
\ee
and we obtain the desired twisted five-theta identity
\be \label{finalfive}
&&
\sum_{\alpha, \beta} \tilde{\eta}_{\alpha \beta} { \thbw{\alpha}{\beta}(z_1) \thbw{\alpha+h_1}{\beta+g_1}(z_2)\thbw{\alpha+h_2}{\beta+g_2}(z_3)\thbw{\alpha+h_3}{\beta+g_3}(z_4) \thbw{\alpha-h_3}{\beta-g_3}(z_5) \over \thbw{\alpha-h_3}{\beta-g_3}(\sum z_i) }  \\
&& =  -2\cdot e^{2\pi i h_3}\cdot
 { \thbw{1/2}{1/2}(z_1') \thbw{1/2-h_3}{1/2-g_3}(z_2')\thbw{1/2+h_2}{1/2+g_2}(z_3')\thbw{1/2+h_1}{1/2+g_1}(z_4') \thbw{1/2-2h_3}{1/2-2g_3}(z_5') \over \thbw{1/2-2h_3}{1/2-2g_3}({\scriptsize{ 1 \over 2}}\sum z_i') }  \; .  \nonumber
\ee

%%%%%%%%%%%%%%%%%%%%%%%%%%%%%%%%%%%%%%%%%%%%%%%%%%%%%%

\section{Spin structure sums}
\label{sec:spin}

In this appendix we explicitly calculate the spin structure sums appearing in the main text. It would be interesting to use the techniques of  \cite{Tsuchiya:1988va} to
systematize the spin structure sums.

\subsection{Four-fermion}
\label{sec:spin_four}

We define
%\be   \label{defZt}
%Z_s^{\vartheta} \equiv \frac{\thbw{\alpha}{\beta}(0) 
% \thbw{\alpha+h_1}{\beta+g_1+\gamma_1}(0) 
% \thbw{\alpha-h_1}{\beta-g_1+\gamma_2}(0)  
% \thbw{\alpha}{\beta+\gamma_3}(0,\tau)}{\thbwp{1/2}{1/2}(0) 
% \thbw{1/2+h_1}{1/2+g_1+\gamma_1}(0) 
% \thbw{1/2-h_1}{1/2-g_1+\gamma_2}(0)  
% \thbw{1/2}{1/2+\gamma_3}(0)} .
%\ee
\be \label{defZt}
 Z_s^\vartheta =
 \frac{
 \vt\ss{\alpha}{\beta}
 \vt\ss{\alpha+h_1}{\beta+\gamma_1+g_1}
 \vt\ss{\alpha+h_2}{\beta+\gamma_2+g_2}
 \vt\ss{\alpha}{\beta+\gamma_3}
 }{
 \vt'\!\ss{\hf}{\hf}
 \vt\ss{\hf+h_1}{\hf+\gamma_1+g_1}
 \vt\ss{\hf+h_2}{\hf+\gamma_2+g_2}
 \vt\ss{\hf}{\hf+\gamma_3}
 }\ .
\ee

We perform the spin structure sum \Ref{absspin} explicitly:
%\be  \label{spin_structure_sum}
%&& 
%\hspace{-1cm} \sum_{{\rm even}} \frac{\eta_{\alpha \beta} \thbw{\alpha}{\beta}(0,\tau) 
% \thbw{\alpha+h_1}{\beta+\gamma_1+g_1}(0,\tau) 
% \thbw{\alpha+h_2}{\beta+\gamma_2+g_2}(0,\tau)  
% \thbw{\alpha}{\beta+\gamma_3}(0,\tau)}{\thbwp{1/2}{1/2}(0,\tau) 
% \thbw{1/2+h_1}{1/2+\gamma_1+g_1}(0,\tau) 
% \thbw{1/2+h_2}{1/2+\gamma_2+g_2}(0,\tau)  
% \thbw{1/2}{1/2+\gamma_3}(0,\tau)} 
%\frac{\thbw{\alpha}{\beta}(\nu,\tau) 
%\thbwp{1/2}{1/2}(0,\tau)}{\thbw{\alpha}{\beta}(0,\tau)
%\thbw{1/2}{1/2}(\nu,\tau)} \times \non
%&& \times \frac{\thbw{\alpha}{\beta+\gamma_3}(\nu,\tau) 
%\thbwp{1/2}{1/2}(0,\tau)}{\thbw{\alpha}{\beta+\gamma_3}(0,\tau)
%\thbw{1/2}{1/2}(\nu,\tau)} \non
%&=&
%\sum_{{\rm even}} 
%\frac{\eta_{\alpha,\beta} \thbw{\alpha+h_1}{\beta+\gamma_1+g_1}(0,\tau) 
% \thbw{\alpha+h_2}{\beta+\gamma_2+g_2}(0,\tau)}{\thbw{1/2+h_1}{1/2+\gamma_1+g_1}(0,\tau) 
% \thbw{1/2+h_2}{1/2+\gamma_2+g_2}(0,\tau)  
% \thbw{1/2}{1/2+\gamma_3}(0,\tau)} 
%\frac{\thbw{\alpha}{\beta}(\nu,\tau) 
%\thbwp{1/2}{1/2}(0,\tau)}{\thbw{1/2}{1/2}(\nu,\tau)}
%\frac{\thbw{\alpha}{\beta+\gamma_3}(\nu,\tau)}{\thbw{1/2}{1/2}(\nu,\tau)}\non [2mm]
%&=&
%\frac{\thbw{1/2}{1/2+\gamma_3}(\nu,\tau)}{\thbw{1/2}{1/2}(\nu,\tau)}
%\frac{\thbwp{1/2}{1/2}(0,\tau)}{\thbw{1/2}{1/2+\gamma_3}(0,\tau)}
%= G^{F}_{(1/2,1/2+\gamma_3)}(\nu,\tau)\ . 
%\ee
\begin{align} \label{spin_structure_sum}
 & \sum_{{\rm even}} \eta_s Z_s^\vartheta G_F^{\gamma_3}(s;\nu) G_F(s;\nu) \, \nonumber \\
 & = \sum_{{\rm even}} \eta_{\alpha \beta} \,
 \frac{
 \vt \ss{\alpha}{\beta}
 \vt\ss{\alpha+h_1}{\beta+\gamma_1+g_1}
 \vt\ss{\alpha+h_2}{\beta+\gamma_2+g_2}
 \vt\ss{\alpha}{\beta+\gamma_3}
 \vt\ss{\alpha}{\beta+\gamma_3+\nu}
 \vt'\!\ss{\hf}{\hf}
 \vt\ss{\alpha}{\beta+\nu}
 \vt'\!\ss{\hf}{\hf}
 }{
 \vt'\!\ss{\hf}{\hf} 
 \vt\ss{\hf+h_1}{\hf+\gamma_1+g_1} 
 \vt\ss{\hf+h_2}{\hf+\gamma_2+g_2}  
 \vt\ss{\hf}{\hf+\gamma_3}
 \vt\ss{\alpha}{\beta+\gamma_3}
 \vt\ss{\hf}{\hf+\nu}
 \vt\ss{\alpha}{\beta}
 \vt\ss{\hf}{\hf+\nu}
 } \\
 & =
 \sum_{{\rm even}} 
 \eta_{\alpha\beta} \,
 \frac{
 \vt\ss{\alpha+h_1}{\beta+\gamma_1+g_1} 
 \vt\ss{\alpha+h_2}{\beta+\gamma_2+g_2}
 \vt\ss{\alpha}{\beta+\gamma_3+\nu}
 \vt\ss{\alpha}{\beta+\nu}
 \vt'\!\ss{\hf}{\hf}
 }{
 \vt\ss{\hf+h_1}{\hf+\gamma_1+g_1} 
 \vt\ss{\hf+h_2}{\hf+\gamma_2+g_2}  
 \vt\ss{\hf}{\hf+\gamma_3}
 \vt\ss{\hf}{\hf+\nu}
 \vt\ss{\hf}{\hf+\nu}} \\
 & =
 \frac{
 \vt\ss{\hf}{\hf+\gamma_3+\nu} \vt'\!\ss{\hf}{\hf}
 }{
 \vt\ss{\hf}{\hf+\gamma_3} \vt\ss{\hf}{\hf+\nu}
 } \\
 & = G_{F}^{\gamma_3}(\nu) \,.
\end{align}

%%%%%%%%%%%%%%%%%%%%%%%%%%%%%%%%%%%%%%%%%%%%%

\subsection{Eight-fermion}
\label{sec:spin_eight}
%%%%%%%%%%%%%%%%%%%%%%%%%%%%%%%%%%%%%%%%%%%%%

\subsubsection*{Simpler spin structure sum}

Note that $\vartheta_s'(0)=0$ for even $s$, so
\be
\partial_{\nu}^2 \ln \vartheta_s(0)= \frac{\vartheta_s''(0)}{\vartheta_s(0)}- 
\frac{\vartheta_s'(0)^2}{\vartheta_s(0)^2}
= \frac{\vartheta_s''(0)}{\vartheta_s(0)}\ .
\ee
Then one can use a general four-theta identity with shifts in both characteristics (for example eq.\ (130) of \cite{Berg:2004ek}) in order to obtain
\be
\sum_{\rm even} \eta_{\alpha\beta} \frac{\t{\alpha}{\beta}''(0)}{\thbwp{1/2}{1/2}(0)} \frac{\prod_{i=1}^3 \t{\alpha+h_i}{\beta+g_i+\gamma_i}(0)}
{\prod_{i=1}^3\t{1/2+h_i}{1/2+g_i+\gamma_i}(0)}= \sum_{i=1}^3 \frac{\thbwp{1/2+h_i}{1/2+g_i+\gamma_i}(0)}{\t{1/2+h_i}{1/2+g_i+\gamma_i}(0)},
\ee
 so
\be \label{eightsumresult}
\sum_{{\rm even} } \eta_s\, Z_s^{\vartheta} \,\partial_{\nu}^2 \ln \vartheta_s(0) = 
 \sum_{i=1}^3 \frac{\thbwp{1/2+h_i}{1/2+g_i+\gamma_i}(0)}{\t{1/2+h_i}{1/2+g_i+\gamma_i}(0)}\ .
\ee

%%%%%%%%%%%%%%%%%%%%%%%%%%%%%%%%%%%%%%%%%%%%%%%%%%

\subsubsection*{More complicated spin structure sum}

Now, using (for even spin structures) $\t{\alpha}{\beta}(\nu,\tau)  = \t{\alpha}{\beta}(-\nu,\tau)$ and 
$
\thbwp{\alpha}{\beta}(\nu,\tau)  = -\thbwp{\alpha}{\beta}(-\nu,\tau) $,
one spin structure sum of interest is
\be
&&\sum_{{\rm even} } \eta_s\, Z_s^{\vartheta}\, \left( {\vartheta_{s}'(\gamma_3) \over \vartheta_{s}(\gamma_3)} \right)^{\!2}= \non
&&=- \sum_{{\rm even}} \frac{\eta_{\alpha \beta}\, \thbw{\alpha}{\beta}(0) 
 \thbw{\alpha+h_1}{\beta+g_1+\gamma_1}(0) 
 \thbw{\alpha-h_1}{\beta-g_1+\gamma_2}(0)  
 \thbw{\alpha}{\beta+\gamma_3}(0)}{\thbwp{1/2}{1/2}(0) 
 \thbw{1/2+h_1}{1/2+g_1+\gamma_1}(0) 
 \thbw{1/2-h_1}{1/2-g_1+\gamma_2}(0)  
 \thbw{1/2}{1/2+\gamma_3}(0)} 
 \frac{\thbwp{\alpha}{\beta+\gamma_3}(0) }{\thbw{\alpha}{\beta+\gamma_3}(0)} 
  \frac{\thbwp{\alpha}{\beta-\gamma_3}(0)}{\thbw{\alpha}{\beta-\gamma_3}(0)} \ \non [2mm]
&&=
-\sum_{{\rm even}} 
\frac{\eta_{\alpha\beta} \, \thbw{\alpha}{\beta}(0) \thbw{\alpha+h_1}{\beta+g_1+\gamma_1}(0) 
 \thbw{\alpha-h_1}{\beta-g_1+\gamma_2}(0)}{\thbwp{1/2}{1/2}(0,\tau) \thbw{1/2+h_1}{1/2+g_1+\gamma_1}(0) 
 \thbw{1/2-h_1}{1/2-g_1+\gamma_2}(0)  
 \thbw{1/2}{1/2+\gamma_3}(0)} 
\frac{\thbwp{\alpha}{\beta+\gamma_3}(0) \thbwp{\alpha}{\beta-\gamma_3}(0)}{\thbw{\alpha}{\beta-\gamma_3}(0)}\ . \label{spinstructuresum}
\ee
 We have the twisted five-theta identity of \refapp{sec:fivetheta} with a $\vartheta$ in the denominator,
 but no derivatives. In order to apply this identity we can rewrite 
 the spin structure dependent part of \eqref{spinstructuresum} as
\bea
&& =-\frac{1}{2}\partial^2_{\nu}\Big|_{\nu=0} 
\sum _{\alpha, \beta} \tilde \eta_{\alpha \beta} { 
\t{\alpha}{\beta}(0)\t{\alpha+h_1}{\beta+g_1+\gamma_1}(0)\t{\alpha-h_1}{\beta-g_1+\gamma_2}(0)}
{\t{\alpha}{\beta+\gamma_3}(\nu) \t{\alpha}{\beta-\gamma_3}(-\nu)
\over  \t{\alpha}{\beta-\gamma_3}(0)} \\
& & + \frac{1}{2} \sum _{\alpha, \beta} \tilde \eta_{\alpha \beta} { 
\t{\alpha}{\beta}(0)\t{\alpha+h_1}{\beta+g_1+\gamma_1}(0)\t{\alpha-h_1}{\beta-g_1+\gamma_2}(0)}
{\t{\alpha}{\beta+\gamma_3}''(0) \t{\alpha}{\beta-\gamma_3}(0) + \t{\alpha}{\beta+\gamma_3}(0) \t{\alpha}{\beta-\gamma_3}''(0)
\over  \t{\alpha}{\beta-\gamma_3}(0)} \nonumber \\
& = &
-\frac{1}{2}\partial^2_{\nu}\Big|_{\nu=0} 
\sum _{\alpha, \beta} \tilde \eta_{\alpha \beta} { 
\t{\alpha}{\beta}(0)\t{\alpha+h_1}{\beta+g_1+\gamma_1}(0)\t{\alpha-h_1}{\beta-g_1+\gamma_2}(0)}
{\t{\alpha}{\beta+\gamma_3}(\nu) \t{\alpha}{\beta-\gamma_3}(-\nu)
\over  \t{\alpha}{\beta-\gamma_3}(0)} 
 \label{L1} \\
& & + \sum _{\alpha, \beta} \tilde \eta_{\alpha \beta} \t{\alpha}{\beta}(0)\t{\alpha+h_1}{\beta+g_1+\gamma_1}(0)\t{\alpha-h_1}{\beta-g_1+\gamma_2}(0)
\t{\alpha}{\beta+\gamma_3}''(0) \equiv L_1 + L_2\ . \label{L2}
\eea
In the equality, we rewrote the second line by simplifying the fraction, using that the sum is effectively only over even spin structures.
Now we can treat the two terms in \eqref{L2} separately. Denoting them by $L_1$ and $L_2$ and using \eqref{5Riemann} (which applies in the supersymmetric case with $\gamma_1+\gamma_2+\gamma_3 = 0$), we obtain
\bea
\xi^{-1} L_1 &=&  -{1 \over 2}\partial^2_{\nu}\Big|_{\nu=0} \left[- 2 \frac{\vartheta_1 (\nu) \vartheta_1 (-\gamma_3) \vartheta_1 (-g_1-h_1\tau +\gamma_2) \vartheta_1 (g_1+h_1 \tau+\gamma_1) \vartheta_1 (-\nu-2 \gamma_3)}{\vartheta_1(-2 \gamma_3)} \right] \nonumber \\
& = &- 2 \frac{\vartheta_1'(0) \vartheta_1 (-\gamma_3) \vartheta_1 (-g_1-h_1\tau +\gamma_2) \vartheta_1 (g_1+h_1 \tau+\gamma_1) 
\vartheta_1'(-2 \gamma_3)}{\vartheta_1(-2 \gamma_3)}\ ,
\eea
where $\xi$ is a phase that arises from moving the shifts in the characteristics into the arguments, using \eqref{period}. It is analogous to the phase $C$ of the previous appendix, \refapp{sec:fivetheta}. The same phase also arises for $L_2$ and for the denominator of \eqref{spinstructuresum}, so that it drops out in the final result, eq.\ \eqref{spinstructure_f1}. The cancellation of the phases of the numerator and denominator of \eqref{spinstructuresum} can be checked with a calculation very analogous to the one we performed in appendix \refapp{sec:fivetheta}. 

In order to evaluate $L_2$ we can make use of the formula (see for instance Appendix A of \cite{Stieberger:2002wk})
\be
\sum _{\alpha, \beta} \tilde \eta_{\alpha \beta} { 
\t{\alpha}{\beta}(z_1)\t{\alpha}{\beta}(z_2)\t{\alpha}{\beta}(z_3)} \t{\alpha}{\beta}(z_4) = 2 \vartheta_1(z_1') \vartheta_1 (z_2') \vartheta_1 (z_3') \vartheta_1 (z_4')\ ,
\ee
where the sum is over all spin structures and we have 
\be
\left( \begin{array}{c}
z_1' \\
z_2' \\
z_3' \\
z_4' \\
\end{array} \right)
 = \frac12 \left(
\begin{array}{cccc}
1 & 1 & 1 & 1 \\
1 & -1 & -1 & 1 \\
1 & -1 & 1 & -1 \\
1 & 1 & -1 & -1 \\
\end{array} \right)
\left( \begin{array}{c}
z_1 \\
z_2  \\
z_3  \\
z_4  \\
\end{array} \right)\ .
\ee
%Using 
%\be
%z_1 = 0\quad , \quad z_2 = kv_1\quad , \quad z_3 = kv_2 \quad , \quad z_4 =  z + kv_3 
%\ee
%we obtain
%\be
%z_1' = \tfrac12 z \quad , \quad z_2' = \tfrac12 z + kv_3\quad , \quad z_3' = -\tfrac12 z + kv_2 \quad , \quad z_4' =  -\tfrac12 z + kv_1  
%\ee
%and, thus,
We obtain
\bea
\xi^{-1} L_2 &=&  \partial_z^2 \Big|_{z=0} \sum _{\alpha, \beta} \tilde \eta_{\alpha \beta} { 
\t{\alpha}{\beta}(0)\t{\alpha}{\beta}(g_1+h_1 \tau+\gamma_1)\t{\alpha}{\beta}(-g_1-h_1 \tau+\gamma_2)}
\t{\alpha}{\beta}(z+\gamma_3) \\
& = & 2 \, \partial_z^2 \Big|_{z=0} \vartheta_1(\tfrac12 z) \vartheta_1 (\tfrac12 z + \gamma_3) \vartheta_1 (-\tfrac12 z- g_1-h_1 \tau+\gamma_2) 
\vartheta_1 (-\tfrac12 z + g_1+h_1 \tau+\gamma_1) \\
&=&  \, \vartheta_1'(0) \Big[ \vartheta_1' (\gamma_3) \vartheta_1 (- g_1-h_1 \tau+\gamma_2) \vartheta_1 (g_1+h_1 \tau+\gamma_1) - 
\vartheta_1 (\gamma_3) \vartheta_1' (- g_1-h_1 \tau+\gamma_2) \vartheta_1 (g_1+h_1 \tau+\gamma_1) - \non
&&\hspace{1cm}-\vartheta_1 (\gamma_3) \vartheta_1 (- g_1-h_1 \tau+\gamma_2) \vartheta_1' (g_1+h_1 \tau+\gamma_1) \Big] \nonumber \; , 
\eea
so we have
\be
 \xi^{-1}(L_1+L_2) &=&  
 \, \vartheta_1'(0) \Big[ \vartheta_1' (\gamma_3) \vartheta_1 (- g_1-h_1 \tau+\gamma_2) \vartheta_1 (g_1+h_1 \tau+\gamma_1) \non
 &&\hspace{1cm}- \vartheta_1 (\gamma_3) \vartheta_1' (- g_1-h_1 \tau+\gamma_2) \vartheta_1 (g_1+h_1 \tau+\gamma_1)  \non
&&\hspace{1cm}-\vartheta_1 (\gamma_3) \vartheta_1 (- g_1-h_1 \tau+\gamma_2) \vartheta_1' (g_1+h_1 \tau+\gamma_1)   \non
&&\hspace{1cm} - 2  \vartheta_1 (\gamma_3) \vartheta_1 (-g_1-h_1\tau +\gamma_2) \vartheta_1 (g_1+h_1 \tau+\gamma_1) 
\frac{\vartheta_1'(2 \gamma_3)}{\vartheta_1(2 \gamma_3)} \Big] ,
\ee 
which, when taking into account also the denominator of \eqref{spinstructuresum}, is the result claimed in the main text, cf.\ eq.\ \eqref{spinstructure_f1}.

%%%%%%%%%%%%%%%%%%%%%%%%%%%%%%%%%%%%%%%%%%%%%%%%%%%%%%%%%

\subsubsection*{Yet another spin structure sum}

Finally we address the sum in \Ref{spinstructtwo}. We have
\be \label{spinsumeightfinal}
&& 
\hspace{-1cm} - \sum_{{\rm even}} \frac{\eta_{\alpha \beta} \thbw{\alpha}{\beta}(0,\tau) 
 \thbw{\alpha+h_1}{\beta+g_1+\gamma_1}(0,\tau) 
 \thbw{\alpha-h_1}{\beta-g_1+\gamma_2}(0,\tau)  
 \thbw{\alpha}{\beta+\gamma_3}(0,\tau)}{\thbwp{1/2}{1/2}(0,\tau) 
 \thbw{1/2+h_1}{1/2+g_1+\gamma_1}(0,\tau) 
 \thbw{1/2-h_1}{1/2-g_1+\gamma_2}(0,\tau)  
 \thbw{1/2}{1/2+\gamma_3}(0,\tau)} 
 \frac{\thbw{\alpha}{\beta+\gamma_3}(\tilde{\nu}_2,\tau) 
\thbwp{1/2}{1/2}(0,\tau)}{\thbw{\alpha}{\beta+\gamma_3}(0,\tau)
\thbw{1/2}{1/2}(\tilde{\nu}_2,\tau)} \times \non
&& \times \frac{\thbw{\alpha}{\beta-\gamma_3}(-\tilde{\nu}_2,\tau) 
\thbwp{1/2}{1/2}(0,\tau)}{\thbw{\alpha}{\beta-\gamma_3}(0,\tau)
\thbw{1/2}{1/2}(-\tilde{\nu},\tau)} \ \non
&=&
-\sum_{{\rm even}} 
\frac{\eta_{\alpha\beta}  \thbw{\alpha}{\beta}(0,\tau) \thbw{\alpha+h_1}{\beta+g_1+\gamma_1}(0,\tau) 
 \thbw{\alpha-h_1}{\beta-g_1+\gamma_2}(0,\tau)}{\thbw{1/2+h_1}{1/2+g_1+\gamma_1}(0,\tau) 
 \thbw{1/2-h_1}{1/2-g_1+\gamma_2}(0,\tau)  
 \thbw{1/2}{1/2+\gamma_3}(0,\tau)} 
\frac{\thbw{\alpha}{\beta+\gamma_3}(\tilde{\nu}_2,\tau)}{\thbw{1/2}{1/2}(\tilde{\nu}_2,\tau)}
\frac{\thbw{\alpha}{\beta-\gamma_3}(-\tilde{\nu}_2,\tau) 
\thbwp{1/2}{1/2}(0,\tau)}{\thbw{\alpha}{\beta-\gamma_3}(0,\tau) \thbw{1/2}{1/2}(-\tilde{\nu}_2,\tau)}
\non [2mm]
&=&
-\sum_{{\rm even}} 
\frac{\eta_{\alpha\beta}  \thbw{\alpha}{\beta}(0,\tau) \thbw{\alpha}{\beta+g_1+\gamma_1}(h_1 \tau,\tau) 
 \thbw{\alpha}{\beta-g_1+\gamma_2}(-h_1 \tau,\tau)}{\thbw{1/2}{1/2+g_1+\gamma_1}(h_1\tau,\tau) 
 \thbw{1/2}{1/2-g_1+\gamma_2}(-h_1 \tau,\tau)  
 \thbw{1/2}{1/2+\gamma_3}(0,\tau)} 
\frac{\thbw{\alpha}{\beta+\gamma_3}(\tilde{\nu}_2,\tau)}{\thbw{1/2}{1/2}(\tilde{\nu}_2,\tau)}
\frac{\thbw{\alpha}{\beta-\gamma_3}(-\tilde{\nu}_2,\tau) 
\thbwp{1/2}{1/2}(0,\tau)}{\thbw{\alpha}{\beta-\gamma_3}(0,\tau) \thbw{1/2}{1/2}(-\tilde{\nu}_2,\tau)}
\non [2mm]
&=& 2\, \frac{\thbw{1/2}{1/2+2 \gamma_3}(\tilde{\nu}_2,\tau) \thbwp{1/2}{1/2}(0,\tau)}{\thbw{1/2}{1/2+2\gamma_3}(0,\tau) \thbw{1/2}{1/2}(\tilde{\nu}_2,\tau)}
=2 \,G_F^{2 \gamma_3}(\tilde{\nu}_2).
\ee
At the beginning we reversed the sign of $\tilde{\nu}_2$ together with $\gamma_3$ for one correlator (this is why there is an overall minus sign), in the second equality we moved the shift $h_i$ in the upper characteristic into the argument using \Ref{period} (note that in this case the phase factors cancel due to the denominator) and we used the
five-theta identity in \refapp{sec:fivetheta} in the third equality (note that in this case one can include the odd spin structure, which vanishes).

%%%%%%%%%%%%%%%%%%%%%%%%%%%%%%%%%%%%%%%%%%%%%

\section{Regularization by partial integration}
\label{sec:delta}

Let us adopt the following general strategy.
Let us write the integrand in \eqref{ps}  as ($\nu = \nu_2 - \nu_1$)
\be \label{Rdelta}
 - 2 \delta \, R_{\delta} \, G_F^\gamma(\nu) \, \bar\p G_F^\gamma(\bar\nu) \,, \quad R_\delta = e^{- \delta G_B^\sigma(\nu_1,\nu_2)}\ .
\ee
We partially integrate the integrand:
\be
 - 2 \delta \, R_{\delta} \, G_F^\gamma(\nu) \, \bar\p G_F^\gamma(\bar\nu) \to
 + 2 \delta \, \bar\p R_{\delta} \, G_F^\gamma(\nu) \, G_F^\gamma(\bar\nu)
 + 2 \delta \, R_{\delta} \, \bar\p G_F^\gamma(\nu) \, G_F^\gamma(\bar\nu)\ .
\ee
Note that the total derivative could be discarded given that we are working on the covering torus which does not have any boundary. 
The $\bar{\partial}G_F$ in the second term yields a delta function. The second term
in fact vanishes when $\delta$ is made large, and therefore by analytic continuation vanishes
in general. This is due to the $R_\delta$ factor which behaves as $R_\delta \rightarrow |\nu|^{2 \delta}$ when $|\nu| \to 0$. Let us now focus on the first term, $2 \delta \, \bar{\partial} R_{\delta} \, G_F^\gamma(\nu) \, G_F^\gamma(\bar\nu)$. 
We first note that the $\bar{\partial}$ brings down a factor of $\delta$. Therefore,
the remaining terms must contribute a pole in $\delta$ to survive the $\delta\rightarrow 0$ limit. 
Now one can focus on an infinitesimally small disk around $\nu=0$ for the same reason as in the 8-fermion case.
To search for such poles, we focus on the  $\nu$ integration in 
%\eqref{amplitude_covering_torus}
\eqref{ps}
over the small disk $|\nu|<\varepsilon$. 
In this region we have 
\be
 \bar \partial R_\delta \rightarrow \delta \frac{|\nu|^{2 \delta}}{\bar \nu} \,, \quad \text{ when } |\nu| \to 0
\ee
so in the limit $|\nu| \to 0$
\be
 + 2 \delta \, \bar\p R_{\delta} \, G_F^\gamma(\nu) \, G_F^\gamma(\bar\nu)
 \stackrel{|\nu|\to0}{\longrightarrow}
 + 2 \delta^2 \, \frac{|\nu|^{2 \delta}}{\bar \nu} 
 \left(\frac{1}{\nu}+c_{0,0} + \ldots \right)
 \left(\frac{1}{\bar \nu} + c_{0,0} + \ldots \right) .
\ee
Then we use the
fact that for positive powers $m$ and $n$, the surface integral in polar coordinates is 
\be
\int_0^{\epsilon} \dd |\nu| \, |\nu| \int_0^{2 \pi} \dd \theta \, \nu^m \bar{\nu}^n \propto 2 \pi \delta_{mn} \,.
\ee
Note that to have the full $2\pi$ angle integral
we need to go through all four adjacent fundamental regions,
as shown in \reffig{figmap}.
Unless we have equal powers of $\nu$ and $\bar{\nu}$ 
%in \eqref{intsing},
i.e.\  unless the integrand can be expressed entirely in $|\nu|$,
 the angle integration makes all the 
other singular terms vanish.  We can then focus on 
\be
\int_{|\nu|<\varepsilon} \dd^2 \nu\, 2\, \delta \, \bar\p R_{\delta} \, G_F^\gamma(\nu) \, G_F^\gamma(\bar\nu)
= 4 \delta^2 \int_0^{2 \pi} \dd \theta \int_0^{\varepsilon} \dd | \nu | \, | \nu |^{2 \delta-1} c_{0,0}
 =  4 \pi \delta c_{0,0} \varepsilon^{2 \delta}
 \stackrel{\delta\rightarrow 0}{\rightarrow} 
4 \pi \delta c_{0,0}\ , \non
\ee
where $c_{0,0}$ is now the residue of $G_F^\gamma(\nu) G_F^\gamma(\bar\nu)$ at $\nu = 0$. Using that
\be
 G_F^\gamma(\nu) = \frac{\vartheta_1(\nu + \gamma) \vartheta'_1(0)}{\vartheta_1(\gamma) \vartheta_1(\nu)}
 = \frac{1}{\nu} + \frac{\vartheta_1'(\gamma)}{\vartheta_1(\gamma)} + \ldots \,,
\ee
we find
\be
 c_{0,0} = \frac{\vartheta_1'(\gamma)}{\vartheta_1(\gamma)} = \p \ln \vartheta_1(\gamma) \,.
\ee
So the result for the integration over vertex positions
is a derivative of the twisted boson propagator. 
Compare also e.g.\ eq.\ (56) in \cite{Conlon:2010qy}.

Thus from \eqref{ps} we have (up to first order in $\delta$)
\be \label{4f}
 \sum_s \eta_s Z_s^{\vartheta} 
 \big\langle V_{Z \bar Z}(p_1) V_{Z \bar Z}(p_2) \big\rangle_\sigma^s \big|_{\substack{\text{Four}~~~~\, \\ \text{Fermions}}}
 = \delta \cdot 8 \pi  \tau_2\,  \frac{\tht_1'(\gamma,\tau_\sigma)}{\tht_1(\gamma,\tau_\sigma)}\ .
\ee
The factor $2 \tau_2$ comes from the $\nu_2$ integral.

%%%%%%%%%%%%%%%%%%%%%%%%%%%%%%%%%%%%%%%%%%%%%

\section{Point splitting} 
\label{sec:point}

Here we give another approach using point splitting. The starting expression is the same as in regularization  by partial integration, but
now we shift the $\nu$-argument of $G_F$ by $\epsilon$, which we are going to choose real for simplicity, and then take the $|\epsilon| \to 0$ limit at the end: 
\be
 2 \delta \, \bar\p R_{\delta} \, G_F^\gamma(\nu-\epsilon) \, G_F^\gamma(\bar\nu)
 + 2 \delta \, R_{\delta} \, \bar\p G_F^\gamma(\nu-\epsilon) \, G_F^\gamma(\bar\nu)\ .
\ee

Let us first look at the first term. Since this is proportional to $\delta^2$, in order to get $1/\delta$ upon $|\nu|$-integration one would have to identify a pole like $|\nu|^{-2}$ from the integrand as done in regularization
by partial integration.  But once we have shifted $\nu$ as above, 
the pole $|\nu|^{-2}$ is absent since point splitting lowers the order of divergence by one: 
\be
\frac{1}{|\nu|^2} \rightarrow \frac{1}{\bar{\nu}\,(\nu-\epsilon)} \approx \left\{ 
  \begin{array}{l l}
    \frac{1}{\bar{\epsilon}} \cdot \frac{1}{\nu-\epsilon} & \quad \text{for  $\nu\to\epsilon$}\\
  -\frac{1}{\bar{\nu}} \cdot \frac{1}{\epsilon}   & \quad \text{for  $\nu\to0$}
  \end{array} \right.
\ee 
Thus one cannot obtain a  $1/\delta$ contribution, so we can discard this term. 

Therefore let us focus on the second term with $\bar{\partial}G_F^\gamma$, which yields a delta function, i.e. $\bar{\partial}G_F^\gamma(\nu)= 2\pi \exp[-\frac{2 \pi i \gamma \Im \nu}{\tau_2}]\delta^2(\nu, \bar{\nu})$. Note that in contrast to regularization by partial integration, here this term does not vanish as we will see now. Since $G_F^\gamma(\bar \nu)$ is non-singular at $\nu=\epsilon$, we can safely set $\delta=0$ for $R_\delta$ 
(i.e $R_\delta=1$), so $\epsilon$ is now the regularization parameter instead of $\delta$.  Thus we obtain
\be
\lim_{\epsilon \to 0}\, 2 \delta \,\int_{\ct} d^2\nu \;  \bar\p G_F^\gamma(\nu-\epsilon) \, G_F^\gamma(\bar\nu)&=&  
4 \pi \delta \;\lim_{\epsilon \to 0} G_F^\gamma(\epsilon) \non
% &=& 4 \pi \delta \left(\frac{1}{\bar{\epsilon}} + c_{0,0}^*\right) \non
&=& 4 \pi \delta \left(\frac{1}{\epsilon} +\frac{\tht_1'(\gamma,\tau)}{\tht_1(\gamma,\tau)}\right)\ . 
\ee
Note that the first (divergent) term cancels out upon adding sectors for each surface because it is sector ($\gamma$ or $k$)-independent, so we can discard it.

Thus from \eqref{ps} we get (up to first order in $\delta$) 
\be
\sum_s \eta_s Z_s^{\vartheta} 
\big\langle V_{Z \bar Z}(p_1) V_{Z \bar Z}(p_2) \big\rangle_\sigma^s \big|_{\substack{\text{Four}~~~~\, \\ \text{Fermions}}}
=\delta \cdot 8\pi\, \tau_2\, \left[\frac{\tht_1'(\gamma,\tau)}{\tht_1(\gamma,\tau)} \right]  . 
\ee
Again, the additional factor $2\, \tau_2$ comes from the $\nu_2$ integral. The result above agrees with analytic continuation and with regularization by  partial integration.
%%%%%%%%%%%%%%%%%%%%%%%%%%%%%%

\section{Eight-fermion: Poles of untwisted correlators for Klein bottle}
\label{stepfunctionappendix}

Here  we recalculate the 8-fermion contribution coming from poles of untwisted correlators for the
Klein bottle as a check on the calculations in the main text.  
For Klein bottle, the untwisted correlator of $f_2$ on the covering torus (i.e. $\nu_1,\nu_2 \in \ct_\ck$) has a pole at 
\be \label{polekleinbottle_step}
\nu_1=I_\ck(\nu_2) - \tau \, \theta\left({\rm Im}\,\nu_2-\frac{\tau_2}{2}\right), 
\ee
where $\theta$ is the  (Heaviside) step function. The reason for the additional term compared to other surfaces is as follows: when ${\rm Im}\,\nu_2>\frac{\tau_2}{2}$, the point $\nu_1=I_\ck(\nu_2)$ is outside the given domain of covering torus since ${\rm Im}\,\nu_1>\tau_2$, but this point is equivalent to $\nu_1-\tau$ up to twist by the periodicity of the torus, so $I_\ck(\nu_2)-\tau$ is the pole that we see within the torus.
Below we suppress the argument of $\theta\left({\rm Im}\,\nu_2-\frac{\tau_2}{2}\right)$.

It is convenient to introduce 
\be 
&&\nu\equiv\nu_1-I_\ck(\nu_2) + \tau \,\theta, \\
&&\tilde{\nu}_2\equiv I_\ck(\nu_2)-\nu_2 - \tau\,\theta= 1-2 {\rm Re}\,\nu_2 +\frac{\tau}{2}(1-2\theta). 
\ee
One can express the arguments of the correlators as follows: 
\be 
&&\nu_1-I_\ck(\nu_2)=\nu - \tau \,\theta,\\
&&I_\ck(\nu_1)-\nu_2=-\bar{\nu}+\tau(1 -\theta),\\ 
&&\nu_1-\nu_2=\nu+\tilde{\nu}_2, \\
&&I_\ck(\nu_2)-I_\ck(\nu_1)= \bar{\nu}+\tilde{\nu}_2-\tau(1-2\theta)\ . 
\ee
Then using perodicity of the propagator along $\tau$ we see
\be
f_2&=&G_F (s; \nu - \tau \,\theta)
 \, G_F \left(s;-\bar{\nu}+\tau(1 -\theta)\right)
\, G_F^\gamma  (s;\nu+\tilde{\nu}_2)
 \, G_F^\gamma \left(s; \bar{\nu}+\tilde{\nu}_2-\tau(1-2\theta)\right) \non
 &=&  e^{2 \pi i \gamma (1-2 \theta)} \,\,G_F  (s; \nu)
 \, G_F (s;-\bar{\nu})
\, G_F^\gamma  (s; \nu+\tilde{\nu}_2)
 \, G_F^\gamma (s; \bar{\nu}+\tilde{\nu}_2). 
\ee 
Therefore near the pole $|\nu|\to0$, we have
\be
&&f_2 \, \stackrel{|\nu|\to0}{\longrightarrow} \, - \,e^{2 \pi i \gamma (1-2 \theta)} \,\,\frac{G_F^\gamma \big(s; \tilde{\nu}_2 \big)^2}{|\nu|^2}, \\
&&G_B^{\ck}(\nu_1,\nu_2)=  G_B^{\ck}(\nu_1,I_\ck(\nu_2))
\stackrel{|\nu|\to0}{\longrightarrow}  |\nu|^{2\delta}
\ee
and the $\nu$-integral over the disk $|\nu_1-I_\sigma(\nu_2)|\leq\varepsilon$ for small $\varepsilon$ gives
\be
\left\{-\delta \int_{\nu\leq\varepsilon} \dd^2 \nu\, e^{-\delta\, G_B^\sigma}\, f_2\right\}_{\delta=0} =
e^{2 \pi i \gamma (1-2 \theta)} \,G_F^\gamma \big(s; \tilde{\nu}_2 \big)^2  \left\{ \delta \int_{\nu\leq\varepsilon} \dd^2 \nu
\frac{ |\nu|^{2 \delta}}{|\nu|^2}\right\}_{\delta=0} = 2 \pi \,e^{2 \pi i \gamma (1-2 \theta)} \,G_F^\gamma \big(s; \tilde{\nu}_2 \big)^2\ . \non  
\ee
The spin structure sum, including the partition function, is the same as in \eqref{sss} with the result
\be
\sum_{{\rm even} } \eta_s\, Z_s^{\vartheta}\, G^\gamma_F(s, \tilde{\nu}_2)^2\,\, =\,\, 2\, G_F^{2\gamma}(\tilde{\nu}_2)\ ,
\ee
cf.\ appendix \refapp{sec:spin}. 

%%%%%%%%%%%%%%%%%%%%%%%%%%%%%%%%%%%%%%%%%%%%%%%%%%%

\subsubsection*{$\nu_2$-integral}

Now we have to compute
\be
\int_{\mathcal{T}_\ck} \dd^2 \nu_2\,\ 2\,e^{2 \pi i \gamma (1-2 \theta)} \,G_F^{2\gamma}(\tilde{\nu}_2) \; . 
\ee
This can be done in analogy with annulus and M\"obius (here $G^F$  also has no twist in upper characteristic and is therefore periodic  in the horizontal direction), but with small changes due to 
the presence of $\theta$, which depends on ${\rm Im}\,\nu_2$:
\be
&&\int_{\mathcal{T}} \dd^2 \nu_2\,2\,e^{2 \pi i \gamma (1-2 \theta)}\, G_F^{2\gamma}(\tilde{\nu}_2)= \non
&& ~~~~~~~~~~~~ =4 \int_0^{\tau_2} \dd\, {\rm Im} \, \nu_2
\int_0^1 \dd\, {\rm Re} \, \nu_2\,e^{2 \pi i \gamma (1-2 \theta)}\, G_F^{2\gamma}(-2\,{\rm Re} \, \nu_2+ \frac{\tau}{2} \, (1-2 \theta)) \\ \non
&& ~~~~~~~~~~~~ =4 \int_0^{\tau_2} \dd\, {\rm Im} \, \nu_2\,e^{2 \pi i \gamma (1-2 \theta)} 
\int_0^1 \dd\, {\rm Re} \, \nu_2\, G_F^{2\gamma}(-2\,{\rm Re} \, \nu_2+ \frac{\tau}{2} \, (1-2 \theta)) \\
&&~~~~~~~~~~~~= 4 \left(\int_0^{\tau_2/2}+\int_{\tau_2/2}^{\tau_2}\right) \dd\, {\rm Im} \, \nu_2\,e^{2 \pi i \gamma (1-2 \theta)} \int_0^1 \dd\, x\,G_F^{2\gamma}(-2x+ \frac{\tau}{2}\, (1-2 \theta)) \\
&&~~~~~~~~~~~~= 4 \left(\int_0^{\tau_2/2}+\int_{\tau_2/2}^{\tau_2}\right) \dd\, {\rm Im} \, \nu_2\,e^{2 \pi i \gamma (1-2 \theta)} \int_0^1 \dd\, x\,G_F^{2\gamma}(-2x+ \frac{\tau}{2} - \tau \theta) \\
&&~~~~~~~~~~~~= 4 \left(\int_0^{\tau_2/2}+\int_{\tau_2/2}^{\tau_2}\right) \dd\, {\rm Im} \, \nu_2\,e^{2 \pi i \gamma (1-2 \theta)} \int_0^1 \dd\, x\,G_F^{2\gamma}(-2x+ \frac{\tau}{2})\,e^{4 \pi i \gamma \theta} \\
&&~~~~~~~~~~~~= 4 \left(\int_0^{\tau_2/2}+\int_{\tau_2/2}^{\tau_2}\right) \dd\, {\rm Im} \, \nu_2\,e^{2 \pi i \gamma} \int_0^1 \dd\, x\,G_F^{2\gamma}(-2x+ \frac{\tau}{2})\\
&&~~~~~~~~~~~~= 4 \,\tau_2 \, e^{2 \pi i \gamma} \int_0^1 \dd\, x\,G_F^{2\gamma}(-2x+ \frac{\tau}{2}) \\
&&~~~~~~~~~~~~= 4 \,\tau_2 \, e^{2 \pi i \gamma} \big[\pi \cot(2 \pi \gamma) - \pi i\big] \\
&&~~~~~~~~~~~~= 4 \pi\,\tau_2\, \sin(2 \pi \gamma)^{-1}\ .
\ee
Here we used the series expansion for $G_F$ as before and 
\be
\int_0^1 dx \cot \left(\pi (- 2 x \pm i \,t)\right) =\mp i, \qquad \text{for any non-zero positive real number $t$}.
\ee
Thus for Klein bottle we get (in agreement with \eqref{nu2integral_untwisted})
\be
\sum_s \eta_s\, Z_s^{\vartheta}\,\lim_{\varepsilon \to 0} \left\{ - \delta \int_{\mathcal{T}_\ck} \dd^2 \nu_2 \int_{|\nu|\leq\varepsilon} \dd^2 \nu_1 \,
 e^{-\delta \, G_B^\ck(\nu_1,\, \nu_2)} \,\, f_2(\nu_1,\nu_2) \right\}_ {\delta=0} =  8\pi^2 \,\tau_2\, \sin(2 \pi \gamma)^{-1} . \non
\ee

%%%%%%%%%%%%%%%%%%%%%%%%%%%%%%%%%%%%%%%%%%%%%%

\section{4 and 8 fermions without lifting}
Here we perform the calculation without using the
lifting trick. The purpose is to check the reliability of the lifting trick used in the main text. 
The main issue has to do with emergence of poles of type $\nu_1=I_\sigma(\nu_2)$ after lifting (in particular for 8 fermions). This type of poles was absent before lifting. This is because before lifting the integration is performed over fundamental region (both for $\nu_1$ and $\nu_2$) and the involution maps a fundamental domain (integration region before lifting) to another one.  Therefore when calculating without lifting there is no issue with poles of type $\nu_1=I_\sigma(\nu_2)$. However we will see the calculation without lifting gives the same results as for lifting, which confirms the reliability of the lifting trick.  We will use regularization 
by partial integration for the 4-fermion term,
and thus we will have to be careful with total derivatives. For 8 fermions we use the same approach as in \refsec{GenArg}.

\subsection{4 fermions}

Let us now consider the four fermion term without lifting to the covering torus. After spin structure summation \refeq{4fermion} takes the form
\begin{align}
 & \sum_s \eta_s Z_s^{\vartheta} 
 \big\langle V_{Z \bar Z}(p_1) V_{Z \bar Z}(p_2) \big\rangle_\sigma^s \big|_\fourfermion
 = - 2 \delta \int_\sigma \dd^2 \nu_1 \int_\sigma \dd^2 \nu_2 \,
 e^{- \delta G_B^\sigma(\nu_1,\, \nu_2)} \nonumber \\
 & \times \Big\{
 (\bar \p_1 \bar \p_2 G_B^\gamma) ( I_\sigma(\nu_1) , I_\sigma(\nu_2) ) \,
 G_F^\gamma \big( \nu_2, \nu_1 \big)
 - (\p_1 \bar \p_2 G_B^\gamma) ( \nu_1 , I_\sigma(\nu_2) ) \,
 G_F^\gamma ( \nu_2 , I_\sigma(\nu_1) ) \nonumber \\ 
 & 
 ~~ - (\bar \p_1 \p_2 G_B^\gamma) ( I_\sigma(\nu_1), \nu_2 ) \,
 G_F^\gamma ( I_\sigma(\nu_2) , \nu_1 )
 + (\p_1 \p_2 G_B^\gamma) ( \nu_1 , \nu_2 ) \,
 G_F^\gamma ( I_\sigma(\nu_2) , I_\sigma(\nu_1) ) \Big\} \,.
\end{align}
As we will see, the second and third terms do not contribute due to the fact that $\nu_1$ and $I_\sigma(\nu_2)$ do not lie in the same fundamental domain, except if $I_\sigma(\nu_2) = \nu_2$, i.e.\ if $\nu_2$ lies on the boundary. Using the arguments of the main text, one sees that these boundary terms do not contribute. Using the results of \refapp{sec:relieving} we can write
\begin{align}
 & \sum_s \eta_s Z_s^{\vartheta} 
 \big\langle V_{Z \bar Z}(p_1) V_{Z \bar Z}(p_2) \big\rangle_\sigma^s \big|_\fourfermion
 = - 2 \delta \int_\sigma \dd^2 \nu_1 \int_\sigma \dd^2 \nu_2 \,
 e^{- \delta G_B^\sigma(\nu_1,\, \nu_2)} \nonumber \\
 & \times \Big\{
 \bar \p_2 G_F^\gamma ( I_\sigma(\nu_1) , I_\sigma(\nu_2) ) \,
 G_F^\gamma \big( \nu_2, \nu_1 \big)
 + \bar \p_2 G_F^\gamma ( \nu_1 , I_\sigma(\nu_2) ) \,
 G_F^\gamma ( \nu_2 , I_\sigma(\nu_1) ) \nonumber \\ 
 & 
 ~~ - \p_2 G_F^\gamma ( I_\sigma(\nu_1), \nu_2 ) \,
 G_F^\gamma ( I_\sigma(\nu_2) , \nu_1 )
 - \p_2 G_F^\gamma ( \nu_1 , \nu_2 ) \,
 G_F^\gamma ( I_\sigma(\nu_2) , I_\sigma(\nu_1) ) \Big\} \,.
\end{align}
Now let us first perform the $\nu_2$ integration using $\delta$-regularization.
First partially integrate, this produces a total derivative and a delta
function term. Then use the divergence theorem to convert
the total derivative to a contour integral along the boundary, which
vanishes when the integrand is made non-singular by letting $\delta$ become
sufficiently large \footnote{Note that in the
main text and in appendix \ref{sec:delta}, we could immediately discard the total derivative due to double periodicity on
the covering torus, but this does not apply here since this integral is over
the fundamental domain of the actual worldsheet surface, not the covering
torus.}. The delta function term also drops out.   Then we are left with
\begin{align}
 & \sum_s \eta_s Z_s^{\vartheta} 
 \big\langle V_{Z \bar Z}(p_1) V_{Z \bar Z}(p_2) \big\rangle_\sigma^s \big|_\fourfermion
 = + 2 \delta \int_\sigma \dd^2 \nu_1 \int_\sigma \dd^2 \nu_2 \, \nonumber \\
 & \times \Big\{
 \bar \p_2 R_\delta(\nu_1,\nu_2) \,
 G_F^\gamma ( I_\sigma(\nu_1) , I_\sigma(\nu_2) ) \,
 G_F^\gamma \big( \nu_2, \nu_1 \big)
 + \bar \p_2 R_\delta(\nu_1,\nu_2) \,
 G_F^\gamma ( \nu_1 , I_\sigma(\nu_2) ) \,
 G_F^\gamma ( \nu_2 , I_\sigma(\nu_1) ) \nonumber \\ 
 & 
 ~~ - \p_2 R_\delta(\nu_1,\nu_2) \,
 G_F^\gamma ( I_\sigma(\nu_1), \nu_2 ) \,
 G_F^\gamma ( I_\sigma(\nu_2) , \nu_1 )
 - \p_2 R_\delta(\nu_1,\nu_2) \,
 G_F^\gamma ( \nu_1 , \nu_2 ) \,
 G_F^\gamma ( I_\sigma(\nu_2) , I_\sigma(\nu_1) ) \Big\} \,. \nonumber
\end{align}
where $R_\delta$ was defined in \eqref{Rdelta}. When the derivative acts on $R_\delta$ we pick up an additional power of $\delta$ resulting in an overall $\delta^2$ factor.
To get a contribution proportional to $\delta$  out of these terms there must be poles. Inside the fundamental domain under consideration, the second and third terms above do not have poles so they do not contribute. Therefore only the first and last terms contribute. The first term is what we considered in \refapp{sec:delta}, however, with one change: after the $\nu_2$-integration we multiply the result by the area of the fundamental domain of $\nu_1$, which is half the area of the covering torus. The fourth term gives an identical contribution and so in the end we find the same result as in the main text.

%%%%%%%%%%%%%%%%%%%%%%%%%%%%%%%%%%%%%%%%%%%%%

\subsection{8 fermions}
Let us now consider \eqref{8fermion}:
\begin{align}
 & \big\langle V_{Z \bar Z}(p_1) V_{Z \bar Z}(p_2) \big\rangle_\sigma^s \big|_\eightfermion
 = - \delta^2 \int_\sigma \dd^2 \nu_1 \int_\sigma \dd^2 \nu_2 \,
 e^{- \delta \, G_B^\sigma(\nu_1,\, \nu_2)} \nonumber \\
 & ~~~~~~~~~~~ \times \Big\{
 \, G_F \big(s; \nu_1, \nu_2 \big)
 \, G_F \big(s; I_\sigma(\nu_1), I_\sigma(\nu_2)\big)
  - G_F \big(s; \nu_1, I_\sigma(\nu_2)\big)
 \, G_F \big(s; I_\sigma(\nu_1), \nu_2\big)
 \Big\} \nonumber \\
 & ~~~~~~~~~~~ \times \Big\{
 \, G_F^\gamma \big(s; \nu_1, \nu_2\big)
 \, G_F^\gamma \big(s; I_\sigma(\nu_2), I_\sigma(\nu_1)\big) \nonumber \\
 & ~~~~~~~~~~~~~
 \, - G_F^\gamma \big(s; I_\sigma(\nu_1), \nu_2\big)
 \, G_F^\gamma \big(s; I_\sigma(\nu_2), \nu_1\big)
 \, e^{2 \pi i \gamma \delta_{\sigma \mathcal{K}}} \nonumber \\
 & ~~~~~~~~~~~~~
  \, - G_F^\gamma \big(s; \nu_1, I_\sigma(\nu_2)\big)
 \, G_F^\gamma \big(s; \nu_2, I_\sigma(\nu_1)\big)
 \, e^{- 2 \pi i \gamma \delta_{\sigma \mathcal{K}}} \nonumber \\
 & ~~~~~~~~~~~~~
 \, + G_F^\gamma \big(s; I_\sigma(\nu_1), I_\sigma(\nu_2)\big)
 \, G_F^\gamma \big(s; \nu_2, \nu_1 \big) \Big\} \,.
\end{align}
For later comparison with the lifted case we reorganize this as
\be
  \big\langle V_{Z \bar Z}(p_1) V_{Z \bar Z}(p_2) \big\rangle_\sigma^s \big|_\eightfermion
 = - \delta^2 \int_\sigma \dd^2 \nu_1 \int_\sigma \dd^2 \nu_2 \,
 e^{- \delta \, G_B^\sigma(\nu_1,\, \nu_2)}  \Big\{ g_1- g_2 - g_3+g_4 \Big\}\ ,
 \ee
 where
\begin{align}
 g_1=& 
 \, G_F \big(s; \nu_1, \nu_2 \big)
 \, G_F \big(s; I_\sigma(\nu_1), I_\sigma(\nu_2)\big) \times  \nonumber\\
 & ~~~ \times
 \Big[ G_F^\gamma \big(s; \nu_1, \nu_2\big)
 \, G_F^\gamma \big(s; I_\sigma(\nu_2), I_\sigma(\nu_1)\big) 
 + G_F^\gamma \big(s; I_\sigma(\nu_1), I_\sigma(\nu_2)\big)
 \, G_F \big(s; \nu_2, \nu_1 \big)\Big]\ , \\
 g_2= &  G_F \big(s; \nu_1, I_\sigma(\nu_2)\big)
 \, G_F \big(s; I_\sigma(\nu_1), \nu_2\big) \times \nonumber\\
 &~~~ \times
 \Big[ G_F^\gamma \big(s; \nu_1, \nu_2\big)
 \, G_F^\gamma \big(s; I_\sigma(\nu_2), I_\sigma(\nu_1)\big)
 + G_F^\gamma \big(s; I_\sigma(\nu_1), I_\sigma(\nu_2)\big)
 \, G_F \big(s; \nu_2, \nu_1 \big)\Big]\ , \\
g_3=&   G_F \big(s; \nu_1, \nu_2 \big)
 \, G_F \big(s; I_\sigma(\nu_1), I_\sigma(\nu_2)\big) \times \nonumber\\
 &~~~ \times
\Big[G_F^\gamma \big(s; I_\sigma(\nu_1), \nu_2\big)
 \, G_F^\gamma \big(s; I_\sigma(\nu_2), \nu_1\big)
 \, e^{2 \pi i \gamma \delta_{\sigma \mathcal{K}}} +  \nonumber\\
 & ~~~~~~~~~~~~~~~~~~~~~~~~~~~~~~~~~~~~~~~~~~~~
  + G_F^\gamma \big(s; \nu_1, I_\sigma(\nu_2)\big)
 \, G_F^\gamma \big(s; \nu_2, I_\sigma(\nu_1)\big)
 \, e^{- 2 \pi i \gamma \delta_{\sigma \mathcal{K}}} \Big]\ , \\
 g_4=&  G_F \big(s; \nu_1, I_\sigma(\nu_2)\big)
 \, G_F \big(s; I_\sigma(\nu_1), \nu_2\big)  \times \nonumber\\
 &~~~ \times
 \big[G_F^\gamma \big(s; I_\sigma(\nu_1), \nu_2\big)
 \, G_F^\gamma \big(s; I_\sigma(\nu_2), \nu_1\big)
 \, e^{2 \pi i \gamma \delta_{\sigma \mathcal{K}}}+  \nonumber\\
 &  ~~~~~~~~~~~~~~~~~~~~~~~~~~~~~~~~~~~~~~~~~~~~
 + G_F^\gamma \big(s; \nu_1, I_\sigma(\nu_2)\big)
 \, G_F^\gamma \big(s; \nu_2, I_\sigma(\nu_1)\big)
 \, e^{- 2 \pi i \gamma \delta_{\sigma \mathcal{K}}} \big] \ .
 \end{align}
From now on we use the same approach as in \refsec{GenArg}: because of the overall factor $\delta^2$ in front of the integral only poles of the integrand  can contribute, so we focus on the infinitesimally small disk around poles. Since our current integration region is a fundamental domain, not a covering torus, we have only poles of type $\nu_1=\nu_2$ on this region because the involution maps a fundamental domain to another one. This means $g_4$ does not contribute and can be discarded. 
Thus we focus on $g_1$, $g_2$ and $g_3$ in turn. We will clarify the relationship to the lifted case by comparing with $f_1$ and $f_2$ (see \eqref{f1} and \eqref{f2}). 

%%%%%%%%%%%%%%%%%%%%%%%%%%%%%%%%%%%%%%%%%%%%%

\subsubsection{$g_1$ versus $f_1$}
Compare $g_1$ with $f_1$ given in \eqref{f1}. It is easy to see that $g_1$ is $f_1$ plus its image by $\nu_1\to I_\sigma(\nu_1)$ and $\nu_2\to I_\sigma(\nu_2)$:
\be
g_1 = f_1 (\nu_1,\nu_2) + f_1\big(I_\sigma(\nu_1),I_\sigma(\nu_2)\big)\ .
\ee
Now we can expand about $\nu_1-\nu_2=\nu=0$, pick up $|\nu|^{-(2-2\delta)}$, integrate over a
small disk around $\nu=0$  and then spin structure sum. But we have to take into account two changes: 1) we have an extra term $f_1\big(I_\sigma(\nu_1),I_\sigma(\nu_2)\big)$, which turns out to give the same result as  $f_1 (\nu_1,\nu_2)$, 
2) the integral over $\nu_2$, being over the fundamental domain instead of over the covering torus, gives a factor of $1/2$ relative to the calculation in the main text. Again these two changes cancel each other and we recover the result of sec.\ \refsec{first_piece}.  

%%%%%%%%%%%%%%%%%%%%%%%%%%%%%%%%%%%%%%%%%%%%%

\subsubsection{$g_2$ versus $f_2$}
Note that $g_2$ also can be written as
\be
g_2 = f_2 (\nu_1,\nu_2) + f_2\big(I_\sigma(\nu_1),I_\sigma(\nu_2)\big)\ .
\ee
Therefore we can adopt the arguments of \refsec{second_piece_nu1_nu2}: Expand about $\nu_1-\nu_2=\nu=0$, pick up $|\nu|^{-(2-2\delta)}$, integrate over a small disk around $\nu=0$  and then spin structure sum. By an argument completely analogous to that of the previous subsection, these two changes cancel each other and we recover the result of \refsec{second_piece_nu1_nu2}.

%%%%%%%%%%%%%%%%%%%%%%%%%%%%%%%%%%%%%%%%%%%%%

\subsubsection{$g_3$ versus $f_2$}
To see to what $g_3$ corresponds to in the lifted case, we note that
\be
g_3=g_2\big(\nu_2 \to I_\sigma(\nu_2)\big) = f_2 \big(\nu_1,I_\sigma(\nu_2)\big) + f_2\big(I_\sigma(\nu_1),I^2_\sigma(\nu_2)\big)\ ,
\ee
so $g_3$ also has to do with $f_2$ as $g_2$ does. Furthermore from the above expression we suspect that $g_3$ should correspond to what we got in \refsec{second_piece_nu1_Inu2} from the pole $\nu_1=I_\sigma(\nu_2)$ of $f_2$, which is the core issue concerning the
lifting trick. So we will pay careful attention to this. 

%To this end, let us consider
%\begin{align}
%g_3=&   G_F \big(s; \nu_1, \nu_2 \big)
% \, G_F \big(s; I_\sigma(\nu_1), I_\sigma(\nu_2)\big) \times \nonumber\\
% &~~~ \times
%\Big[G_F^\gamma \big(s; I_\sigma(\nu_1), \nu_2\big)
% \, G_F^\gamma \big(s; I_\sigma(\nu_2), \nu_1\big)
% \, e^{2 \pi i \gamma \delta_{\sigma \mathcal{K}}} +  \nonumber\\
% & ~~~~~~~~~~~~~~~~~~~~~~~~~~~~~~~~~~~~~~~~~~~~
%  + G_F^\gamma \big(s; \nu_1, I_\sigma(\nu_2)\big)
% \, G_F^\gamma \big(s; \nu_2, I_\sigma(\nu_1)\big)
% \, e^{- 2 \pi i \gamma \delta_{\sigma \mathcal{K}}} \Big]. 
%\end{align}  
In the same way as $g_1$ and $g_2$ we expand $g_3$ about  $\nu\equiv\nu_1-\nu_2=0$ and pick up terms with $|\nu|^{-2}$. Using $\tilde{\nu}_2\equiv I_\sigma(\nu_2)-\nu_2$ this amounts to picking up 
\be
\lim_{\nu\to0}\,  g_3 = - \frac{1}{|\nu|^2} \times \Big[
 \, G_F^\gamma (s; \tilde{\nu}_2)^2
 \, e^{2 \pi i \gamma \delta_{\sigma \mathcal{K}}} 
  + G_F^\gamma (s; - \tilde{\nu}_2)^2
 \, e^{- 2 \pi i \gamma \delta_{\sigma \mathcal{K}}} \Big]\ .
\ee
After the spin structure sum, one gets (see \eqref{spin_sum_G})
\be
\sum_{{\rm even} } \eta_s\, Z_s^{\vartheta}\,\lim_{\nu\to0}\,  g_3 =
- \frac{2}{|\nu|^2} \times \Big[
 \, G_F^{2\gamma} (\tilde{\nu}_2)
 \, e^{2 \pi i \gamma \delta_{\sigma \mathcal{K}}} 
  + G_F^{2\gamma} (- \tilde{\nu}_2)
 \, e^{- 2 \pi i \gamma \delta_{\sigma \mathcal{K}}} \Big]\ .
\ee
By performing the integral over $\nu_1$ (or equivalently $\nu$) using 
\be
\left\{\delta \int_{|\nu|\leq\varepsilon} \dd^2 \nu  \frac{|\nu|^{2\delta}}{|\nu|^2}\right\}_{\delta=0} = 2 \pi
\ee  
we get ($\nu_2=\nu_2^R + i \, \nu_2^I$, $\tilde{\nu}_2= 1 - 2 \nu_2^R+ \tau\, \delta_{\sigma \mathcal{K}}/2$)
\be
&&\sum_s \eta_s\, Z_s^{\vartheta}\,\lim_{\varepsilon \to 0} \left\{ \delta \int_\sigma \dd^2 \nu_2 \int_{|\nu|<\varepsilon} \dd^2 \nu \, e^{-\delta \, G_B^\sigma(\nu_1,\, \nu_2)} \,\, g_3\right\}_ {\delta= 0} = \nonumber\\
&&~~~~~~~~ =- 4 \pi \int_\sigma \dd^2\nu_2\, 
 \Big[ G_F^{2\gamma} (\tilde{\nu}_2)
 \, e^{2 \pi i \gamma \delta_{\sigma \mathcal{K}}} 
  + G_F^{2\gamma} (- \tilde{\nu}_2)
 \, e^{- 2 \pi i \gamma \delta_{\sigma \mathcal{K}}} \Big] \nonumber \\
 && ~~~~~~~~ = - 8 \pi \int_\sigma \dd \nu_2^R\, \dd \nu_2^I\,
 \Big[ G_F^{2\gamma} (1 -2 \nu_2^R + \frac{\tau}{2} \delta_{\sigma \mathcal{K}})
 \, e^{2 \pi i \gamma \delta_{\sigma \mathcal{K}}} 
  + G_F^{2\gamma} (1+2 \nu_2^R - \frac{\tau}{2} \delta_{\sigma \mathcal{K}})
 \, e^{- 2 \pi i \gamma \delta_{\sigma \mathcal{K}}} \Big]  \nonumber \\
&& ~~~~~~~~ = - 8 \pi\, \tau_2^\sigma  \int_{\sigma^R} \dd \nu_2^R\, 
 \Big[ G_F^{2\gamma} (-2 \nu_2^R + \frac{\tau}{2} \delta_{\sigma \mathcal{K}})
 \, e^{2 \pi i \gamma \delta_{\sigma \mathcal{K}}} 
  + G_F^{2\gamma} (2 \nu_2^R - \frac{\tau}{2} \delta_{\sigma \mathcal{K}})
 \, e^{- 2 \pi i \gamma \delta_{\sigma \mathcal{K}}} \Big]\ , \label{g3}
 \ee
 where $\tau_2^\sigma$ is the imaginary part of the modulus of the fundamental domain, not  the covering torus, in particular $\tau_2^\mathcal{K}= \tau_2/2$ (see \reffig{fig:ws}, but note that in this figure, the Klein bottle's fundamental domain has an imaginary part $\nu^I$ that runs from 0 to $\tau_2/2$, not to $\tau_2$). Moreover, $\sigma^R$ is ${\rm Re}(\sigma)$, for instance $(0,1/2)$ for annulus, $(1/2,1)$ for M\"obius and $(0,1)$ for Klein bottle.
Now, we use the expansion
\be
\frac{\vartheta_1(y+z)\vartheta_1'(0)}{\vartheta_1(y)\vartheta_1(z)}=\pi \cot(\pi y)+\pi\cot(\pi z)+
4 \pi \sum_{m=1}^\infty\sum_{n=1}^\infty q^{mn}\sin(2 \pi m y+2 \pi n z),
\ee 
and 
\be
\int_0^1 dx \cot \left(\pi (\pm 2 x + \frac{\tau}{2}\, \delta_{\sigma \mathcal{K}})\right) = - i \,\delta_{\sigma \mathcal{K}}\,\,{\rm sign}({\rm Im}(\tau)).
\ee
So for annulus and M\"obius we get
\be
 \int_{\sigma^R} \dd \nu_2^R\, 
 \Big[ G_F^{2\gamma} (-2 \nu_2^R) 
  + G_F^{2\gamma} (2 \nu_2^R)
  \Big] =  \pi  \cot(2 \pi \gamma) 
\ee 
and for Klein bottle
\be
&& \int_{\sigma^R} \dd \nu_2^R\, 
 \Big[ G_F^{2\gamma} (-2 \nu_2^R + \frac{\tau}{2})
 \, e^{2 \pi i \gamma} 
  + G_F^{2\gamma} (2 \nu_2^R - \frac{\tau}{2})
 \, e^{- 2 \pi i \gamma} \Big] = \nonumber \\
&&  ~~~~~~~ = 
\pi \big[\cot(2 \pi \gamma)-i\big] e^{2 \pi i \gamma}+ \pi \big[\cot(2 \pi \gamma)+i\big] e^{- 2 \pi i \gamma}
   \\
&& ~~~~~~~ = 2 \pi \sin(2 \pi \gamma)^{-1}\ . 
\ee 

Thus we have
\be
\eqref{g3} &=&- 8 \pi^2 \tau_2 \cot(2 \pi \gamma) ~~~~~~~~~~~~~~~~~~ \text{for annulus and M\"obius} \\ 
&=& - 8 \pi^2 \tau_2  \sin(2 \pi \gamma)^{-1} ~~~~~~~~~~~~~~~ \text{for Klein bottle,} 
\ee
in agreement with the main text (see \eqref{nu2integral_untwisted}). 

%%%%%%%%%%%%%%%%%%%%%%%%%%%%%%%%%%%%%%%%%%%%%

\section{The generalized Mellin transform}

\label{sec:gmt}

\subsection{General considerations}

\begin{figure}
\begin{center}
\subfigure[]{\includegraphics[width=0.49\textwidth]{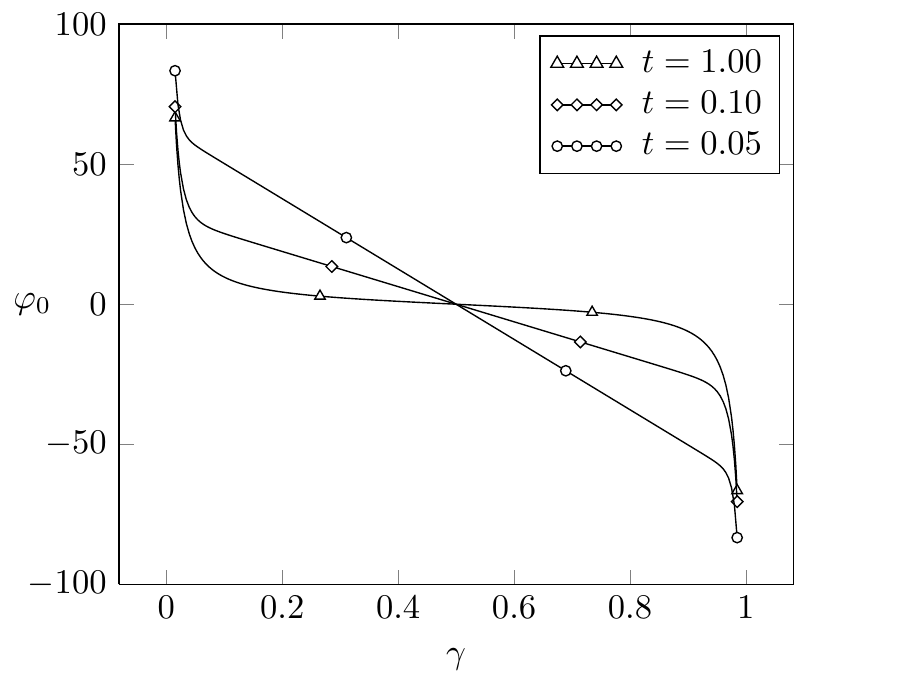}\label{interpolation}}
\subfigure[]{\includegraphics[width=0.49\textwidth]{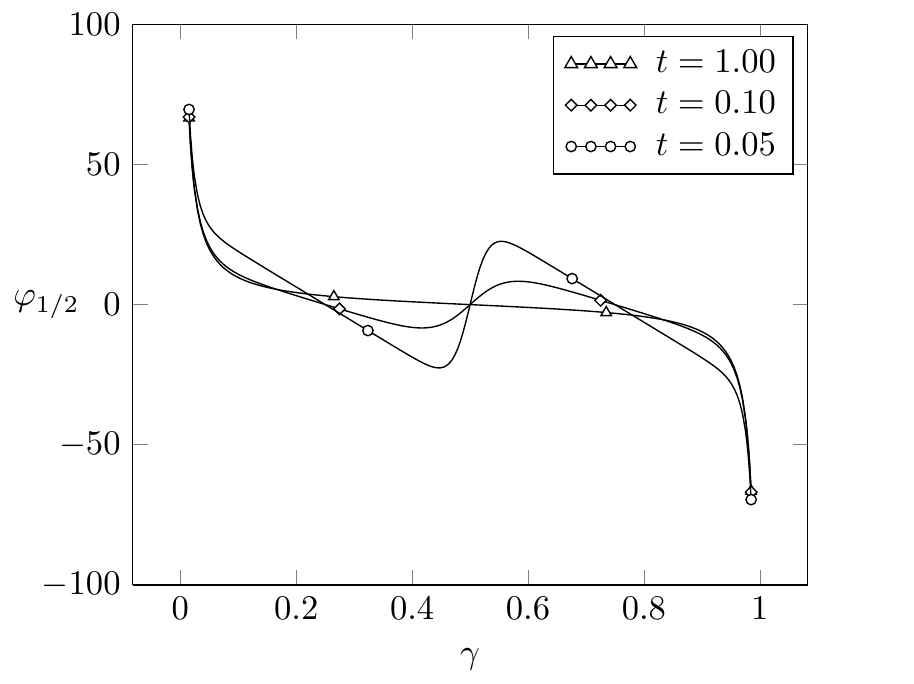}\label{interpolation_mob}}
\subfigure[]{\includegraphics[width=0.49\textwidth]{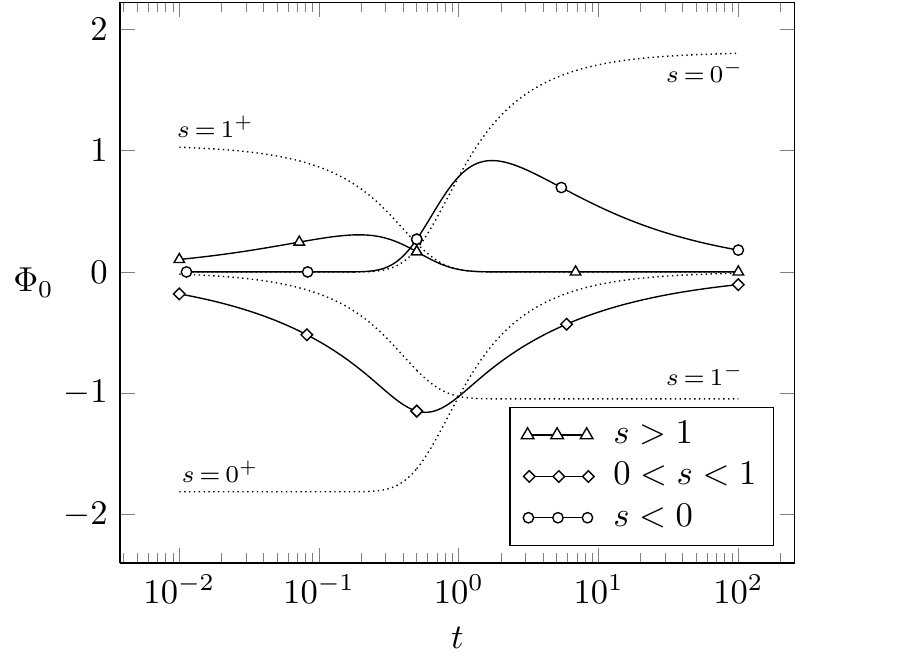}\label{bump_function}}
\subfigure[]{\includegraphics[width=0.49\textwidth]{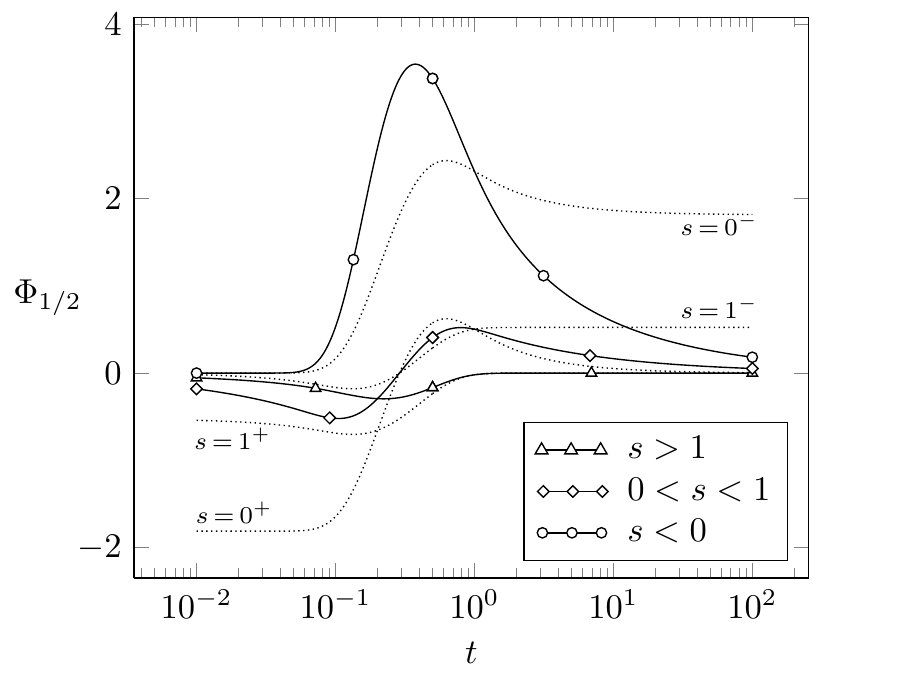}\label{bump_function_mob}}
\caption{In \ref{interpolation} and \ref{interpolation_mob} we plot $\varphi_0 = \p_\gamma \ln \vartheta_1(\gamma,it)$ and $\varphi_{1/2} = \p_\gamma \ln \vartheta_1(\gamma,\hf+it)$, respectively, as functions of $\gamma$ for varying $t$. For $t$ large $\varphi_0 \sim \pi \cot \pi \gamma$ and $\varphi_{1/2} \sim \pi \cot \pi \gamma$ while for $t$ small $\varphi_0 \sim \pi (1-2\gamma) t^{-1}$ (with $0<\gamma<\hf$) and $\varphi_{1/2} \sim \frac{\pi}{2} (1-4\gamma) t^{-1}$, assuming $0<\gamma<1$ and $0<\gamma<\hf$ respectively. In \ref{bump_function} and \ref{bump_function_mob} we plot $\Phi_0(t) = t^s [\varphi_0(t) - \theta(s) \pi \cot \pi \gamma - \theta(1-s) \pi (1-2\gamma) t^{-1}]$ and $\Phi_{1/2}(t) = t^s [\varphi_{1/2}(t) - \theta(s) \pi \cot \pi \gamma - \theta(1-s) \frac{\pi}{2} (1-4\gamma) t^{-1}]$, respectively, (i.e. the integrands of the generalized Mellin transform) as functions of $t$ for varying real values of $s$ and $\gamma=1/3$ fixed. The solid lines correspond to three representative choices of $s$ ($=\frac{3}{2},\hf,-\hf$ for $\triangle,\Diamond,\,$\textcircled{{}}) for which the integrand decays in both limits $t \to 0$ and $t \to \infty$. The dotted lines correspond to the four singular choices $s=0,1$ (from either above or below) for which the integrands fail to decay rapidly enough to define the generalized Mellin transform.}
\end{center}
\end{figure}

Let $\varphi(t;\gamma)$ be any of the two functions
\begin{alignat}{3}
 \varphi_0 (t;\gamma) & : = \p_\gamma \ln \vartheta_1(\gamma,it) \,, & \quad & 0 < \gamma < 1 \,, \label{integrandzero} \\
 \varphi_{1/2} (t;\gamma) & : = \p_\gamma \ln \vartheta_1(\gamma,\thf+it) \,, & \quad \quad & 0 < \gamma < \thf \,. \label{integrandhalf}
\end{alignat}
Since the two functions are both periodic with period $1$ and odd in $\gamma$ we can,
without loss of generality, restrict our attention to $0<\gamma<\hf$.
This choice will be found useful when studying $\varphi_{1/2}$. When studying $\varphi_0$ we will find the choice $0<\gamma<1$ to do just fine. Now, consider the Mellin transform
\be
 \int_0^\infty \dd t \, t^{s-1} \varphi(t;\gamma) \,.
\ee
The naive Mellin transform is not well-defined for any value of $s$,
since the integrand has power law tails at both ends of the integration region
\begin{alignat}{2}
 t \to 0 : && \quad\quad \varphi(t;\gamma)
 & = a(\gamma) t^{-1} + \text{rapid decay} \,, \\
 t \to \infty : && \quad\quad \varphi(t;\gamma) & = b(\gamma) + \text{rapid decay} \,,
\end{alignat}
where $a_0(\gamma) = \pi(1-2\gamma)$ and $b_0(\gamma) = \pi \cot \pi \gamma$ for $\varphi_0$ (assuming $0< \gamma < 1$) and $a_{1/2}(\gamma) = \ts{\frac{\pi}{2}} (1-4\gamma)$ and $b_{1/2}(\gamma) = \pi \cot \pi \gamma$ for $\varphi_{1/2}$ (assuming $0< \gamma < \thf$), see also \reffig{interpolation} and \reffig{interpolation_mob}.
To remedy the situation we pick an arbitrary $T \in (0,\infty)$ and define
\begin{alignat}{3}
  \wt \varphi_{\leq T} (s;\gamma) & : = \int_0^T \dd t \, t^{s-1} \big[ \varphi(t;\gamma) - a(\gamma) t^{-1} \big] + \frac{a(\gamma) T^{s-1}}{s-1} \,, & \quad\quad s & \in \mathbb{C} \backslash \{1\} \,, \\
  \wt \varphi_{\geq T} (s;\gamma) & : = \int_T^\infty \dd t \, t^{s-1} \big[ \varphi(t;\gamma) -b(\gamma) \big] - \frac{b(\gamma) T^s}{s} \,, & \quad\quad s & \in \mathbb{C} \backslash \{0\} \,.
\end{alignat}
Having subtracted off the tails, the above integrals are perfectly finite for any value of $s \in \mathbb{C}$. The point of the above definition is that
\begin{alignat}{3}
 \wt \varphi_{\leq T} (s;\gamma) & = \int_0^T \dd t \, t^{s-1} \varphi(t;\gamma) \,, & \quad \quad \Re s & > 1 \,, \\
 \wt \varphi_{\geq T} (s;\gamma) & = \int_T^\infty \dd t \, t^{s-1} \varphi(t;\gamma) \,, & \quad\quad \Re s & <0 \,.
\end{alignat}
This observation leads us, following Zagier \cite{Zagier},  to define the generalized Mellin transform
of $\varphi(t;\gamma)$ in \refeqs{integrandzero}{integrandhalf} as
\be
 \wt \varphi(s;\gamma) : = \wt \varphi_{\leq T} (s;\gamma) + \wt \varphi_{\geq T} (s;\gamma) \,, \quad
 s \in \mathbb{C} \backslash \{0,1\} \,. \label{generalizedmellintransform}
\ee
The generalized Mellin transform is a meromorphic function on the whole complex $s$-plane with single poles at $s=1$ and $s=0$, with residues $\Res_{s=1} \wt \varphi = a$ and $\Res_{s=0} \wt \varphi = -b$, respectively. Furthermore, even though the individual terms $\wt \varphi_{\leq T}$ and $\wt \varphi_{\geq T}$ depend on $T$, their sum defining $\wt \varphi$ is independent of $T$.\footnote{A change $T \to T + \delta T$ results in a change $\delta \tilde \varphi_{\leq T}(s) = \delta T \, T^{s-1} \varphi(T)$ and $\delta \tilde \varphi_{\geq T}(s) = - \delta T \, T^{s-1} \varphi(T)$, thus $\delta \tilde \varphi (s) = 0$.} E.g. taking the limit $T \to 0$, for $\Re s>1$, and $T \to \infty$, for $\Re s < 0$, we have, see also \reffig{bump_function} and \reffig{bump_function_mob},
\begin{alignat}{3}
 \wt \varphi(s;\gamma) & = \int_0^\infty \dd t \, t^{s-1} \big[ \varphi(t;\gamma) - b(\gamma) \big] \,, & \quad \Re s & > 1 \,, \\
 \wt \varphi(s;\gamma) & = \int_0^\infty \dd t \, t^{s-1} \big[ \varphi(t;\gamma) - a(\gamma) t^{-1} \big] \,, & \quad \Re s & < 0 \,.
\end{alignat}
These equations identify the generalized Mellin transform with the finite part obtained using a cut-off. Indeed, in the limits $\Lambda \to \infty$ and $\lambda \to 0$ we have
\begin{alignat}{3}
 \int_0^\Lambda \dd t \, t^{s-1} \, \varphi(t;\gamma) & = \wt \varphi(s;\gamma) +\frac{b(\gamma) \Lambda^s}{s} \,, & \quad \Re s & > 1 \,, \\
 \int_\lambda^\infty \dd t \, t^{s-1} \, \varphi(t;\gamma) & = \wt \varphi(s;\gamma) - \frac{a(\gamma)}{(s-1)\lambda^{1-s}} \,, & \quad \Re s & < 0 \,.
\end{alignat}
Furthermore, the infinities obtained using a cut-off are determined by the residues of the generalized Mellin transform $a = \Res_{s=1} \wt \varphi$ and $b =- \Res_{s=0} \wt \varphi$.

%%%%%%%%%%%%%%%%%%%%%%%%%%%%%%%%%%%%%%%%%%%%%

\subsection{General results}
\label{appendixclausen}

In \refsecs{sec:casezero}{sec:casehalf} we evaluate the generalized Mellin transforms in terms of generalized Clausen functions and Hurwitz zeta functions, which are defined as the analytic continuations of the (for $\Re s > 1$) absolutely convergent series\footnote{Note that we used a different symbol in the main text for the Clausen function with $s=2$, i.e.\ ${\rm Cl}_2$, \refeq{clausenrelation}. It is common to use different symbols depending on whether $s$ is an integer or not, cf.\ \url{http://en.wikipedia.org/wiki/Clausen_function}.}
\begin{alignat}{3}
 \Si_s(2 \pi a) & = \sum_{n=1}^\infty \frac{\sin 2 \pi n a}{n^s} \,, & \quad & \Re s > 1 \,, \label{clausendef} \\
 \zeta(s,a) & = \sum_{n=0}^\infty \frac{1}{(n+a)^s} \,, & \quad & \Re s > 1 \,, \quad a \neq -1,-2,\ldots \,. \label{hurwitzdef}
\end{alignat}
For $\Re s > 1$ we obtain (see \refeq{casezerorightresult} and \refeq{casehalfrightresult})
\begin{align}
 \wt \varphi_0(s;\gamma)
 & = 2(2\pi)^{1-s} \Gamma(s) \zeta(s) \Si_s(2 \pi \gamma) \,, \label{clausenzero} \\
 \wt \varphi_{1/2}(s;\gamma)
 & = (4 \pi)^{1-s} \Gamma(s) \zeta(s) \big[ \Si_s(2 \pi \gamma) + (2^s-1) \Si_s(2 \pi \gamma+\pi) \big] \,.\label{clausenhalf}
\end{align}
For $\Re s < 0$, with $0<\gamma<1$ for $\wt \varphi_0$ and $0 < \gamma < \hf$ for $\wt \varphi_{1/2}$, we obtain (see \refeq{casezeroleftresult} and \refeq{casehalfleftresult})
\begin{align}
 \wt \varphi_0(s;\gamma)
 & = (2\pi)^s \Gamma(1-s) \zeta(1-s) \big[ \zeta(1-s,\gamma) - \zeta(1-s,1-\gamma) \big] \,, \label{hurwitzzero} \\
 \wt \varphi_{1/2}(s;\gamma)
 & = \pi^s \Gamma(1-s) \zeta(1-s) \big[ \zeta(1-s,\gamma) - \zeta(1-s,1-\gamma) \label{hurwitzhalf} \\
 & + (2^s-1) \big[ \zeta(1-s,\thf+\gamma) - \zeta(1-s,\thf-\gamma) \big] \big] \,. \nonumber
\end{align}
Through analytic continuation, any of the two sets of equations, \refeqs{clausenzero}{clausenhalf} or \refeqs{hurwitzzero}{hurwitzhalf}, determines the generalized Mellin transforms on all of $\mathbb{C} \backslash \{0,1\}$. So, it is reassuring to notice that \refeqs{clausenzero}{clausenhalf} and \refeqs{hurwitzzero}{hurwitzhalf} are completely equivalent. Indeed, the generalized Clausen function can be expressed in terms of Hurwitz zeta function through the following reflection formula
\be
 \frac{\Gamma(s) \zeta(s)}{(2 \pi)^s} \Si_s(2 \pi a) = \frac{\Gamma(1-s) \zeta(1-s)}{2(2 \pi)^{1-s}} \big[ \zeta(1-s,a) - \zeta(1-s,1-a) \big] \,, \label{hurwitzclausenreflectionformula}
\ee
for $\Re s > 1$ and $0 < a < 1$ (using NIST (\href{http://dlmf.nist.gov/25.11.E9}{25.11.9})). Furthermore, the pole structures of \refeqs{clausenzero}{clausenhalf} and \refeqs{hurwitzzero}{hurwitzhalf} agree with those of $\refeq{generalizedmellintransform}$: finiteness at $s=2,3,4,\ldots$ is clear from \refeqs{clausenzero}{clausenhalf} and finiteness at $s=-1,-2,-3,\ldots$ is clear from \refeqs{hurwitzzero}{hurwitzhalf}. Using $\zeta(0,\gamma) = \hf - \gamma$ we reproduce the correct residue at $s=1$ and using $\zeta(1-s,\gamma) = - s^{-1} - \psi(\gamma) + \mathscr{O}(s)$, together with $\psi(\gamma)-\psi(1-\gamma) = - \pi \cot \pi \gamma$, we reproduce the correct residue at $s=0$.\footnote{Furthermore, for $s=1+\epsilon$ and using \eqref{clausendef} and \eqref{clausenzero} we have $- \frac{1}{\pi 2^{1+\epsilon}} \wt \varphi_0(1+\epsilon,\gamma) = - \frac{1}{\pi} \sum_{n=1}^\infty \frac{\sin 2 \pi n \gamma}{n(4\pi n)^\epsilon} \Gamma(1+\epsilon) \zeta(1+\epsilon)$ in agreement with (3.5) of \cite{Akerblom:2007np}.}

%%%%%%%%%%%%%%%%%%%%%%%%%%%%%%%%%%%%%%%%%%%%%

\subsection{Special results}

For all negative integers $s=-n$, with $n=1,2,3,\ldots$, the result can be expressed in terms of polygamma functions $\psi^{(n)}(\gamma)$. Using $\zeta(1+n,\gamma) = \frac{(-1)^{n+1}}{n!} \psi^{(n)}(\gamma)$ we have, with $0<\gamma<1$ for $\wt \varphi_0$ and $0 < \gamma < \hf$ for $\wt \varphi_{1/2}$,
\begin{align}
 \wt \varphi_0(-n;\gamma) & = (-1)^{n+1} \frac{\zeta(1+n)}{(2\pi)^n} \big[ \psi^{(n)}(\gamma) - \psi^{(n)}(1-\gamma) \big] \,, \label{annulusintegerresult} \\
 \wt \varphi_{1/2}(-n;\gamma) & = (-1)^{n+1} \frac{\zeta(1+n)}{\pi^n} \big[ \psi^{(n)}(\gamma) - \psi^{(n)}(1-\gamma)
 - \ts{\frac{2^n-1}{2^n}} \big[ \psi^{(n)}(\thf + \gamma) - \psi^{(n)}(\thf-\gamma) \big] \big] \,. \label{mobiussintegerresult}
\end{align}
For positive odd ($\neq 1$) integers  $s=2n+1$, with $n=1,2,3,\ldots$, the result can be expressed in terms of Bernoulli polynomials $B_{2n+1}(\gamma)$. Using $\zeta(-2n,\gamma) = -\frac{B_{2n+1}(\gamma)}{2n+1}$, with $B_{2n+1}(1-\gamma) = - B_{2n+1}(\gamma)$, together with the limiting value $\lim_{s \to 2n+1} \Gamma(1-s) \zeta(1-s) = \frac{\zeta'(-2n)}{(2n)!} = \pi (-1)^n \frac{\zeta(2n+1)}{(2\pi)^{2n+1}}$ we have
\begin{align}
 \wt \varphi_0(2n+1;\gamma) & = 2 \pi (-1)^{n+1} \frac{\zeta(2n+1)}{2n+1} B_{2n+1}(\gamma) \,, \\
 \wt \varphi_{1/2}(2n+1;\gamma) & = 2 \pi (-1)^{n+1} \frac{\zeta(2n+1)}{2^{2n+1}(2n+1)} \big[ B_{2n+1}(\gamma) + (2^{2n+1}-1) B_{2n+1}(\gamma+\thf) \big] \,.
\end{align}

%%%%%%%%%%%%%%%%%%%%%%%%%%%%%%%%%%%%%%%%%%%%%

\subsection{The generalized Mellin transform of $\varphi_0(t;\gamma) = \p_\gamma \ln \vartheta_1(\gamma,it)$}

\label{sec:casezero}

%%%%%%%%%%%%%%%%%%%%%%%%%%%%%%%%%%%%%%%%%%%%%

\subsubsection{Case $\Re s>1$}

\label{sec:casezeroright}

For $\Re s>1$ the generalized Mellin transform of $\varphi_0(t;\gamma) = \p_\gamma \ln \vartheta_1(\gamma,it)$ takes the form
\be
 \wt \varphi_0(s;\gamma) = \int_0^\infty \dd t \, t^{s-1} \left[ \varphi_0(t;\gamma) - \pi \cot \pi \gamma \right] \,.
\ee
Series expanding the integrand as
$\varphi_0(t;\gamma) - \pi \cot \pi \gamma = 4 \pi \sum_{n=1}^\infty \frac{\sin 2 \pi n \gamma}{e^{2 \pi n t}-1}$
and changing variables $x = 2 \pi n t$, for each term in the sum, we obtain
\be
 \wt \varphi_0(s;\gamma)
 = 4 \pi (2 \pi)^{-s} \sum_{n=1}^\infty \frac{\sin 2 \pi n \gamma}{n^s} \int_0^\infty \frac{\dd x \, x^{s-1}}{e^x-1} \,.
\ee
The sum is identified with the generalized Clausen function $\Si_s(2 \pi \gamma)$, defined in \refeq{clausendef}, while the integral is identified with $\Gamma(s) \zeta(s)$, using the integral representation of Riemann's zeta function $\zeta(s) = \frac{1}{\Gamma(s)} \int_0^\infty \frac{\dd x \, x^{s-1}}{e^{x} - 1}$, for $\Re s>1$. Thus we conclude
\be
 \label{casezerorightresult}
 \wt \varphi_0(s;\gamma) = 2 (2 \pi)^{1-s} \Gamma(s) \zeta(s) \Si_s(2 \pi \gamma) \,, \quad \Re s >1 \,.
\ee

%%%%%%%%%%%%%%%%%%%%%%%%%%%%%%%%%%%%%%%%%%%%%

\subsubsection{Case $\Re s <0$}

\label{sec:casezeroleft}

For $\Re s<0$ the generalized Mellin transform of $\varphi_0(t;\gamma) = \p_\gamma \ln \vartheta_1(\gamma,it)$ takes the form
\be
 \wt \varphi_0(s;\gamma) = \int_0^\infty \dd t \, t^{s-1} \left[ \varphi_0(t;\gamma) - \pi(1-2\gamma) t^{-1} \right] \,, \quad
 0 < \gamma < 1 \,.
\ee
By first modular transforming the integrand $t \to 1/t = \ell$
\be
  \frac{\vartheta_1'(\gamma,it)}{\vartheta_1(\gamma,it)}
 = - 2 \pi \gamma \ell - i \ell \frac{\vartheta_1'(- i\gamma \ell,i \ell)}{\vartheta_1(-i\gamma \ell,i\ell)} \,,
\ee
and then series expanding
\be
 \frac{\vartheta_1'(- i\gamma \ell,i \ell)}{\vartheta_1(-i\gamma \ell,i\ell)}
 = \pi i \coth \pi\gamma\ell - 4 \pi i \sum_{n=1}^\infty \frac{\sinh 2 \pi \gamma \ell}{e^{2\pi n \ell} - 1} \,,
\ee
we have
\be
 \label{annulusfullintegral}
 \wt \varphi_0(s;\gamma) =
 \int_0^\infty \dd \ell \,\, \ell^{-s}
 \left[ \pi (\coth \pi \gamma \ell - 1) - 4 \pi \sum_{n=1}^\infty \frac{\sinh 2 \pi n \gamma \ell}{e^{2 \pi n \ell} - 1} \right] \,.
\ee
The change of variables $x = 2 \pi \gamma \ell$ ($\gamma>0$) for the first term and $x= 2 \pi n \ell$ for the second term give 
\be
 \wt \varphi_0(s;\gamma) =
 (2\pi)^s \int_0^\infty \dd x \, x^{-s}
 \left[ \frac{\gamma^{s-1}}{2} \left( \coth \frac{x}{2} - 1 \right) - \zeta(1-s) \frac{2 \sinh \gamma x}{e^{x} - 1} \right] \,.
\ee
Using $\frac{1}{2} (\coth \frac{x}{2} - 1) = \frac{1}{e^x-1}$, the first integral gives $\int_0^\infty \frac{\dd x \, x^{-s}}{e^x - 1} = \Gamma(1-s)\zeta(1-s)$, valid for $\Re s<0$. The second integral can be evaluated in terms of Hurwitz zeta function, using NIST (\href{http://dlmf.nist.gov/25.11.E25}{25.11.25})
\be
 \int_0^\infty \dd x \, x^{-s} \, \frac{ 2 \sinh \gamma x}{e^x-1}
 = \Gamma(1-s) \big[ \zeta(1-s,1-\gamma) - \zeta(1-s,1+\gamma) \big] \,, \quad \Re s < 0 \,, \quad 0 < \gamma < 1 \,.
\ee
Using the recurrence relation $\zeta(1-s,1+\gamma) = \zeta(1-s,\gamma) - \gamma^{s-1}$ when adding up, we conclude
\be
 \label{casezeroleftresult}
 \wt \varphi_0(s;\gamma) = (2\pi)^s \Gamma(1-s) \zeta(1-s) \big[ \zeta(1-s,\gamma) - \zeta(1-s,1-\gamma) \big] \,, \quad
 \Re s < 0 \,, \quad 0 < \gamma < 1 \,.
\ee

%%%%%%%%%%%%%%%%%%%%%%%%%%%%%%%%%%%%%%%%%%%%%

\subsection{The generalized Mellin transform of $\varphi_{1/2}(t;\gamma) = \p_\gamma \ln \vartheta_1(\gamma,\hf+it)$}

\label{sec:casehalf}

\subsubsection{Case $\Re s>1$}

\label{sec:casehalfright}

For $\Re s > 1$ the generalized Mellin transform of $\varphi_{1/2}(t;\gamma) = \p_\gamma \ln \vartheta_1(\gamma,\hf+it)$ takes the form
\be
 \wt \varphi_{1/2}(t;\gamma) = \int_0^\infty \dd t \, t^{s-1} \big[ \varphi_{1/2}(t;\gamma) - \pi \cot \pi \gamma \big] \,.
\ee
Series expand the integrand
\be
 \varphi_{1/2}(t;\gamma) - \pi \cot \pi \gamma
 = 4 \pi \sum_{n=1}^\infty \frac{\sin 2 \pi n \gamma}{(-1)^n e^{2 \pi n t} - 1}
 = 4 \pi \left[ \sum_{m=1}^\infty \frac{\sin 4 \pi m \gamma}{e^{4 \pi m t} - 1} - \sum_{m=0}^\infty \frac{\sin 2 \pi (2m+1) \gamma}{e^{2 \pi (2m+1) t} + 1} \right] \,.
\ee
Make the change of variables $x = 4 \pi m t$ for the even sum and $x= 4 \pi (m+\hf) t$ for the odd sum
\be
 \wt \varphi_{1/2}(t;\gamma)
 = (4 \pi)^{1-s} \left[ \sum_{m=1}^\infty \frac{\sin 4 \pi m \gamma}{m^s} \int_0^\infty \frac{\dd x \, x^{s-1}}{e^x-1}
 - \sum_{m=0}^\infty \frac{\sin 4 \pi (m+\hf) \gamma}{(m+\hf)^s} \int_0^\infty \frac{\dd x \, x^{s-1}}{e^x+1} \right] \,.
\ee
The integrals contribute as
$\int_0^\infty \frac{\dd x \, x^{s-1}}{e^x-1} = \Gamma(s)\zeta(s)$
and
$\int_0^\infty \frac{\dd x \, x^{s-1}}{e^x+1} = \Gamma(s)\eta(s)$,
where the Dirichlet eta function is related to the Riemann zeta function as  $\eta(s) = 2^{1-s}(2^{s-1}-1) \zeta(s)$.
The sums are identified with generalized Clausen functions, using $\sin(x+n\pi) = (-1)^n \sin(x)$,
\begin{align}
 \sum_{m=1}^\infty \frac{\sin 4 \pi m \gamma}{m^s} & = 2^{s-1} [\Si_s(2 \pi \gamma) + \Si_s(2 \pi \gamma + \pi)] \,, \\
 \sum_{m=0}^\infty \frac{\sin 4 \pi (m+\hf) \gamma}{(m+\hf)^s} & = 2^{s-1} [ \Si_s(2 \pi \gamma) - \Si_s(2 \pi \gamma + \pi)] \,.
\end{align}
Collecting terms we conclude
\be
 \label{casehalfrightresult}
 \wt \varphi_{1/2}(t;\gamma) = (4 \pi)^{1-s} \Gamma(s)\zeta(s) \big[ \Si_s(2 \pi \gamma) + (2^s-1) \Si_s(2 \pi \gamma+\pi) \big] \,, \quad \Re s > 1.
\ee

%%%%%%%%%%%%%%%%%%%%%%%%%%%%%%%%%%%%%%%%%%%%%

\subsubsection{Case $\Re s<0$}

\label{sec:casehalfleft}

For $\Re s < 0$ the generalized Mellin transform of $\varphi_0(t;\gamma) = \p_\gamma \ln \vartheta_1(\gamma,\hf+it)$ takes the form
\be
 \wt \varphi_{1/2}(s;\gamma) = \int_0^\infty \dd t \, t^{s-1} \left[ \varphi_{1/2}(t;\gamma) - \ts{\frac{\pi}{2}}(1-4 \gamma) t^{-1} \right] \,, \quad
 0<\gamma<\thf \,.
\ee
The sequence of modular transformations $S T^2 S$ gives for $\ell = 1/t$
\be
  \frac{\vartheta_1'(\gamma,\hf + it)}{\vartheta_1(\gamma,\hf + it)}
 = - 2 \pi \gamma \ell
 - \frac{i \ell}{2} \frac{\vartheta_1'(-\frac{i\gamma \ell}{2},-\frac{1}{2}+ \frac{i\ell}{4})}{\vartheta_1(-\frac{i\gamma \ell}{2},-\frac{1}{2}+ \frac{i\ell}{4})} \,.
\ee
Series expanding
\be
  \frac{\vartheta_1'(-\frac{i\gamma \ell}{2},-\hf + \frac{i \ell}{4})}{\vartheta_1(-\frac{i\gamma \ell}{2},-\hf + \frac{i \ell}{4})}
 = i \pi \coth \frac{\pi \gamma \ell}{2} - 4 \pi i \sum_{n=1}^\infty \frac{\sinh \pi n \gamma \ell}{(-1)^n e^{\frac{\pi n \ell}{2}} - 1} \,,
\ee
and splitting the sum into even and odd terms give
\be
 \wt \varphi_{1/2}(s;\gamma) = \int_0^\infty \dd \ell \, \ell^{-s} \left[ \frac{\pi}{2} \left( \coth \frac{\pi \gamma \ell}{2} -1 \right) -2 \pi \sum_{n=1}^\infty \frac{\sinh 2 \pi n \gamma \ell}{e^{\pi n \ell} - 1} + 2 \pi \sum_{n=0}^\infty \frac{\sinh 2\pi (n+\hf) \gamma \ell}{e^{\pi(n+\hf)\ell} + 1} \right] \,.
\ee
Make the change of variables $x = \pi \gamma \ell$ ($\gamma > 0$) in the first term, $x = \pi n \ell$ in the second, and $x = \pi (n+\hf) \ell$ in the third, and identify the appropriate zeta functions
\be
 \wt \varphi_{1/2}(s;\gamma) = \pi^s \int_0^\infty \dd x \, x^{-s} \left[ \frac{\gamma^{s-1}}{2} \left( \coth \frac{x}{2} -1 \right) - \zeta(1-s) \frac{2\sinh 2 \gamma x}{e^{x} - 1} + \zeta(1-s,\thf) \frac{2 \sinh 2 \gamma x}{e^{x} + 1} \right] .
\ee
Just as earlier, the first integral gives
$\int_0^\infty \dd x \, \frac{x^{-s}}{2} \left( \coth \frac{x}{2} - 1 \right) = \Gamma(1-s)\zeta(1-s)$.
The second integral is evaluated using NIST (\href{http://dlmf.nist.gov/25.11.E25}{25.11.25}) and NIST (\href{http://dlmf.nist.gov/25.11.E15}{25.11.15}) for $\Re s<0$ and $|{\Re \gamma}| < \thf$
\begin{align}
 & -\int_0^\infty \dd x \, x^{-s} \, \frac{2 \sinh 2 \gamma x}{e^{x} - 1} \\
 & = \Gamma(1-s) \big[ \zeta(1-s,1+2\gamma) - \zeta(1-s,1-2\gamma) \big] \,, \nonumber \\
 & = 2^{s-1} \Gamma(1-s) \big[ \zeta(1-s,\gamma) - \zeta(1-s,1-\gamma) + \zeta(1-s,\thf+\gamma) - \zeta(1-s,\thf-\gamma) - \gamma^{s-1} \big] \,.\nonumber
\end{align}
The third integral is evaluated using NIST (\href{http://dlmf.nist.gov/25.11.E35}{25.11.35}) for $\Re s < 0$ and $|{\Re \gamma}| < \thf$ 
\begin{align}
 & \int_0^\infty \dd x \, x^{-s} \, \frac{2 \sinh 2 \gamma x}{e^{x} + 1} \\
 & = 2^{s-1} \Gamma(1-s) \big[ \zeta(1-s,\gamma) - \zeta(1-s,1-\gamma) - \zeta(1-s,\thf + \gamma) + \zeta(1-s,\thf - \gamma) - \gamma^{s-1} \big] \,.\nonumber
\end{align}
Putting together the results, using $\zeta(1-s,\hf) = (2^{1-s}-1)\zeta(1-s)$, we have
\begin{align}
 & - \int_0^\infty \dd x \, x^{-s} \, \frac{2 \sinh 2 \gamma x}{e^{x} - 1} + (2^{1-s}-1) \int_0^\infty \dd x \, x^{-s} \, \frac{2 \sinh 2 \gamma x}{e^{x} + 1} \\
 & = \Gamma(1-s) \big[ \zeta(1-s,\gamma) - \zeta(1-s,1-\gamma) + (2^s-1)[\zeta(1-s,\thf+\gamma) - \zeta(1-s,\thf-\gamma)] - \gamma^{s-1} \big] \,.\nonumber
\end{align}
When combining the above results the $\gamma^{s-1}$ terms cancel and we conclude
\begin{align}
 \label{casehalfleftresult}
 \wt \varphi_{1/2}(s;\gamma)
 & = \pi^s \Gamma(1-s) \zeta(1-s) \big[ \zeta(1-s,\gamma)-\zeta(1-s,1-\gamma) \\
 & + (2^s-1) \big[ \zeta(1-s,\thf+\gamma) - \zeta(1-s,\thf-\gamma) \big] \big] \,, \quad \Re s < 0 \,, \quad 0 < \gamma < \thf \,. \nonumber
\end{align}

%%%%%%%%%%%%%%%%%%%%%%%%%%%%%%%%%%%%%%%%%%%%%

\subsection{The case $\Re s <0$ using the methods of L\"ust and Stieberger}

\label{sec:lustandstieberger}

In this section we revisit the calculation of $\wt\varphi_0(s)$ for $\Re s < 0$ using the powerful techniques developed in \cite{Lust:2003ky}. Starting with the expansion for $|{\Im \nu}| < \Im \tau$ (with $\tau = it$) and Taylor expanding the sine function we have (cf.\ eq.\ \eqref{wwseries})
\be
 \p_\nu \ln \vartheta_1(\nu,\tau) - \pi \cot \pi \nu
 = 4 \pi \sum_{n=1}^\infty \frac{q^n}{1-q^n} \sin 2 \pi n \nu
 = - \sum_{k=1}^\infty \nu^{2k-1} \left[ \frac{2(2\pi i)^{2k}}{(2k-1)!} \sum_{n=1}^\infty \frac{n^{2k-1} q^n}{1-q^n} \right] \,.
\ee
Rearranging the last sum as $\sum_{n=1}^\infty \frac{n^{2k-1} q^n}{1-q^n} = \sum_{n=1}^\infty \sigma_{2k-1}(n) q^n$, where $\sigma_{2k-1}(n)=\sum_{d|n} d^{2k-1}$ is the divisor function, gives the starting point of \cite{Lust:2003ky} (cf.\ their formula (3.7))
\be
 \label{stiebergerrep}
 \p_\nu \ln \vartheta_1(\nu,\tau)
 - \pi \cot \pi \nu = - \sum_{k=1}^\infty \nu^{2k-1} [ G_{2k}(\tau) - 2 \zeta(2k) ] \,.
\ee
Here $G_{2k}(\tau)$ denotes the Eisenstein series, see for example \cite{Apostol}
\be
 G_{2k}(\tau) = 2\zeta(2k) + \frac{2 (2\pi i)^{2k}}{(2k-1)!} \sum_{n=1}^\infty \sigma_{2k-1}(n) q^n \,, \quad k=1,2,\ldots
\ee
For $k=2,3,\ldots$ the $G_{2k}$ are modular forms of weight $2k$ but for $k=1$ the $G_{2}$ transforms anomalously
\be
 G_{2k}(-1/\tau) =
 \begin{cases}
  \tau^{2k} G_{2k}(\tau) & k = 2,3, \ldots \\
  \tau^2 G_2(\tau) - 2 \pi i \tau & k = 1 \,.
 \end{cases}
\ee

Now, modular transforming the integrand $t \to 1/t = \ell$,
and using the series expansion \refeq{stiebergerrep},
the generalized Mellin transform for $\Re s < 0$ and $|\gamma|<1$ takes the form
\be
 \label{annulusmellinintegral}
 \tilde\varphi_0(s;\gamma)
 = \int_0^\infty \dd \ell \, \ell^{-s} \left[ \pi (\coth \pi \gamma \ell - 1) - \sum_{k=1}^\infty (-i)^{2k} \gamma^{2k-1} \ell^{2k-1} \big[ G_{2k}(i \ell) - 2 \zeta(2k) \big] \right] \,.
\ee
The first integral gives (after a change of variables $x = 2 \pi \gamma \ell$ and assuming $\Re s < 0$ and $0 < \gamma < 1$)
\be
 \int_0^\infty \dd \ell \, \ell^{-s} \pi (\coth \pi \gamma \ell - 1) = (2 \pi)^s \gamma^{s-1} \int_0^\infty \dd x \, \frac{x^{-s}}{2} (\coth \frac{x}{2} - 1) = (2\pi)^s \Gamma(1-s) \zeta(1-s) \gamma^{s-1} \,.
\ee
Let us analyze the second integral. Using the modular properties of $G_{2k}$, for $k=1,2,3,\ldots$, we have
\begin{align}
 \ell \to 0 : \quad G_{2k}(i \ell) - 2\zeta(2k) & = 2 \zeta(2k) \ell^{-2k} + \ldots \ ,\\
 \ell \to \infty : \quad G_{2k}(i \ell) - 2 \zeta(2k) & = e^{-2 \pi \ell} + \ldots\ .
\end{align}
Thus each term $\ell^{2k-s-1} [G_{2k}(i\ell) - 2 \zeta(2k) ]$ in the sum can be integrated under the assumption $\Re s<0$. Below we show that the resulting series is absolutely convergent for $\Re s<0$, thus we can safely interchange sum and integral. Notice how the modular transformation $t \to 1/t = \ell$ brought us just the appropriate factor of $\ell^{2k-1}$, for each $k=1,2,3,\ldots$, to make the integral finite for fixed $\Re s <0$. This is to be contrasted with \cite{Lust:2003ky} where a different $\Re s$ would need to be chosen for each term in the sum.

We now show that
\be
 \label{eisensteinintegral}
 \int_0^\infty \dd \ell \, \ell^{2k-s-1} \left[ G_{2k}(i\ell) - 2 \zeta(2k) \right]
 = (2\pi)^{s-2k} \Gamma(2k-s) \widetilde{G}_{2k}(2k-s) \,,
\ee
where $\wt G_{2k}(\varsigma)$ is the Dirichlet series associated with $G_{2k}(\tau)$ defined as, see for example \cite{Apostol}
\be
 \widetilde{G}_{2k}(\varsigma)
 = \frac{2(2\pi i)^{2k}}{(2k-1)!} \sum_{n=1}^\infty \frac{\sigma_{2k-1}(n)}{n^\varsigma} \,, \quad \Re \varsigma > 2k \,.
\ee
Since $\sigma_{2k-1}(n) = \mathscr{O}(n^{2k-1})$ the series converges absolutely for $\Re \varsigma > 2k$ and in fact evaluates to, using Prob. 13 Ch. 6 of \cite{Apostol},
\be \label{dirichletseries}
 \sum_{n=1}^\infty \frac{\sigma_{2k-1}(n)}{n^\varsigma} = \zeta(\varsigma) \zeta(\varsigma - 2k + 1) \,, \quad \Re \varsigma > 2k \,.
\ee
In our case $\varsigma = 2k-s$, so as promised, the sum converges absolutely for $\Re s<0$
\begin{align}
 \int_0^\infty \dd \ell \, \ell^{2k-s-1} \left[ G_{2k}(i\ell) - 2 \zeta(2k) \right]
 & = \frac{2(2\pi i)^{2k}}{(2k-1)!} \sum_{n=1}^\infty \sigma_{2k-1}(n) \int_0^\infty \dd \ell \, \ell^{2k-s-1} e^{-2 \pi n \ell}  \\
 & = (2\pi)^{s-2k} \Gamma(2k-s) \frac{2(2\pi i)^{2k}}{(2k-1)!} \sum_{n=1}^\infty \frac{\sigma_{2k-1}(n)}{n^{2k-s}} \,,
\end{align}
which gives the sought result. Using \eqref{dirichletseries} we have (in agreement with (A.14) of \cite{Lust:2003ky})
\be
 \int_0^\infty \dd \ell \, \ell^{2k-s-1} \left[ G_{2k}(i\ell) - 2 \zeta(2k) \right]
 = (2\pi)^s \zeta(1-s) \frac{2 i^{2k}}{(2k-1)!} \Gamma(2k-s) \zeta(2k-s) \,.
\ee
Thus we conclude for $\Re s < 0$ and $0<\gamma<1$
\be
 \wt \varphi_0(s;\gamma) = (2 \pi)^s \zeta(1-s) \Gamma(1-s) \left[ \gamma^{s-1} - \frac{1}{\Gamma(1-s)}\sum_{k=1}^\infty \frac{2 \gamma^{2k-1}}{(2k-1)!} \Gamma(2k-s) \zeta(2k-s) \right] \,.
\ee
Even though this was derived assuming $\Re s < 0$ the sum converges for all $s \neq 0$ and $|\gamma|<1$ using NIST (\href{http://dlmf.nist.gov/25.11.E10}{25.11.10}), with the result
\be
 \wt \varphi_0(s;\gamma) = (2 \pi)^s \zeta(1-s) \Gamma(1-s) \left[ \gamma^{s-1} + \zeta(1-s,1+\gamma) - \zeta(1-s,1-\gamma)\right] \,.
\ee
Using the recurrence relation for Hurwitz zeta function we conclude
\be
 \wt \varphi_0(s;\gamma) = (2 \pi)^s \zeta(1-s) \Gamma(1-s) \left[ \zeta(1-s,\gamma) - \zeta(1-s,1-\gamma)\right] \,, \quad
 \Re s < 0 \,, \quad 0 < \gamma < 1 \,.
\ee

As a consistency check we notice how the result $\wt \varphi_0(s;\gamma)$ vanishes at $\gamma = \hf$, just as the integrand $\varphi_0(t,\gamma)$ and the counter terms $a_0(\gamma) = \pi(1-2\gamma)$, $b_0(\gamma)=\pi \coth \pi \gamma$ do.

%%%%%%%%%%%%%%%%%%%%%%%%%%%%%%%%%%%%%%%%%%%%%%%%%%%%%%%%
\section{Direct computation of the $t$ integral}
\label{sec:direct} 

In this appendix we give a direct version of computing the integral \eqref{singularmodintegral}, that produces the same results as in \refsec{tint} using the generalized Mellin transform. 

\subsection{Annuls and Klein bottle}

We first compute the following $t$ integral for the annulus:
\be
I=\int_{0}^{\infty} \frac{dt}{t^2} \, \frac{\tht_1'(\gamma,i t/2)}{\tht_1(\gamma,i t/2)}\ .
\ee
By modular transformation of the Jacobi theta function, it is easy to see that
\be \label{modul}
\frac{\tht_1'(\gamma,i t/2)}{\tht_1(\gamma,i t/2)}=-4 \pi \gamma l - 
2 i l \frac{\tht_1'(- 2 i \gamma l, 2 i l)}{\tht_1(-2 i \gamma l,2 i l)}\ ,
\ee
where $ l \equiv \frac{1}{t}$. 
Using the representation 
\be \label{abram}
\frac{\tht_1'(z)}{\tht_1(z)} &=&  \pi \cot \pi z + 4 \pi \sum_{n=1}^{\infty} \frac{q^n}{1-q^n} \sin 2 \pi n z \\
&=& \pi \cot \pi z + 4 \pi \sum_{n,m=1}^{\infty} q^{n m} \sin 2 \pi n z,
\ee 
we arrive at
\be \label{treeA}
I=2 \pi \int_{0}^{\infty} dl\,l\,\left(- 2\gamma+\coth(2\pi \gamma l)
-4 \sum_{n,m=1}^{\infty} e^{-4 \pi l n m} \sinh(4 \pi n \gamma l)\right)\ .
\ee
Clearly there is no IR divergence (i.e.\ at $l=0$). 
The first two terms in the bracket do contribute a UV divergence as $l \to \infty$ 
(so one should cut off the $l$-integral at an upper bound $\Lambda$), 
whereas the last one does not (so one can interchange
the order of summation and integration).
Let us first focus on the last term:
\be
I_1&=&-8 \pi \sum_{n,m=1}^{\infty} \int_{0}^{\infty} dl\,l\, e^{-4 \pi l n m} \sinh(4 \pi n \gamma l)
=- \pi \sum_{n,m=1}^{\infty} \frac{\gamma m}{(\gamma^2-m^2)^2 n^2 \pi^2} \nonumber \\
&&= \frac{\pi}{24}[\psi'(1+\gamma)-\psi'(1-\gamma)] 
= \frac{\pi}{24}\left[\psi'(\gamma)-\psi'(1-\gamma)-\frac{1}{\gamma^2}\right].
\ee
Here $\psi'(x)$ denotes the trigamma function and in the last line we used $\psi'(1+\gamma)=\psi'(\gamma)-1/\gamma^2$. 

Now let us look at the first and second term in \eqref{treeA}:
\be
I_2=2 \pi \int_{0}^{\Lambda} dl \,(-2 \gamma l) = - 2 \pi \gamma \Lambda^2\ ,
\ee 
\be
I_3=2\pi \int_{0}^{\Lambda} dl \,l\,\coth(2\pi \gamma l)
&=&\frac{\pi}{24 \gamma^2}+\pi \Lambda^2+\frac{\Lambda \log(1-e^{-4\gamma\Lambda\pi})}{\gamma}
-\frac{Li_2(e^{-4\gamma\Lambda\pi})}{4 \gamma^2 \pi} \nonumber \\
&\stackrel{\Lambda\to\infty}{=}& \frac{\pi}{24 \gamma^2}+\pi \Lambda^2.
\ee
Here $Li_2(z)$ is a  polylogarithm function. In the second equality it has been used that the third and last term go to zero as $\Lambda\to\infty$. 

In total ($I=I_1+I_2+I_3$) we have
\be 
I_\ca(\gamma)=\int_{1\over \Lambda}^\infty \frac{dt}{t^2}  \, \frac{\tht_1'(\gamma,i t/2)}{\tht_1(\gamma,i t/2)}=
\pi(1-2\gamma) \Lambda^2 + \frac{\pi}{24} \left[ \psi'(\gamma) - \psi'(1-\gamma) \right] .
\ee

In the same way we get for the Klein bottle 
\be
I_\ck(\gamma)=\int_{1\over 4 \Lambda}^\infty \frac{dt}{t^2}  \, \frac{\tht_1'(\gamma,2 i t)}{\tht_1(\gamma,2 i t)}=4\pi(1-2\gamma) \Lambda^2 + \frac{\pi}{6} \left[ \psi'(\gamma) - \psi'(1-\gamma) \right].
\ee

\subsection{$t$-integral for M\"obius}  
\label{sec:tintM}
Now let us consider the  M\"obius integral, i.e. 
\be
I=\int_{0}^{\infty} \frac{dt}{t^2} \, \frac{\tht_1'(\gamma,\tau_{\cm})}{\tht_1(\gamma, \tau_{\cm})}\ .
\ee
Here $\tau_{\cm}=\frac{i t}{2}+\frac{1}{2}$.
Then we perform the sequence $ST^2S$ of modular transformations: 
\be
\tau_{\cm}= {i t  \over 2} + \frac{1}{2} \;\to\; -\frac{1}{\tau_{\cm}} \;\to\; 
- \frac{1}{\tau_{\cm}} +2  \;\to\; \left(\frac{1}{\tau_{\cm}} - 2  \right)^{-1} =  2 i l - \frac{1}{2}
=: l_{\cm} \ .
\ee 
Here $l=\frac{1}{4 t}$. The $S$ transformation reads
\be
\frac{\tht_1'(\gamma,\tau)}{\tht_1(\gamma,\tau)}=-\frac{2 \pi \gamma i}{\tau} + 
\frac{1}{\tau} \frac{\tht_1'(\gamma/\tau, -1/\tau)}{\tht_1(\gamma/\tau, -1/\tau)}\ .
\ee
The result of $ST^2S$ modular transformations is
\be
\frac{\tht_1'(\gamma,\tau_{\cm})}{\tht_1(\gamma, \tau_{\cm})} \stackrel{l=\frac{1}{4t}}{=} -16 \pi \gamma l -
4 i l \frac{\tht_1'( - 4 i \gamma l, 2 i l - \frac{1}{2})}{\tht_1(- 4 i \gamma l, 2 i l - \frac{1}{2})}\ .
\ee 
Using the representation   
\be
\frac{\tht_1'(z)}{\tht_1(z)} &=&  \pi \cot \pi z + 4 \pi \sum_{n=1}^{\infty} \frac{q^n}{1-q^n} \sin 2 \pi n z \\
&=& \pi \cot \pi z + 4 \pi \sum_{n,m=1}^{\infty} q^{n m} \sin 2 \pi n z\ ,
\ee 
we get (assuming $\gamma<1/2$)  
\be \label{treeM}
I=16 \pi \int_{0}^{\infty} dl\,l\,\left(- 4\gamma+\coth(4 \pi \gamma l)
-4 \sum_{n,m=1}^{\infty} (-1)^{n m}e^{-4 \pi l n m} \sinh(8 \pi n \gamma l)\right).
\ee
Clearly there is no IR divergence (i.e.\ at $l=0$). 
The first two terms in the bracket contribute a UV divergnce as $l \to \infty$ 
(requiring a cut off $\Lambda$ on $l$ again as above), 
whereas the last one doesn't (so one can interchange
the order of summation and integration).

Let us look at the first and second term:
\be
I_2=16 \pi \int_{0}^{\Lambda} dl \,(-4 \gamma l) = - 32 \pi \gamma \Lambda^2\ ,
\ee 
\be
I_3=16 \pi \int_{0}^{\Lambda} dl \,l\,\coth(4 \pi \gamma l)
&=&\frac{\pi}{12 \gamma^2}+ 8 \pi \Lambda^2+4 \frac{\Lambda \log(1-e^{-8\gamma\Lambda\pi})}{\gamma}
-\frac{Li_2(e^{-8\gamma\Lambda\pi})}{2 \gamma^2 \pi} \nonumber \\
&\stackrel{\Lambda\to\infty}{=}& \frac{\pi}{12 \gamma^2}+8 \pi \Lambda^2.
\ee
In the second equality it has been used that the third and fourth term go to zero as $\Lambda\to\infty$ assuming $\gamma>0$.

Now let us look at the last term and we get  
\be
I_1&=&-64 \pi \sum_{n,m=1}^{\infty} \int_{0}^{\infty} dl\,l\, (-1)^{n m} e^{-4 \pi l n m} \sinh(8 \pi n \gamma l)
=- \sum_{n,m=1}^{\infty} (-1)^{n m} \frac{16 \gamma m}{(4\gamma^2-m^2)^2 n^2 \pi} \nonumber \\
&=& - \sum_{m=1}^{\infty} \frac{16 \gamma m \, Li_2\left((-1)^m \right)}{(4\gamma^2-m^2)^2 \pi}\ .
\ee
Note that the integral converges provided that $2 |\gamma|\leq m$. Now we split the sum into sums over even and odd $m$:
\be
I_1&=&- \sum_{k=1}^{\infty} \left[\frac{16 \gamma (2k) \, Li_2 (1)}{(4\gamma^2-(2k)^2)^2 \pi}\right]
- \sum_{k=0}^{\infty} \left[\frac{16 \gamma (2k+1) \, Li_2 (-1)}{(4\gamma^2-(2k+1)^2)^2 \pi}\right] \nonumber \\
&=& \frac{\pi}{12}[\psi'(1+\gamma)-\psi'(1-\gamma)]  \nonumber \\
&& -\frac{\pi}{48 \gamma}\left[(1-2 \gamma)\,\psi'\left(\frac{1}{2}-\gamma\right)+ (1+2 \gamma)\,\psi'\left(\frac{1}{2}+\gamma\right)- \frac{\pi^2}{\cos^2(\pi\gamma)}\right] \nonumber \\
&=&\frac{\pi}{12}\left[\psi'(\gamma)-\psi'(1-\gamma)-\frac{1}{\gamma^2}- \frac{1}{2} \psi'\left(\frac{1}{2}+\gamma\right) 
+ \frac{1}{2} \psi'\left(\frac{1}{2}-\gamma\right) \right] \ .
\ee
In the third equality  we used  $\psi'(1+\gamma)=\psi'(\gamma)-1/\gamma^2$ and 
$\psi'(1-z)=-\psi'(z)+\pi^2/ \sin^2(\pi z)$ with $z=\frac{1}{2}+\gamma$. 

Altogether 
\be
I=I_1+I_2+I_3= 8\pi(1-4\gamma) \Lambda^2 + \frac{\pi}{12} \left[ \psi'(\gamma) - \psi'(1-\gamma)- \thf \psi'(\thf+\gamma) + \thf \psi'(\thf-\gamma) \right] \ .
\ee
Note that the above result is valid for $0<\gamma<\frac{1}{2}$.

Thus we have  
% The calculation for M\"obius  is similar and performed in \ref{sec:tintM}, with the result
\be
I_{\M}(\gamma)= \int_{1\over 4 \Lambda}^\infty \frac{dt}{t^2}  \, \frac{\tht_1'(\gamma, \frac{i t}{2} +\frac{1}{2})}{\tht_1(\gamma,\frac{i t}{2} +\frac{1}{2})} 
=8\pi(1-4\gamma) \Lambda^2 + \frac{\pi}{12} \left[ \psi'(\gamma) - \psi'(1-\gamma)- \thf \psi'(\thf+\gamma) + \thf \psi'(\thf-\gamma) \right] . \non
\ee

%%%%%%%%%%%%%%%%%%%%%%%%%%%%%%%%%%%%%%%%%%%%%%%%%%%%%%%%

\section{Tadpoles}
\label{sec:tadpoles}
In $\mathbb{Z}_6'$ the tadpoles arise from the vacuum amplitudes summarized in \reftab{tab:tadpoles}, cf.\ \cite{Aldazabal:1998mr}.

\begin{table}[h]
\renewcommand{\arraystretch}{1.5}
 \[
\begin{array}{llll}
\toprule
\textbf{SUSY} & \textbf{Amplitudes} & \textbf{Volume factors} & \textbf{Conditions}  \\
\midrule
{\cal N}=4 & \ck_{\Theta^0}^{(0)} + \cm_9^{(0)} + \ca_{99}^{(0)} & V_4 V_1 V_2 V_3 & \gamma^T_{\Omega,9}=\gamma_{\Omega, 9}, n_9=32 \\
 &  \ck_{\Theta^0}^{(3)} + \cm_5^{(3)} + \ca_{55}^{(0)} &  V_4 V_3 / V_1 V_2 &  \gamma^T_{\Omega_3,5}=\gamma_{\Omega_3, 5}, n_5=32\\  
 & & &  \gamma^T_{\Omega_3,9}=-\gamma_{\Omega_3, 9}, \gamma_9^{(6)} = -1 = \gamma_5^{(6)} \\
\midrule
{\cal N}=2 & \ca_{99}^{(3)} + \ca_{55}^{(3)} + \ca_{59}^{(3)} & V_4 V_3 & {\rm Tr} \, \gamma_{9}^{(3)} = 0 = {\rm Tr} \, \gamma_{5}^{(3)}\\
 & \ck_{\Theta^0}^{(4)} + \cm_9^{(4)} + \ca_{99}^{(2)} & V_4 V_2 & {\rm Tr} \, \gamma_{9}^{(2)}=-8 \\
 & \ck_{\Theta^0}^{(2)} + \cm_9^{(2)} + \ca_{99}^{(4)}& V_4 V_2 & {\rm Tr} \, \gamma_{9}^{(4)}=8 \\
 &  \ck_{\Theta^0}^{(1)} + \cm_5^{(1)} + \ca_{55}^{(2)}  & V_4 / V_2 &  {\rm Tr} \, \gamma_{5}^{(2)}=-8\\ 
 &  \ck_{\Theta^0}^{(5)} + \cm_5^{(5)} + \ca_{55}^{(4)}  & V_4 / V_2 &  {\rm Tr} \, \gamma_{5}^{(4)}=8\\
\midrule
{\cal N}=1 & \ca_{99}^{(k)} + \ca_{55}^{(k)} + \ca_{59}^{(k)} , k = 1,5 & V_4 & {\rm Tr} \, \gamma_{9}^{(k)} = 0 = {\rm Tr} \gamma_{5}^{(k)}, k=1,5 \\
\bottomrule
\end{array}
\]
\caption{Tadpoles when all D5-branes are at the origin. $V_4$ stands for a regularizing volume of the external space-time (which one would have to take to infinity for a non-compact space-time) and $V_1-V_3$ stand for the volumes to the three internal complex dimensions.}
\label{tab:tadpoles}
\end{table}

In this appendix we verify those $\cn = 2$ tadpoles that involve the Klein bottle, in order to determine the relative sign between annulus and M\"obius strip that we will also need for the $\cn = 1$ sectors of the two-point function discussed in the main text. One word of warning: Compared to sec.\ \refsec{sec:addingup}, we use a slightly different notation here in that we do not separate CP factors from the amplitudes and also include the overall factor $(4 \pi^2 \alpha' t)^{-2}$. 

%%%%%%%%%%%%%%%%%%%%%%%%%%%%%%%%%%%%%%%%%%%%%%%%%%%%%%%%%%%%%%%%%%%%%%%

\subsection{Annulus amplitudes}
For $k=2,4$ we have \cite{Antoniadis:1999ge}%
\be \label{a99_startingpoint}
\ca_{99}^{(k=2,4)} = \frac{1}{(4 \pi^2 \alpha' t)^2} \sum_{\alpha, \beta} \eta_{\alpha \beta} \frac{\thba{\alpha}{\beta}}{\eta^3}  \frac{\thba{\alpha}{\beta + k v_2}}{\eta^3} \prod_{i=1,3} (2 \sin (\pi k v_i))\frac{\thba{\alpha}{\beta+k v_i}}{\thba{\frac{1}{2}}{\frac{1}{2}+k v_i}} \Big({\rm tr} \gamma_9^{(k)}\Big)^2 \tht\ba{\vec 0}{\vec 0}(0,it (G^{(2)})^{-1})\ ,
\ee
where $G^{(2)}$ is the metric on the second torus. Now we use \eqref{modtransmoment}, \eqref{modtransannulus}, $t^{-1} = \ell$ (i.e.\ $-1/\tau = 2 i \ell$) and $v_1 + v_2 +v_3 = 0$ to obtain
\be \label{a99_1}
\ca_{99}^{(k=2,4)} &=& \frac{\sqrt{G^{(2)}} \ell}{(4 \pi^2 \alpha')^2}  \sin (\pi k v_1) \sin (\pi k v_3) \Big({\rm tr} \gamma_9^{(k)}\Big)^2 \tht\ba{\vec 0}{\vec 0}(0,i \ell G^{(2)}) \non
&& \sum_{\alpha, \beta} \eta_{\alpha \beta} \frac{e^{8 \pi i \alpha \beta}}{e^{\pi i (1 - k v_2)}} \frac{\thba{\beta}{\alpha}(2 i \ell) \prod_{i=1}^3 \thba{\beta - k v_i}{\alpha}(2 i \ell)}{\eta^6(2i\ell)  \thba{1/2 - k v_1}{1/2}(2 i \ell) \thba{1/2 - k v_3}{1/2}(2 i \ell)} \ ,
\ee
where we also used that \eqref{shift} implies
\be
\thba{-\beta - k v_i}{\alpha} = \thba{\beta - k v_i}{\alpha} \ , \quad {\rm for}\ \beta = 0, 1/2\ .
\ee
For $\alpha, \beta \in \{ 0, 1/2 \}$ one has $e^{8 \pi i \alpha \beta} = 1$. Now we would like to obtain the contribution to the tadpole by expanding \eqref{a99_1} for large $\ell$, using \eqref{shift} and \eqref{limits}. Moreover, we facilitate the extraction of the tadpole by only looking at the R-sector, as the NS-sector is related to the R-sector by supersymmetry. For the different worldsheets the R-sector is given by
\be \label{rsector}
({\rm R-sector}): \quad  \ca, \ck \rightarrow (\alpha, \beta) = (0, 1/2) \quad , \quad \cm  \rightarrow (\alpha, \beta) = (1/2, 0)\ .
\ee
Thus we specialize to $(\alpha, \beta) = (0, 1/2)$ in \eqref{a99_1} and use 
\be
 \eta_{\alpha \beta} = (-1)^{2 (\alpha + \beta + 2 \alpha \beta)}\quad , \quad (v_1, v_2, v_3) = (1/6, -1/2, 1/3)
\ee
and the notation 
\be \label{qtilde}
\tilde q = e^{-4 \pi \ell}
\ee
to obtain 
\be \label{a99_2}
\ca_{99}^{(2)} & \stackrel{\ell \rightarrow \infty}{\longrightarrow}& - \frac{\sqrt{G^{(2)}} \ell}{(4 \pi^2 \alpha')^2}  \sin (\pi/3) \sin (2 \pi/3) \Big({\rm tr} \gamma_9^{(2)}\Big)^2 \frac{1}{e^{2 \pi i}} \ \frac{2 \tq^{1/8} \tq^{1/72} 2 \tq^{1/8} \tq^{1/72}}{\tq^{1/4}  \tq^{1/72} e^{\pi i/6} \tq^{1/72} e^{-\pi i/6}} \non
&  & = - \frac{\sqrt{G^{(2)}} \ell}{(4 \pi^2 \alpha')^2}  \frac34  \Big({\rm tr} \gamma_9^{(2)}\Big)^2 4  =  - 3 \frac{\sqrt{G^{(2)}} \ell}{(4 \pi^2 \alpha')^2} \Big({\rm tr} \gamma_9^{(2)}\Big)^2 \ , \non
\ca_{99}^{(4)} & \stackrel{\ell \rightarrow \infty}{\longrightarrow}& - \frac{\sqrt{G^{(2)}} \ell}{(4 \pi^2 \alpha')^2} (-\frac34) \Big({\rm tr} \gamma_9^{(4)}\Big)^2 (-4) =   - 3 \frac{\sqrt{G^{(2)}} \ell}{(4 \pi^2 \alpha')^2} \Big({\rm tr} \gamma_9^{(4)}\Big)^2\ .
\ee

The 55-annulus works very similarly. The starting point, eq.\ \eqref{a99_startingpoint}, is almost the same as it does not matter for the oscillator contributions whether the boundary conditions are NN or DD. The only difference is that the momentum modes are replaced by winding modes, i.e.\ $\sqrt{G^{(2)}}$ is replaced by $\sqrt{G^{(2)}}^{-1}$. Thus
\be \label{a55}
\ca_{55}^{(2)} & \stackrel{\ell \rightarrow \infty}{\longrightarrow}& - 3 \frac{\ell}{(4 \pi^2 \alpha')^2 \sqrt{G^{(2)}}} \Big({\rm tr} \gamma_5^{(2)}\Big)^2 \ , \non
\ca_{55}^{(4)} & \stackrel{\ell \rightarrow \infty}{\longrightarrow}& - 3 \frac{\ell}{(4 \pi^2 \alpha')^2 \sqrt{G^{(2)}}} \Big({\rm tr} \gamma_5^{(4)}\Big)^2\ .
\ee
%

%%%%%%%%%%%%%%%%%%%%%%%%%%%%%%%%%%%%%%%%%%%%%%%%%%%%%%%%%%%%%%%%%%%%%%%

\subsection{M\"obius amplitudes}

For $k=2,4$ we have \cite{Antoniadis:1999ge}
\be \label{m9_startingpoint}
\cm_{9}^{(k=2,4)} = - \frac{1}{(4 \pi^2 \alpha' t)^2} \sum_{\alpha, \beta} \eta_{\alpha \beta} \frac{\thba{\alpha}{\beta}}{\eta^3}  \frac{\thba{\alpha}{\beta + k v_2}}{\eta^3} \prod_{i=1,3} (2 \sin (\pi k v_i))\frac{\thba{\alpha}{\beta+k v_i}}{\thba{\frac{1}{2}}{\frac{1}{2}+k v_i}} {\rm tr} \gamma_9^{(2k)} \tht\ba{\vec 0}{\vec 0}(0,it (G^{(2)})^{-1})\ .
\ee
This looks formaly very similar to \eqref{a99_startingpoint} (up to the overall sign and the CP-trace), but one has to remember that the theta functions in \eqref{a99_startingpoint} depend on $\tau_\ca = i t /2$ whereas those in \eqref{m9_startingpoint} depend on $\tau_\cm = 1/2 + i t /2$.

Similar to above, we use \eqref{modtransmoment}, \eqref{modtransmoebius}, $t^{-1} = 4 \ell$ (i.e.\ $\frac{\tau}{1 - 2 \tau} = 2 i \ell - \frac12 \equiv  i \ell_\cm$) and $v_1 + v_2 +v_3 = 0$ to obtain
\be \label{m9_1}
\cm_{9}^{(k=2,4)} &=& \frac{16 \sqrt{G^{(2)}} \ell}{(4 \pi^2 \alpha')^2}  \sin (\pi k v_1) \sin (\pi k v_3)  {\rm tr} \gamma_9^{(2k)} \tht\ba{\vec 0}{\vec 0}(0,4 i \ell G^{(2)}) e^{-2 \pi i k v_2 (1 + k v_2)} \non
&& \sum_{\alpha, \beta} \eta_{\alpha \beta} \frac{\thba{\alpha}{\beta}(i \ell_\cm) \prod_{i=1}^3 \thba{\alpha + 2 k v_i}{\beta + k v_i}(i \ell_\cm)}{\eta^6(i\ell_\cm)  \thba{1/2 + 2 k v_1}{1/2 + k v_1}(i \ell_\cm) \thba{1/2 + 2 k v_3}{1/2 + k v_3}(i \ell_\cm)} \ .
\ee
Specializing to $(\alpha, \beta) = (1/2, 0)$ (cf.\ \eqref{rsector}), we get
\be \label{m9_2}
\cm_{9}^{(2)} & \stackrel{\ell \rightarrow \infty}{\longrightarrow}& -\frac{16 \sqrt{G^{(2)}} \ell}{(4 \pi^2 \alpha')^2}  \sin (\pi/3) \sin (2 \pi/3) {\rm tr} \gamma_9^{(4)} \frac{2 (- \tq)^{1/8} (-\tq)^{1/72} e^{\pi i / 9} 2 (-\tq)^{1/8} \cos(\pi) (-\tq)^{1/72} e^{-2 \pi i / 9}}{(-\tq)^{1/4} (-\tq)^{1/72} e^{5\pi i/18} (-\tq)^{1/72} e^{-7 \pi i/18}} \non
&  & = -\frac{16 \sqrt{G^{(2)}} \ell}{(4 \pi^2 \alpha')^2}  \frac34   {\rm tr} \gamma_9^{(4)} (-4)  =  3 \frac{16 \sqrt{G^{(2)}} \ell}{(4 \pi^2 \alpha')^2} {\rm tr} \gamma_9^{(4)} \ , \non
\cm_{9}^{(4)} & \stackrel{\ell \rightarrow \infty}{\longrightarrow}& - \frac{16 \sqrt{G^{(2)}} \ell}{(4 \pi^2 \alpha')^2} (-\frac34) {\rm tr} \gamma_9^{(8)} 4 =   - 3 \frac{16 \sqrt{G^{(2)}} \ell}{(4 \pi^2 \alpha')^2} {\rm tr} \gamma_9^{(2)}\ ,
\ee
where we used $\gamma_9^{(6)}= -1$ in the last step.

The $\cm_5$ amplitudes are related to the $\cm_9$ amplitudes in a simple manner. The D5-branes are wrapped around the third torus and, thus, the 5-5 strings have DD-boundary conditions along the first and second torus. This leads to an extra minus sign (relative to $\cm_9$) in the action of $\Omega$ on the oscillators in those directions, cf.\ (7.3.5) and (7.3.6) of \cite{Kiritsis:2007zza}. This extra minus sign leads to the fact that the lower characteristic in the contribution to the partition function of the oscillators from the first and second torus is shifted by $+1/2$ or $-1/2$, respectively, relative to the corresponding $\cm_9$-amplitude.\footnote{Which one is shifted by $+1/2$ and which one by $-1/2$ is a matter of convention, but the shifts have to have opposite sign so that shifting does not introduce an unwanted relative sign between the $\alpha=0$ and $\alpha=1/2$ terms, cf.\ eq.\ (112) in \cite{Angelantonj:2002ct}.} Thus, we have for $k=1,5$ (cf.\ (B.3) in \cite{Antoniadis:1999ge} with the above modification)
\be \label{m5_startingpoint}
\cm_{5}^{(k=1,5)} &=& \frac{1}{(4 \pi^2 \alpha' t)^2} \sum_{\alpha, \beta} \eta_{\alpha \beta} \frac{\thba{\alpha}{\beta}}{\eta^3} \Big(-2 \sin (\pi (k v_1+\tfrac12))\Big)\frac{\thba{\alpha}{\beta+k v_1+\frac12}}{\thba{\frac{1}{2}}{\frac{1}{2}+k v_1+\frac12}} 
\Big(-2 \sin (\pi (k v_2-\tfrac12))\Big)\frac{\thba{\alpha}{\beta+k v_2-\frac12}}{\thba{\frac{1}{2}}{\frac{1}{2}+k v_2-\frac12}} \non
&& \Big(-2 \sin (\pi k v_3)\Big)\frac{\thba{\alpha}{\beta+k v_3}}{\thba{\frac{1}{2}}{\frac{1}{2}+k v_3}} 
{\rm tr} \gamma_5^{(2k)} \tht\ba{\vec 0}{\vec 0}(0,it G^{(2)})\ ,
\ee
where we used that the winding mode sum survives in the $k=1,5$ sectors of $\cm_5$ (cf.\ (9.14.48) in \cite{Kiritsis:2007zza}) and the different overall sign compared to $\cm_9$ arises from \cite{Aldazabal:1998mr}
\be
\tr(\gamma_{\Omega_{k+3}, 9}^{-1} \gamma_{\Omega_{k+3}, 9}^{T}) = \tr \gamma_9^{(2k)} \quad , \quad \tr(\gamma_{\Omega_k, 5}^{-1} \gamma_{\Omega_k, 5}^{T}) = - \tr \gamma_5^{(2k)}\ .
\ee
Note that \eqref{m5_startingpoint} is not well defined for $k=1,5$ as it is written, given that $v_2=-1/2$. Rather we have to take the limit
\be
k=1,5: \quad \lim_{\epsilon \rightarrow 0} \Big(-2 \sin (\pi (kv_2-\tfrac12+\epsilon))\Big)\frac{1}{\thba{\frac{1}{2}}{\frac{1}{2}+k v_2-\frac12+\epsilon}} &=&  \lim_{\epsilon \rightarrow 0} \Big(-2 \sin (\pi (-1+\epsilon))\Big)\frac{1}{\thba{\frac{1}{2}}{-\frac{1}{2}+\epsilon}} \non
& = &  \lim_{\epsilon \rightarrow 0} \Big(2 \sin (\pi (-1+\epsilon))\Big)\frac{1}{\thba{\frac{1}{2}}{\frac{1}{2}+\epsilon}} \non
& = &  \lim_{\epsilon \rightarrow 0} \Big(-2 \sin (\pi \epsilon)\Big)\frac{1}{\thba{\frac{1}{2}}{\frac{1}{2}+\epsilon}} = \frac{1}{\eta^3}
\ee
Moreover, using $\sin(a+\pi/2)=\cos(a)$, we end up with
\be \label{m5_startingpoint_2}
\cm_{5}^{(k=1,5)} &=& \frac{1}{(4 \pi^2 \alpha' t)^2} \sum_{\alpha, \beta} \eta_{\alpha \beta} \frac{\thba{\alpha}{\beta}}{\eta^3} \Big(-2 \cos (\pi k v_1)\Big)\frac{\thba{\alpha}{\beta+k v_1+\frac12}}{\thba{\frac{1}{2}}{\frac{1}{2}+k v_1+\frac12}} 
\frac{\thba{\alpha}{\beta+k v_2-\frac12}}{\eta^3} \non
&& \Big(-2 \sin (\pi k v_3)\Big)\frac{\thba{\alpha}{\beta+k v_3}}{\thba{\frac{1}{2}}{\frac{1}{2}+k v_3}} 
{\rm tr} \gamma_5^{(2k)} \tht\ba{\vec 0}{\vec 0}(0,it G^{(2)})\ .
\ee
This leads to
\be \label{m9_2_2}
\cm_{5}^{(1)} & \stackrel{\ell \rightarrow \infty}{\longrightarrow}& - 3 \frac{16 \ell}{(4 \pi^2 \alpha')^2  \sqrt{G^{(2)}}} {\rm tr} \gamma_5^{(2)}\ , \non
\cm_{5}^{(5)} & \stackrel{\ell \rightarrow \infty}{\longrightarrow}&  3 \frac{16 \ell}{(4 \pi^2 \alpha')^2 \sqrt{G^{(2)}}} {\rm tr} \gamma_5^{(4)}\ .
\ee
Together with the results for annulus and Klein bottle, these results reproduce the tadpole conditions in \reftab{tab:tadpoles}, which indicates that our rule to obtain $\cm_5$ from $\cm_9$ is correct. 

Before proceeding with the Klein bottle, we would like to use this prescription to determine the form of $\cm_5^{(k=2,4)}$ which will be needed for the $\cn = 1$ amplitudes. Note that \eqref{m5_startingpoint} would still hold for $k=2,4$ and, thus we obtain (using also $\sin(a-\pi/2)=-\cos(a)$)
\be \label{m5_k24}
\cm_{5}^{(k=2,4)} &=& -\frac{1}{(4 \pi^2 \alpha' t)^2} \sum_{\alpha, \beta} \eta_{\alpha \beta} \frac{\thba{\alpha}{\beta}}{\eta^3} \Big(2 \cos (\pi k v_1)\Big)\frac{\thba{\alpha}{\beta+k v_1+\frac12}}{\thba{\frac{1}{2}}{\frac{1}{2}+k v_1+\frac12}} 
\Big(2 \cos (\pi k v_2)\Big)\frac{\thba{\alpha}{\beta+k v_2-\frac12}}{\thba{\frac{1}{2}}{\frac{1}{2}+k v_2-\frac12}} \non
&& \Big(-2 \sin (\pi k v_3)\Big)\frac{\thba{\alpha}{\beta+k v_3}}{\thba{\frac{1}{2}}{\frac{1}{2}+k v_3}} 
{\rm tr} \gamma_5^{(2k)}\ ,
\ee
which confirms the factor $\tilde \chi=-1$ for both $\cm_9$ and $\cm_5$ in the main text. 

%%%%%%%%%%%%%%%%%%%%%%%%%%%%%%%%%%%%%%%%%%%%%%%%%%%%%%%%%%%%%%%%%%%%%%%

\subsection{Klein bottle amplitudes}

The Klein bottle amplitude can be read off from (7.3) in \cite{Aldazabal:1998mr} and it reads
\be \label{kb_startingpoint}
\ck_{\Theta^0}^{(k=2,4)} = \frac{1}{(4 \pi^2 \alpha' t)^2} \sum_{\alpha, \beta} \eta_{\alpha \beta} \frac{\thba{\alpha}{\beta}}{\eta^3}  \frac{\thba{\alpha}{\beta + 2 k v_2}}{\eta^3} \prod_{i=1,3} (2 \sin (2 \pi k v_i))\frac{\thba{\alpha}{\beta+2 k v_i}}{\thba{\frac{1}{2}}{\frac{1}{2}+2 k v_i}} \tht\ba{\vec 0}{\vec 0}(0,it (G^{(2)})^{-1})
\ee
and the same for $k=1,5$ with the KK-momentum sum replaced by a winding sum, cf.\ the discussion in sec.\ 9.14.2 of \cite{Kiritsis:2007zza}. Here, $\tau_\ck = 2 i t$. Now we use \eqref{modtransmoment}, \eqref{modtransannulus}, $t^{-1} = 4 \ell$ (i.e.\ $-1/\tau = 2 i \ell$) and $v_1 + v_2 +v_3 = 0$ to obtain
\be \label{kb_1}
\ck_{\Theta^0}^{(k=2,4)} &=& \frac{64 \sqrt{G^{(2)}} \ell}{(4 \pi^2 \alpha')^2} \sin (2 \pi k v_1) \sin (2 \pi k v_3) \tht\ba{\vec 0}{\vec 0}(0,4 i \ell G^{(2)}) \non
&& \sum_{\alpha, \beta} \eta_{\alpha \beta} \frac{1}{e^{\pi i (1 - 2 k v_2)}} \frac{\thba{\beta}{\alpha}(2 i \ell) \prod_{i=1}^3 \thba{\beta - 2 k v_i}{\alpha}(2 i \ell)}{\eta^6(2i\ell)  \thba{1/2 - 2 k v_1}{1/2}(2 i \ell) \thba{1/2 - 2 k v_3}{1/2}(2 i \ell)} \ .
\ee
Specializing to $(\alpha, \beta) = (0, 1/2)$ (cf.\ \eqref{rsector}), we get
\be \label{kb_2}
\ck_{\Theta^0}^{(k=2,4)} & \stackrel{\ell \rightarrow \infty}{\longrightarrow}& -\frac{64 \sqrt{G^{(2)}} \ell}{(4 \pi^2 \alpha')^2} (-\frac34) (-4) =  - 3 \frac{64 \sqrt{G^{(2)}} \ell}{(4 \pi^2 \alpha')^2} \ ,
\ee
and similarly
\be \label{kb_2_2}
\ck_{\Theta^0}^{(k=1,5)} & \stackrel{\ell \rightarrow \infty}{\longrightarrow}& -\frac{64 \ell}{(4 \pi^2 \alpha')^2\sqrt{G^{(2)}}} \frac34 4 =  - 3 \frac{64 \ell}{(4 \pi^2 \alpha')^2 \sqrt{G^{(2)}}} \ .
\ee

\end{document}